\documentclass[structabstract]{aa}  

\usepackage{graphicx}
\usepackage{txfonts}
\usepackage{natbib}
  \def\ProDiMo{{\sc ProDiMo\ }}
  
  \def\abl#1#2{\frac {d #1}{d #2}}
  \def\pabl#1#2{\frac {\partial #1}{\partial #2}}
  \def\Td{T_{\hspace*{-0.2ex}\rm d}}
  \def\Tg{T_{\hspace*{-0.2ex}\rm g}}
  \def\cT2{c_T^2}
  \def\pabl#1#2{\frac{\partial #1}{\partial #2}}
  \def\Rin{R_{\rm in}}
  \def\Rout{R_{\rm out}}
  \def\Nel{N_{\rm el}}
  \def\Nsp{N_{\rm sp}}
  \def\HH{{\rm\langle H\rangle}}
  \def\nH{n_{\HH}}
  \def\etal{${\rm \hspace*{0.8ex}et\hspace*{0.7ex}al.\hspace*{0.7ex}}$}
  \def\plus{${\rm \hspace*{0.7ex}\&\hspace*{0.7ex}}$}
  \def\ie{i.\,e.\ }
  \def\eg{e.\,g.\ }
  \def\nn{\vec{n}}
  
  \def\r0{\vec{r}_0}
  \def\kabs{\kappa_\nu^{\rm abs}}
  \def\ksca{\kappa_\nu^{\rm sca}}
  \def\kext{\kappa_\nu^{\rm ext}}
  \def\amin{{a_{\rm min}}}
  \def\amax{{a_{\rm max}}}
  \def\apow{{a_{\rm pow}}}
  \def\fconst{f_{\rm const}}
  \def\Icont{I^{\rm cont}}
  \def\Jcont{J^{\rm cont}}
  \def\Jquer{\overline{J}}
  \def\pe{p^{\rm \,esc}}
  \def\Pe{P^{\rm \,esc}}
  \def\Pp{P^{\rm \,pump}}
  \def\tauL{\tau^{\rm ver}}
  \def\SL{S^{\rm L}}

\newcommand{\hh}[0]{H$_2$}

\begin{document}

   \title{Radiation thermo-chemical models of protoplanetary disks}

   \subtitle{I. Hydrostatic disk structure and inner rim}

   \author{P. Woitke
          \inst{1,2}
          \and
          I. Kamp\inst{3}
          \and
          W.-F. Thi\inst{4}
          }

   \institute{UK Astronomy Technology Centre, Royal Observatory, Edinburgh,
              Blackford Hill, Edinburgh EH9 3HJ, UK
         \and
             School of Physics \& Astronomy, University of St.~Andrews,
             North Haugh, St.~Andrews KY16 9SS, UK
	 \and
             Kapteyn Astronomical Institute, Postbus 800,
             9700 AV Groningen, The Netherlands
         \and
	     SUPA\thanks{The Scottish Universities Physics Alliance}, 
             Institute for Astronomy, University of Edinburgh,
             Royal Observatory, Blackford Hill, Edinburgh EH9 3HJ, UK
             }

   \date{Received February 10, 2009; accepted April 2, 2009}

  \abstract
   {
    Emission lines from protoplanetary disks originate mainly
    from the irradiated surface layers, where the gas is generally
    warmer than the dust. Therefore, the interpretation of emission
    lines requires detailed thermo-chemical models, which are
    essential to convert line observations into understanding disk physics.}
   {We aim at hydrostatic disk models that are valid from 0.1\,AU to
    1000\,AU to interpret gas emission lines from UV to sub-mm. In
    particular, our interest lies in the interpretation of far IR gas
    emission lines as will be observed by the Herschel satellite,
    related to the {{\sc Gasps}} open time key program. This paper
    introduces a new disk code called {\sc ProDiMo}.}
   {We combine frequency-dependent 2D dust continuum radiative
    transfer, kinetic gas-phase and UV photo-chemistry, ice formation,
    and detailed non-LTE heating \& cooling with the consistent
    calculation of the hydrostatic disk structure. We include 
    Fe\,{\sc ii} and CO ro-vibrational line heating/cooling relevant for the
    high-density gas close to the star, and apply a modified escape
    probability treatment. The models are characterized by a high
    degree of consistency between the various physical, chemical and
    radiative processes, where the mutual feedbacks are solved
    iteratively.}
   {In application to a T\,Tauri disk extending from 0.5\,AU to
    500\,AU, the models show that the dense, shielded and cold
    midplane ($z/r\!\la\!0.1$, $\Tg\!\approx\!\Td$) is surrounded by a
    layer of hot ($\Tg\!\approx\!5000\,$K) and thin
    ($\nH\!\approx\!10^{\,7}$ to $10^{\,8}\rm\,cm^{-3}$) atomic gas
    which extends radially to about 10\,AU, and vertically up to
    $z/r\!\approx\!0.5$.  This layer is predominantly heated by the
    stellar UV (\eg PAH-heating) and cools via Fe\,{\sc ii} semi-forbidden
    and O{\sc i}\,630\,nm optical line emission. The dust grains in this
    ``halo'' scatter the star light back onto the disk which impacts
    the photo-chemistry.  The more distant regions are characterized
    by a cooler flaring structure. Beyond $r\!\ga\!100\,$AU, $\Tg$
    decouples from $\Td$ even in the midplane and reaches values of
    about $\Tg\!\approx\!2\Td$.}
   {Our models show that the gas energy balance is the key to
    understand the vertical disk structure. Models calculated with the
    assumption $\Tg\!=\!\Td$ show a much flatter disk structure. The
    conditions in the close regions ($<\!10\,$AU) with densities
    $\nH\!\approx\!10^{\,8}$ to $10^{\,15}\rm\,cm^{-3}$ resemble those
    of cool stellar atmospheres and, thus, the heating and cooling is
    more stellar-atmosphere-like. The application of heating and
    cooling rates known from PDR and interstellar cloud research alone
    can be misleading here and more work needs to be invested to identify
    the leading heating and cooling processes.}

   \keywords{ Astrochemistry; circumstellar matter; stars: formation; 
              Radiative transfer; Methods: numerical; line: formation }

   \maketitle


\section{Introduction}

The structure and composition of protoplanetary disks play a key role
in understanding the process of planet formation. From thermal and
scattered light observations, we know that protoplanetary disks are
ubiquitous in star forming regions and that the dust in these disks
evolves on timescales of $10^6$~yr \citep{Haisch2006,Watson2007}.
However, dust grains represent only about 1\% of the mass in these
disks -- 99\% of their mass is gas.

While observations of the dust in these systems have a long history
\citep[e.g.][]{Beckwith1990}, gas observations are intrinsically more
difficult and have focused until recently on rotational lines of
abundant molecules such as CO, HCN, HCO$^+$
\citep[e.g.][]{Beckwith1986,Koerner1993,Dutrey1997,vanZadelhoff2001,Thi2004}.
These lines generally trace the outer cooler regions of protoplanetary
disks and probe a layer at intermediate heights, where the stellar UV
radiation is sufficiently shielded to suppress photodissociation, but
still provides enough ionization to drive a rich ion-molecule
chemistry \citep{Bergin2007}. However, the sensitivity of current
radio telescopes allows only observations of small samples
\citep{dent2005} and in a few cases detailed studies of individual
objects \citep[e.g.][]{qi2003,semenov2005,qi2008}. The gas temperature
in the disk surface down to continuum optical depth of $\sim\!1$
decouples from the dust temperature and ranges from a few thousand to
a few hundred Kelvin \citep{Kamp2004,Jonkheid2004,Dullemond2007}. The
hot inner disk can show ro-vibrational line emission in the near-IR
either due to fluorescence (at the disk surface) and/or thermal
excitation \citep[e.g.][]{Bary2003,Brittain2007,Bitner2007}. More
recently, near-IR gas lines have also been detected in Spitzer IRS
spectra, revealing the presence of water, H$_2$ and the importance of
X-rays \citep{Pascucci2007,Lahuis2007,Salyk2008}. The launch of the
Herschel satellite in 2009, opens yet another window to study the gas
component of protoplanetary disks through the dominant cooling lines
[O\,{\sc i}], [C{\sc ii}] at the disk surface as well as many
additional molecular tracers of the warmer inner disk such as water
and CO. The study of the gas in protoplanetary disks is the main topic
of the Herschel open time Key Program ``Gas in Protoplanetary
Systems'' ({\sc Gasps}, PI: Dent). Other guaranteed and open time Key
Programs, such as ``Water in Star Forming Regions with Herschel''
({\sc Wish}, PI: van Dishoeck) and ``HIFI Spectral Surveys of Star
Forming Regions'' (PI: Ceccarelli), will also observe gas lines in a
few disks.

Disk structure modeling was initially driven by dust observations and
developed from a simple two-layer disk model
\citep[e.g.][]{Chiang1997} into detailed dust continuum radiative
transfer models that are coupled with hydrostatic equilibrium
\citep[e.g.][]{Dalessio1998,Dullemond2002,Dullemond2004,Pinte2006}. The
assumption in all these models is that gas and dust are well coupled
and the hydrostatic scale height then follows from the dust
temperature. However, the gas temperature decouples from the dust
temperature and the vertical disk structure will adjust to the gas
scale height, forcing the dust to follow if it is dynamically
coupled. This approach has been followed by \citet{Nomura2005} and
\citet{Gorti2004,Gorti2008}. Nomura \& Millar use a small chemical
network (only CO, C$^+$ and O) and a limited number of heating/cooling
processes namely photoelectric heating, gas-grain collisions and line
cooling from [O\,{\sc i}], [C{\sc ii}] and CO. Gorti \& Hollenbach use
an extended set of reactions (84 species, $\sim 600$ reactions) and
the relevant low-density heating/cooling processes drawn from photo
dissociation region (PDR) physics. Other models do not solve for the
vertical hydrostatic disk structure \citep[\eg][]{Kamp2004,
Meijerink2008, Woods2009}. Kamp \& Dullemond use a chemical reaction
network of $\sim$ 250 reactions among 48 species and a set of
heating/cooling processes comparable to \citet{Gorti2004}.  The models
of Meijerink\etal focus entirely on the X-ray irradiation of the disk,
thus excluding UV processes; the chemical reaction network is limited
to 25 species and 125 reactions. Woods\plus Willacy use again a
standard set of PDR heating/cooling processes, but also account for
X-rays. Their chemical network includes 475 gas and ice species
connected through $\sim$ 8000 gas phase and surface reactions.

This paper presents a new disk code that includes additional
heating/cooling processes relevant for the high densities and high
temperatures present in the inner parts of the disk, resembling the
conditions in tenuous atmospheres of cool stars.  The models are
characterized by a high degree of consistency between the various
physical, chemical and radiative processes. In particular, the results
of a full 2D dust continuum radiative transfer are used as input for
the UV photo-processes and as radiation background for the non-LTE 
modelling of atoms and molecules to calculate the line heating and cooling
rates.  This allows the models to extend closer to the star and
include modelling of the so-called inner rim.

The paper is structured as follows. Section~2 introduces the new code
\ProDiMo and presents the concept of global iterations.  Section~3
describes the assumptions used to calculate the hydrostatic disk
structure including ``soft edges''. In Sect.~4, we present the 2D dust
continuum radiative transfer with scattering and band-mean
opacities. Section~5 summarizes the gas-phase and photo-chemistry
dependent on the UV continuum transfer results. In Sect.~6, we outline
the heating and cooling rates included in our model and present a
modified escape probability method. Section~7 closes the theory part
of the paper with the calculation of the sound speeds as preparation
of the next calculation the the disk structure.
We apply \ProDiMo to a standard T\,Tauri-type protoplanetary disk with
disk mass $0.01M_\odot$ which extends from 0.5\,AU to 500\,AU in
Sect.~8. The resulting physical and chemical structure of the disk is
shown and compared to a model where we assume $\Tg\!=\!\Td$.  We
conclude the paper in Sect.~9 with an outlook to future applications.



\section{ProDiMo}

\ProDiMo is an acronym for \underline{Pro}toplanetary \underline{Di}sk
\underline{Mo}del. It is based on the thermo-chemical models of Inga
Kamp \citep{Kamp2000,Kamp2001,Kamp2004}, but completely re-written to
be more flexible and to include more physical processes.

\begin{figure}
  \centering
  \includegraphics[width=6cm]{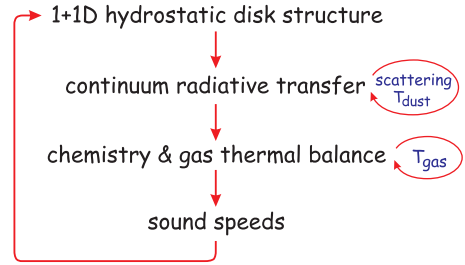} 
  \caption{Concept of global iterations in {\sc ProDiMo}. The circular 
    arrows on the r.h.s.\ indicate sub-iterations. For example, the
    dust temperature structure needs to be iterated in the continuum 
    radiative transfer.}
  \label{fig:iteration}
\end{figure}

\ProDiMo uses global iterations to consistently calculate the
physical, thermal and chemical structure of protoplanetary disks. The
iterations involve 2D dust continuum radiative transfer, 
gas-phase and photo-chemistry, thermal energy balance of the gas, and
the calculation of the hydrostatic disk structure in axial symmetry
(see Fig.~\ref{fig:iteration}). The different components will be
explained separately in the forthcoming sections.

Physical processes not yet included are X-ray heating, X-ray
chemistry, spatially dependent dust properties, and
PAH-chemistry. These processes will be addressed in future
papers. \ProDiMo is under current development. The code can be
downloaded from {\sffamily\mdseries
https://forge.roe.ac.uk/trac/ProDiMo}, start at
{\sffamily\mdseries
https://forge.roe.ac.uk/trac/ROEforge/wiki/NewUserForm} to get a
\ProDiMo user account.

\section{Hydrostatic disk structure}
\label{sec:struc}

We consider the hydrostatic equation of motion in axial symmetry with
rotation around the $z$-axis, but $v_r\!=\!0$ and $v_z\!=\!0$
\begin{eqnarray}
  \frac{v_\phi^2}{r} &=& \frac{1}{\rho} \pabl{p}{r} + \pabl{\Phi}{r}
  \label{eq:motion1}\\ 
                   0 &=& \frac{1}{\rho} \pabl{p}{z} + \pabl{\Phi}{z}
                  \ ,
  \label{eq:motion2}
\end{eqnarray}
where $v_r$, $v_z$ and $v_\phi$ are the three components of the
velocity field, $p$ the gas pressure and $\rho$ the mass density,
respectively.  $r\!=\!(x^2+y^2)^{1/2}$ is the distance from the
symmetry axis and $z$ is the distance from the midplane. Neglecting
self-gravity, the gravitational potential is given by
\begin{equation}
  \Phi(r,z) = -\,\frac{G M_\star}{\sqrt{r^2+z^2}}
\end{equation}
where $M_\star$ is the stellar mass and $G$ the gravitational
constant.  We follow the idea of ``1+1D'' modelling
\citep[][]{Dalessio1998,Malbet2001,Dullemond2002} by assuming that
the radial pressure gradient $1/\rho(\partial p/\partial r)$ in
Eq.~(\ref{eq:motion1}) is small compared to centrifugal acceleration
and gravity, in which case the radial and the vertical components of
the equation of motion decouple from each other. The radial component
then simply results in circular Keplerian orbits
\begin{equation}
  v_\phi = \left( \frac{ r^2\,GM_\star}{\big(r^2+z^2\big)^{3/2}}
           \right)^{1/2} \ ,
  \label{eq:vKepler}
\end{equation}
leaving the radial distribution of matter undetermined, as it is in
fact mostly determined by the actual distribution of angular momentum
in the disk.  Consequently, the vertical component of
the equation of motion (Eq.~\ref{eq:motion2}) can be solved
independently for every vertical column in the disk
\begin{equation}
  \frac{1}{\rho}\frac{dp}{dz} = -\,\frac{z\;GM_\star}{\big(r^2+z^2\big)^{3/2}}
  \label{eq:hydrostat} \ .
\end{equation}
Equation~(\ref{eq:hydrostat}) is integrated from the midplane upwards
by substituting the density for the pressure via $p\!=\!\cT2\,\rho$,
and assuming that the isothermal sound speed $c_T$ is a known function
of $z$. Numerically, we perform this integration by means of an
ordinary differential equation (ODE) solver, using a simple pointwise
linear interpolation of $\cT2(z)$ between calculated grid points
$\cT2(r_j,z_k)$. Since for known $c_T(z)$ the solution $p(z)$ has a
free factor, we put $p(0)\!=\!1$ and scale the results later to
achieve any desired column density at distance $r$
\begin{equation}
  \Sigma(r) = 2\!\!\int\limits_0^{z_{\rm max}(r)}\!\!\rho(r,z)\,dz \ ,
\end{equation}
where the factor 2 is because of the lower half of the disk,
which is assumed to be symmetric. In this paper, we assume a powerlaw
distribution of the column density 
\begin{equation}
  \Sigma(r) = \Sigma_0\;r^{\,-\epsilon}
  \label{eq:coldens}
\end{equation}
in the main part of the disk, except for the ``soft edges'' (see
Sect.~\ref{sec:SoftEdge}), and determine $\Sigma_0$ from the specified
disk mass $M_{\rm disk}$
\begin{equation}
  M_{\rm disk} = 2\pi \int\limits_{\Rin}^{\Rout} \Sigma(r)\,r\,dr \ ,
\end{equation}
where $\Rin$ is the inner radius and $\Rout$ the outer radius of the
disk. In summary, supposed that $\cT2(r,z)$ is known, the disk
structure is determined by the parameters $M_{\rm disk}$, $M_\star$,
$\Rin$, $\Rout$, and $\epsilon$.

\subsection{Soft Edges}
\label{sec:SoftEdge}

The application of a radial surface density powerlaw
(Eq.\,\ref{eq:coldens}) in the disk between $\Rin$ and $\Rout$ is,
although widely used, obviously quite artificial and even
unphysical. Equation (\ref{eq:motion1}) demonstrates that an abrupt
radial cutoff would produce an infinite force because of the radial
pressure gradient $\partial p/\partial r$, which would push gas inward
at $\Rin$, and outward at $\Rout$, respectively, causing a smoothing
of the radial density structure at the boundaries.

Let us consider an abrupt cutoff in the beginning and study the motion
of the gas as it is pushed inward due to the radial pressure gradient
at the inner boundary. Since the specific angular momentum
$L_3(\Rin)\!=\!\Rin\,v_\phi(\Rin)\!=\!r\,v_\phi(r)$ is conserved
during this motion, the gas will spin up as it is pushed inward, until
the increased centrifugal force balances the radial pressure gradient
(+\,gravity). According to (Eq.\,\ref{eq:motion1}) the force
equilibrium in this relaxed state is given by
\begin{equation}
 \frac{L_3(\Rin)^2}{r^3} = \frac{1}{\rho}\pabl{p}{r} + \pabl{\Phi}{r}
\end{equation}
which provides an equation for the desired density structure $\rho(r)$. 
Using $p(r)=\cT2\,\rho(r)$ and assuming $\cT2\!=\!\rm const$, the result is
\begin{equation}
  \ln\frac{\rho(r_0)}{\rho(\Rin)} \,=\, 
  -\frac{1}{c_T^2}\left[\frac{L_3(\Rin)^2}{2r^2}
                        +\Phi(r)\right]^{\,r_0}_{\,\Rin}
  \label{eq:softedge}
\end{equation}
where $r_0$ is an arbitrary point inside $\Rin$. Generalizing
Eq.\,(\ref{eq:softedge}) to column densities (with $\cT2$ measured in
the midplane) we write
\begin{equation}
  \Sigma(r_0) \,\approx\, \Sigma(\Rin)\,\exp\left(-\frac{1}{\cT2}
                \left[\frac{L_3(\Rin)^2}{2r^2}
                     +\frac{G
                M_\star}{r}\right]^{\,r_0}_{\,\Rin}\right)
  \label{eq:softedge2}
\end{equation}
A similar expression can be found for the column density outside of
the outer boundary. The CO observations of \citet{Hughes2008} show
that such treatments can improve model fits. However, we have
chosen to apply our approach for soft edges only to the inner boundary
in this paper.

To summarize, if angular momentum is transported inside-out in the
disk, the density structure may decrease more gradually or even
increase further inward \citep{Hartmann1998}. However, it is
hard to figure out any circumstances where the column density could
decrease more rapidly at the inner rim as compared to
Eq.\,(\ref{eq:softedge2}).


\section{Continuum radiative transfer}
\label{sec:RT}

The chemistry and the heating \& cooling balance of the gas in the
disk (see Sects.~\ref{sec:chem} and \ref{sec:heatcool}) depend on the
local continuous radiation field $J_\nu(r,z)$ and the local dust
temperature $\Td(r,z)$ which is a result thereof. These dependencies
include

\begin{itemize}
\item[1.] thermal accommodation between gas and dust, which is usually
  the dominant heating/cooling process for the gas in the midplane
  ($\to\!\Td$),
\item[2.] photo-ionization and photo-dissociation of molecules, as
  well as heating by absorption of UV photons, \eg photo-electric
  heating ($\to \!J_{\rm UV}$),
\item[3.] radiative pumping of atoms and molecules by continuum
  radiation which alters the non-LTE population and cooling rates,
  sometimes turning cooling into heating ($\to\!J_\nu$),
\item[4.] surface chemistry on grains, in particular the H$_2$-formation,
  and ice formation and desorption ($\to\!\Td,J_{\rm UV}$).   
\end{itemize}

\noindent Previous chemical models have often treated these couplings
by means of simplifying assumptions and approximate formula
\citep[e.g.][]{Kamp2000,Hollenbach1991,Nomura2005}. For a 
rigorous solution, a full 2D continuum radiative transfer must
be carried out, which provides $\Td(r,z)$ and $J_\nu(r,z)$, including
the UV part, at every location in the disk.

\ProDiMo solves the 2D dust continuum radiative transfer of irradiated
disks by means of a simple, ray-based, long-characteristic,
accelerated $\Lambda$-iteration method.  From each grid point in the
disk, a number of rays (typically about 100) are traced backward along
the photon propagation direction, while solving the radiative transfer
equation
\begin{equation}
  \abl{I_\nu}{\tau_\nu} = S_\nu - I_\nu
  \label{eq:RT}
\end{equation}
assuming LTE and coherent isotropic scattering
\begin{equation}
  S_\nu = \frac{\kabs B_\nu(\Td)+ \ksca J_\nu}{\kext} \ .
  \label{eq:source}
\end{equation}
$I_\nu$ is the spectral intensity, $J_\nu\!=\!\frac{1}{4\pi}\int I_\nu
\,d\Omega$ the mean intensity, $S_\nu$ the source function, $B_\nu$
the Planck function, and $\kabs$, $\ksca$ and $\kext=\kabs+\ksca$
$[\rm cm^{-1}]$ are the dust absorption, scattering and extinction
coefficients, respectively.

The dust grains of various sizes at a certain location in the disk are
assumed to have a unique temperature $\Td$ in modified radiative
equilibrium
\begin{equation} 
  \Gamma_{\rm dust} \,+ \int \kabs J_\nu\,d\nu
  \,=\, \int \kabs B_\nu(\Td)\,d\nu \ ,
  \label{eq:RE}
\end{equation}
where the additional heating rate $\Gamma_{\rm dust}$ accounts for
non-radiative heating (negative for cooling) processes like thermal
accommodation with gas particles and frictional heating.  An
accelerated $\Lambda$ scheme is used to get converged results
concerning $J_\nu(r,z)$ and $\Td(r,z)$.  The details of this method
will be described in the following sections.

\subsection{Geometry of rays}

Let $\r0\!=\!(x_0, y_0, z_0)$ denote a point in the disk where the mean
intensities $J_\nu(\r0)$ are to be calculated.  The direction of a ray
starting from $\r0$ is specified by a unit vector which points 
in the reverse direction of the photon propagation
\begin{equation}
    \left(\begin{array}{c} n_1 \\ n_2 \\ n_3 \end{array}\right)
  = \left(\begin{array}{c} \sin\theta\cos\phi \\ 
                     \sin\theta\sin\phi \\
                     \cos\theta              \end{array}\right) 
  \label{eq:angles}
\end{equation}
as specified in a local coordinate system where $(0,0,1)$ points
toward the star.  One ray is reserved for the solid angle occupied by
the star $\Omega_\star$ as seen from point $\r0$, subsequently called
the ``core ray''. All other $I\times J$ rays represent the remainder
of the $4\pi$ solid angle by a 2D-mesh of angular grid points
$\{\theta_i\,|\,i=1,...,I\}$ and $\{\phi_j\,|\,j=1,...,J\}$
\begin{eqnarray}
  \theta_i &=& \theta_\star + (\pi-\theta_\star)
               \left(\frac{i-1}{I-1}\right)^{p} \\
  \phi_j   &=& \pi\, \frac{j-1}{J-1} \ ,
\end{eqnarray}
where $\theta_\star\!= \arcsin(R_\star/d)$ is the half angular
diameter of the star as seen from point $\r0$, $R_\star$ the stellar
radius, and $d\!=\!(x_0^2+y_0^2+z_0^2)^{1/2}$ the radial
distance. $\phi$ only ranges from 0 to $\pi$, because the disk problem
is symmetric 
$I_\nu(\r0,\theta,+\phi)\!=\!I_\nu(\r0,\theta,-\phi)$.  A power index
$p\!>\!1$ ($p\!\approx\!1.5$) assures that there are more rays
pointing toward the hot inner regions than toward the cooler
interstellar side. The integration over solid angle is carried out as
\begin{eqnarray}
  4\pi &=& \Omega_\star + \sum_{i=1}^{I-1}\sum_{j=1}^{J-1} d\Omega_{ij}\\
  \Omega_\star &=& 2\pi\,(1-\cos\theta_\star)\\
  d\Omega_{ij} &=&
  2(\phi_{j+1}-\phi_j)\,(\cos\theta_i-\cos\theta_{i+1}) \ .
\end{eqnarray}
The central direction of solid angle interval $d\Omega_{ij}$ is given
by $\bar{\theta_i}\!=\!(\theta_{i+1}+\theta_i)/2$ and
$\bar{\phi_j}\!=\!(\phi_{j+1}+\phi_j)/2$, and these are the angles
actually considered in Eq.\,(\ref{eq:angles}). In order let the core
ray with $\theta\!=\!0$ point toward the star, we apply the following
rotation matrix
\begin{equation}
  \left(\begin{array}{c} n_x \\ n_y \\ n_z \end{array}\right)
   = \left(\begin{array}{ccc}  
            \cos\alpha & 0 & -\sin\alpha \\
                 0     & 1 &      0      \\
            \sin\alpha & 0 & \cos\alpha  \end{array}\right)
     \left(\begin{array}{c} n_1 \\ n_2 \\ n_3 \end{array}\right)
\end{equation}
where $\alpha=\beta+90^o$ and $\tan\beta=z_0/(x_0^2+y_0^2)^{1/2}$.

\subsection{Solution of the radiative transfer equation}

From every grid point $\r0$ along each ray in direction
$\nn\!=\!(n_x,n_y,n_z)$ we solve the radiative transfer equation
(Eq.\,\ref{eq:RT}) backward to the photon propagation direction.  The
optical depth along the ray is given by
\begin{equation}
  \tau_\nu(s) = \int_0^s \kext(\r0+s'\nn)\,ds'
  \label{eq:tau}
\end{equation}
The formal solution of the transfer equation Eq.\,(\ref{eq:RT}) is
\begin{equation}
  I_\nu \,=\, I^{\rm inc}_\nu e^{-\tau_\nu(s_{\rm max})}
        \,+\! \int_0^{s_{\rm max}}\hspace*{-1mm}
          \kext(s)\,S_\nu(s)\,e^{-\tau_\nu(s)}\;ds
  \label{eq:Inu}
\end{equation} 
where $I^{\rm inc}_\nu$ is the intensity incident from the end of the
ray at $s_{\rm max}$. We start a ray at $s\!=\!0$ with $\tau_\nu\!=\!0$ and
$I_\nu\!=\!0$ and choose a suitable spatial step size $\Delta s$. For
each step, the opacities and source functions (all wavelengths) at the
start point and the end point of the step, $\r0\!+\!s\,\nn$ and
$\r0\!+\!(s\!+\!\Delta s)\nn$, are interpolated from the pre-calculated
values on the grid points using a 2D-interpolation in cylinder
coordinates $((x^2+y^2)^{1/2},|z|)$.  For the numerical integration of
Eqs.\,(\ref{eq:tau}) and (\ref{eq:Inu}), we assume $\kext\!=\!A+Bs$
and $S_\nu\!=\!C\exp(Ds)$, where the coefficients $A,B,C,D$ are
determined by the start and end point values. Simplifying the
exponent by putting $\bar{\kappa}^{\rm ext}_\nu\!\approx\!A+B\Delta
s/2$ yields
\begin{eqnarray}
  \tau_\nu(s\!+\!\Delta s) &=& \tau_\nu(s) 
                 + \!\int_0^{\Delta s}\hspace*{-1mm}(A+Bs')\,ds'   \\
  I_\nu(s\!+\!\Delta s)    &=& I_\nu(s) 
                 + C\,e^{-\tau_\nu(s)}\!\!
                   \int_0^{\Delta s}\hspace*{-1mm}  (A+Bs')\;
                   e^{\,(D\,-\,\bar{\kappa}^{\rm ext}_\nu)\,s'}\,ds' \ .
\end{eqnarray}
The numerical integration is carried out with analytic expressions for
these integrals.  The procedure is repeated for two half steps of size
$\Delta s/2$.  If the results differ too much, the step size $\Delta
s$ is reduced and the step is re-calculated. In case of small
differences, the step size is increased for the following step.

At the end of each ray, the attenuated incident intensities $I^{\rm
inc}_\nu e^{-\tau_\nu(s_{\rm max})}$ are added according to
Eq.~(\ref{eq:Inu}), where for the core ray the stellar intensity
$I^{\rm inc}_\nu=I_\nu^\star$ is used, and for all other rays the
interstellar intensity $I^{\rm inc}_\nu=I_\nu^{\rm ISM}$ is
applied. Non-core rays may temporarily leave the disk, but
re-enter the disk after some large distance. These ``passages'' are
treated with large, exactly calculated $\Delta s$ and zero opacities.

For the 2D-interpolation, it turned out to be important to use a
log-interpolation for the source function $S_\nu(r,z)$ which can
change by orders of magnitude, \eg across a shadow, within one
step. In case of linear interpolation, the numerical radiative
transfer shows much more numerical diffusion.

\subsection{Irradiation}

The radiation field in (and around) passive disks is completely
determined by the stellar and interstellar irradiation, and the
geometry of the dust opacity structure. Therefore, setting the
irradiation as realistic as possible is of prime importance.

\begin{figure}
  \centering
  \includegraphics[width=8.5cm,height=6cm]{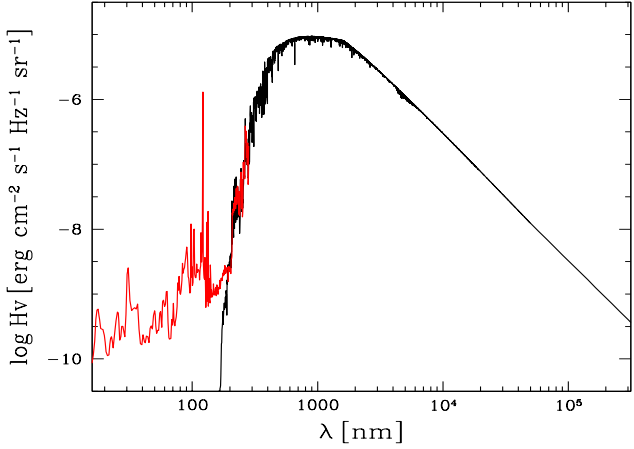}
  \caption{Incident stellar intensity compiled from
         two sources: a {\sc Phoenix} solar model spectrum with $T_{\rm
         eff}\!=\!5800\,$K, $\log\,g\!=\!4.5$, $Z\!=\!1$ (black line) and 
         the chromospheric flux of ``young sun'' HD 129333 
         \citep[][red line]{Dorren1994}. Note that
         the ionizing and photo-dissociating flux is assumed to be
         restricted to the interval $[\rm 91.2\,nm,205\,nm]$ which
         includes Ly$\alpha$ at 121.6\,nm.}
  \label{fig:Iinc}
  \vspace*{-1mm}
\end{figure}

\paragraph{Stellar irradiation}
\label{sec:irradiation}

For the incident stellar irradiation, a model spectrum from stellar
atmosphere codes is used, \eg a {\sc Phoenix}-model\footnote{see
{ftp://ftp.hs.uni-hamburg.de/pub/outgoing/phoenix/GAIA/}}. Neglecting
limb-darkening, the incident stellar intensities are related to the
surface flux at the stellar radius via
\begin{equation}
  I_\nu^\star = \frac{1}{\pi} F_\nu^\star(R_\star) 
              \,=\, 4\,H_\nu^\star(R_\star) 
\end{equation}
where $F_\nu^\star$ is the spectral flux and $H_\nu$ the Eddington flux. Young
stars, which are active and possibly accreting, have excess UV
as compared to model atmospheres, in particular cool
stars. This is of central importance for \ProDiMo, because it is just
this radiation that ionizes and photo-dissociates the atoms and
molecules in the disk. Therefore, we add extra UV-flux as \eg reduced 
from observations or given by other recipes, see Fig.~\ref{fig:Iinc}.

\paragraph{Interstellar irradiation}

Assuming an isotropic interstellar radiation field, all incident
intensities for non-core rays are approximated by a highly diluted
20000\,K-black-body field plus the 2.7\,K-cosmic background.
\begin{eqnarray}
  I_\nu^{\rm ISM} 
        = \chi^{\rm ISM}\cdot 1.71\cdot W_{\rm dil}\,B_\nu(20000{\rm\,K})
        \,+ B_\nu(2.7{\rm\,K})
\end{eqnarray}
The applied dilution factor $W_{\rm dil}=9.85357\times10^{-17}$ is
calculated from the normalization $\chi\!=\!1$ according to
Eq.\,(\ref{eq:chi}), which is close to the value given by
\citet{Draine1996}. $\chi^{\rm ISM}$ is a free parameter which
describes the strength of the UV field with respect to standard
interstellar conditions.

\subsection{Iteration and dust temperature determination}

\begin{figure}
  \centering
  \hspace*{-3mm}\includegraphics[width=9.5cm]{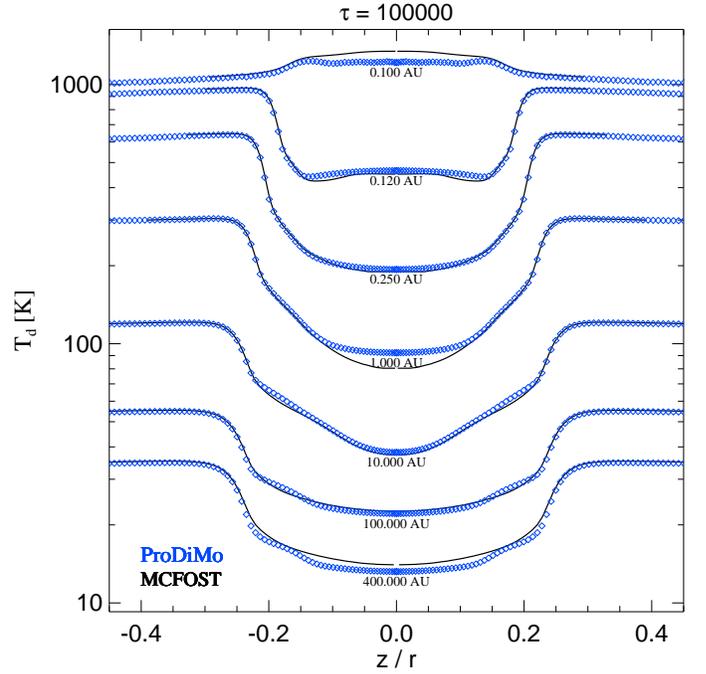} 
  \caption{Benchmark for the dust continuum radiative transfer
    part. Vertical cuts of the calculated dust temperature structure
    $\Td(r,z)$ are shown at different radii of a disk with
    midplane optical depth $\tau_{\rm 1\mu{\rm m}}\!=\!10^5$ at
    $1\,\mu$m. The {\sc ProDiMo} results are shown with blue diamonds
    in comparison to the MCFOST results \citep{Pinte2006} with black
    lines.}
  \label{fig:RTbench}
  \vspace*{-1mm}
\end{figure}

In order to solve the condition of the dust radiative equilibrium
(Eq.\,\ref{eq:RE}) and the scattering problem, a simple $\Lambda$-type
iteration is applied. The source functions are pre-calculated on the
grid points according to Eq.\,(\ref{eq:source}), with
$J_\nu\!=\!J_\nu^{\rm old}$ and $\Td\!=\!\Td^{\rm old}$, and fixed
during one iteration step.  After having solved all rays from all
points for all frequencies, the mean intensities are updated as
\begin{equation}
  J_\nu(\r0) = \frac{1}{4\pi}\left( I_\nu(\r0,0,0)\,\Omega_\star + 
       \sum_{i=1}^{I-1}\sum_{j=1}^{J-1} I_\nu(\r0,\bar{\theta_i},\bar{\phi_j})
               \,d\Omega_{ij} \right)\ ,
\end{equation}
and the dust temperatures are renewed according to Eq.\,(\ref{eq:RE}).
If the maximum relative change $|J_\nu-J_\nu^{\rm old}|/(J_\nu+J_{\rm
small})$ (all points, all frequencies, $J_{\rm small}\!=\!10^{-30}\rm
erg\, cm^{-2} s^{-1} Hz^{-1} sr^{-1}$) is larger than some threshold
($\sim$\,0.01), the source functions are re-calculated and the
radiative transfer is solved again.  In order to accelerate the
convergence, we apply the procedure of \citet{Auer1984}. 
We benchmarked results of our radiative transfer method against
results of other Monte-Carlo and ray-based methods in
\citep{Pinte2009}.  The convergence in optically thick disks is
tricky, but we can manage test problems up to a midplane optical
depths of about $\tau\!=\!10^5$ with this code (see
Fig~\ref{fig:RTbench}).

\subsection{Spectral bands and band-mean quantities}

The main purpose of the continuum radiative transfer in {\sc ProDiMo} is to
calculate certain frequency integrals, \eg solving the
condition of radiative equilibrium for the dust grains
(Eq.\,\ref{eq:RT}) or calculating the local strength of the UV
radiation field $\chi$ (Eq.\,\ref{eq:chi}). The incident stellar
spectrum is strongly varying in frequency space, especially in the
blue and UV (see Fig.~\ref{fig:Iinc}) and the evaluation of these
integrals, in principle, requires a large number of frequency grid points 
$\nu_k$, which is computationally expensive.

However, the incident radiation interacts with quite smooth
and often completely flat dust opacities in the disk. Thus, it makes
sense to ``interchange'' the order of radiative transfer and frequency
integration, and to switch from a monochromatic treatment to a
treatment with spectral bands.

We consider a coarse grid of frequency points $\{\nu_k\,|\,k=0,...,K\}$
(\eg $K\!=\!12$) which covers the whole SED, ranging from
$100\,$nm to $1000\,\mu$m. Instead of $B_\nu(T)$ we consider band
means as
\begin{equation}
  B_k(T) = \frac{1}{\Delta\nu_k}\int_{\nu_{k-1}}^{\nu_k} B_\nu(T)\,d\nu 
\end{equation}
where $\Delta\nu_k=\nu_k-\nu_{k-1}$. In a similar way, we treat the
intensities, mean intensities and opacities
\begin{eqnarray}
  I_k &=& \frac{1}{\Delta\nu_k}\int_{\nu_{k-1}}^{\nu_k} I_\nu\,d\nu \\
  J_k &=& \frac{1}{\Delta\nu_k}\int_{\nu_{k-1}}^{\nu_k} J_\nu\,d\nu \\
  \kappa_k &=& \frac{1}{\Delta\nu_k}
                 \int_{\nu_{k-1}}^{\nu_k} \kappa_\nu\,d\nu 
  \label{eq:kband}
\end{eqnarray}
Henceforth, we exchange the index $\nu$ by the index $k$ in all equations in
this section, and retrieve the recipes for the band-averaged continuum
radiative transfer. This is of course not an exact treatment, because
it ignores all non-linear couplings, but an approximation that allows
us to use fewer frequency grid points without loosing too much
accuracy.

\subsection{Dust kind, abundance, size distribution, and opacities}

We assume a uniform dust abundance and size distribution throughout the disk.
The dust particle density is given by
\begin{equation}
  n_d = \int_\amin^\amax\!\!\! f(a)\,da \ ,
\end{equation}
where $a$ is the particle radius and $f(a)$ is the dust size
distribution function $\rm [cm^{-4}]$, which is assumed to be given by
a powerlaw as $f(a)\!=\!\fconst a^{-\apow}$. The moments of the size
distribution are
\begin{equation}
  \langle a^j \rangle = \frac{1}{n_d} 
                        \int_\amin^\amax\!\!\! a^j\,f(a)\,da \ .
\end{equation}
The constant in the powerlaw size distribution $\fconst$ is determined
by the requirement that the dust mass density
\begin{equation}
  \rho_d = n_d\,\rho_{\rm gr} \frac{4\pi}{3} \langle a^3 \rangle 
\end{equation}
is given by a specified fraction of the gas mas density
$\rho_d/\rho$. $\rho_{\rm gr}$ is the dust material mass density.

The dust opacities are calculated from effective medium theory
\citep{Bruggeman1935} and Mie theory \citep[{\sc Miex} from
S.~Wolf, according to][]{Voshchinnikov2002}. Any uniform volume
mix of solid materials with known optical constants can be used.
The dust opacities are calculated as
\begin{equation}
  \kappa_\lambda^{\rm ext} = \int_\amin^\amax\!\!\! 
                             \pi a^2\,Q_{\rm ext}(a,\lambda)\,f(a)\,da \ ,
  \label{eq:kappa_dust}
\end{equation}
where $Q_{\rm ext}(a,\lambda)$ is the extinction efficiency. Similar
formula apply for absorption and scattering opacities,
$\kappa_\lambda^{\rm abs}$ and $\kappa_\lambda^{\rm sca}$, where $Q_{\rm
ext}(a,\lambda)$ is replaced by $Q_{\rm abs}(a,\lambda)$ and
$Q_{\rm sca}(a,\lambda)$, respectively.

\begin{table}
\centering
\caption{Elements and chemical species}
\label{tab:Species}
\begin{tabular}{c|p{6.6cm}}
\\[-4.5ex]
\hline 
 9 elements & H, He, C, N, O, Mg, Si, S, Fe\\
\hline 
\hline 
 71 species & H, H$^+$, H$^-$,
He, He$^+$, C, C$^+$, O, O$^+$, S, S$^+$, Si, Si$^+$, Mg, Mg$^+$, Fe,
Fe$^+$, N, N$^+$, H$_2$, H$_2^+$, H$_2^\star$, H$_3^+$, OH, OH$^+$,
H$_3$O$^+$, H$_2$O, H$_2$O$^+$, CO, CO$^+$, HCO, H$_2$CO, HCO$^+$,
O$_2$, O$_2^+$, CO$_2$, CO$_2^+$, CH, CH$^+$, CH$_2$, CH$_2^+$,
CH$_3$, CH$_3^+$, CH$_4$, CH$_4^+$, CH$_5^+$, SiO, SiO$^+$, SiH,
SiH$^+$, SiH$_2^+$, SiOH$^+$, NH, NH$^+$, NH$_2$, NH$_2^+$, NH$_3$,
NH$_3^+$, N$_2$, HN$_2^+$, CN, CN$^+$, HCN, HCN$^+$, NO, NO$^+$, CO\#,
H$_2$O\#, CO$_2$\#, CH$_4$\#, NH$_3$\#\\
\hline 
\end{tabular}\\[1mm]
\hspace*{11mm}\begin{minipage}{6cm}
\footnotesize
Ice species are denoted with ``\#''.\\
H$_2^\star$ designates vibrationally excited H$_2$.
\end{minipage}
\end{table}


\section{Chemistry}
\label{sec:chem}

The chemistry part of \ProDiMo is written in a modular form that makes
it possible to consider any selection of elements and chemical
species.  In the models presented in this paper, we consider chemical
reactions involving $\Nel\!=\!9$ elements among $\Nsp\!=\!71$ atomic,
ionic, molecular and ice species as listed to Table~\ref{tab:Species}.

The rate coefficients $R$ are mostly taken from the {\sc
Umist\,2006} data compilation \citep{UMIST2007}. Among the 
species listed in Table~\ref{tab:Species} we find 911 {\sc Umist}
chemical reactions, 21 of them have multiple $\Tg$-fits.  We add
39 further reactions which
are either not included in {\sc Umist} or are treated in a more
sophisticated way, as explained in Sects.~\ref{sec:specialUV} to
\ref{sec:ice}. Among the altogether 950 reactions, there are
74 photo reactions, 177 neutral-neutral and 299 ion-neutral reactions,
209 charge-exchange reactions, 46 cosmic ray and cosmic ray particle
induced photo reactions, and 26 three-body reactions.
The net formation rate of a chemical species $i$ is calculated as
\begin{eqnarray}
  \frac{d n_i}{dt} &=& 
   \sum\limits_{jk\ell} R_{jk\to i\ell}(\Tg)\,n_j n_k
\,+\sum\limits_{j\ell} \big(R^{\rm ph}_{j\to i\ell}
                           +R^{\rm cr}_{j\to i\ell}\big)\,n_j \nonumber\\
&-\,n_i&\!\left(\sum\limits_{jk\ell} R_{i\ell\to\!jk}\,n_\ell
   \,+\sum\limits_{jk} \big(R^{\rm ph}_{i\to\!jk}
                           +R^{\rm\,cr}_{i\to\!jk}\big)\right)
  \label{eq:chem}
\end{eqnarray}
where $R_{jk\to i\ell}$ designates (two-body) gas phase reactions
between two reactants $j$ and $k$, forming two products $i$ and
$\ell$.  $R^{\rm ph}_{i\to\!jk}$ indicates a photo-reaction which
depend on the local strength of the UV radiation field, and
$R^{\rm\,cr}_{j\to i\ell}$ a cosmic ray induced reaction.

\begin{figure*}
  \centering
  \begin{tabular}{ccc}
    \hspace*{-5mm}\includegraphics[width=6.4cm]{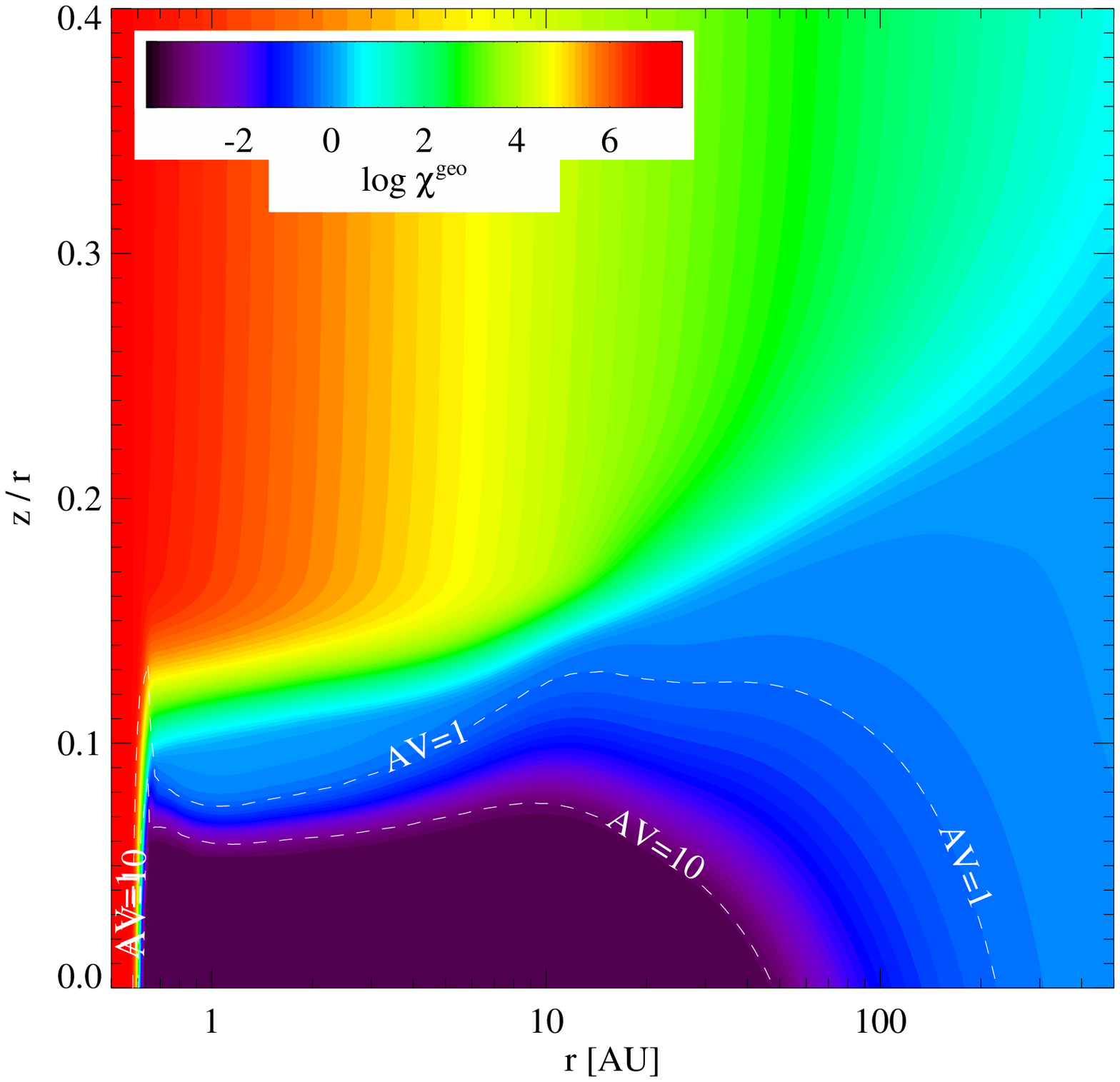} &
    \hspace*{-6mm}\includegraphics[width=6.4cm]{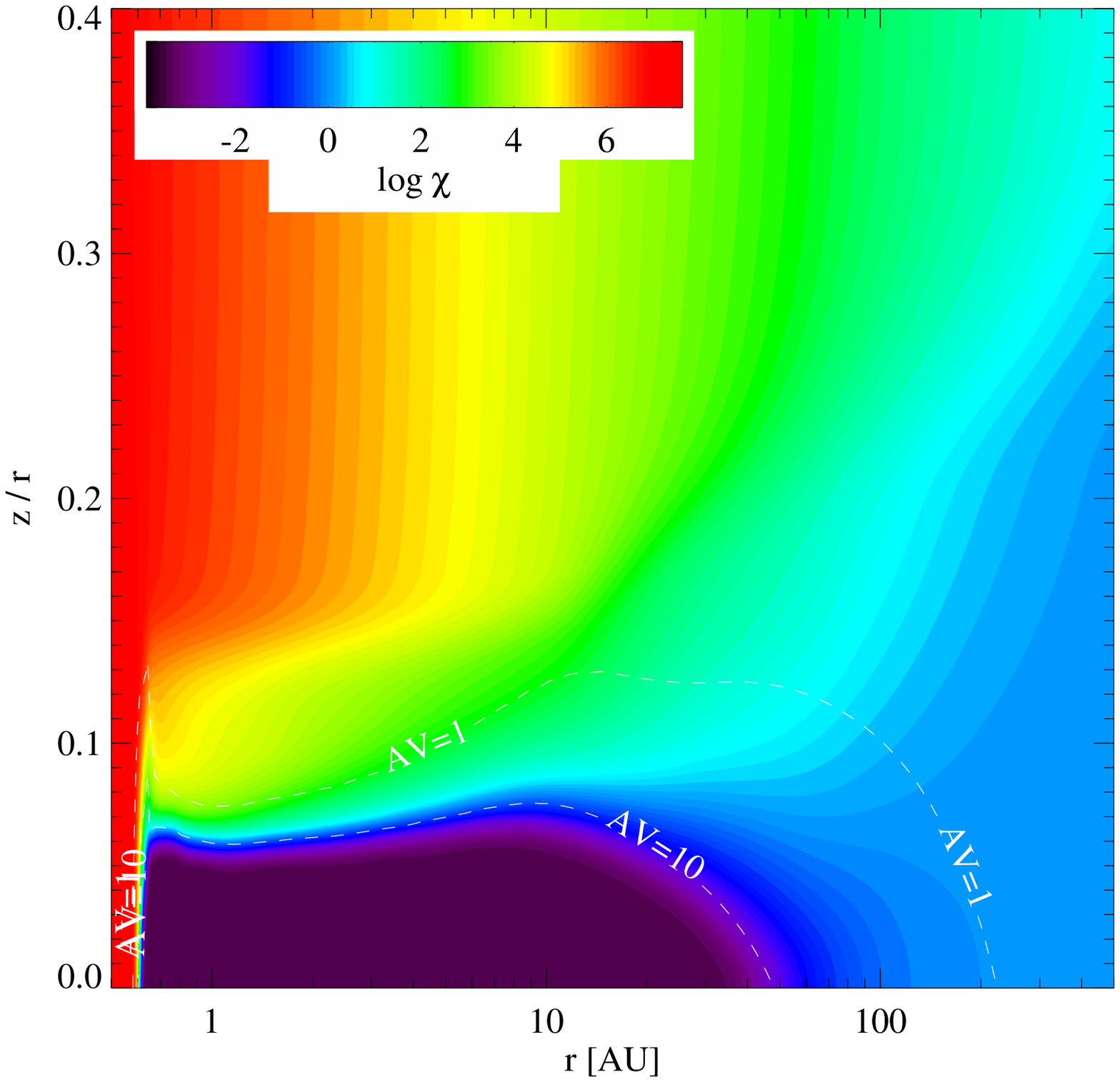} &
    \hspace*{-6mm}\includegraphics[width=6.4cm]{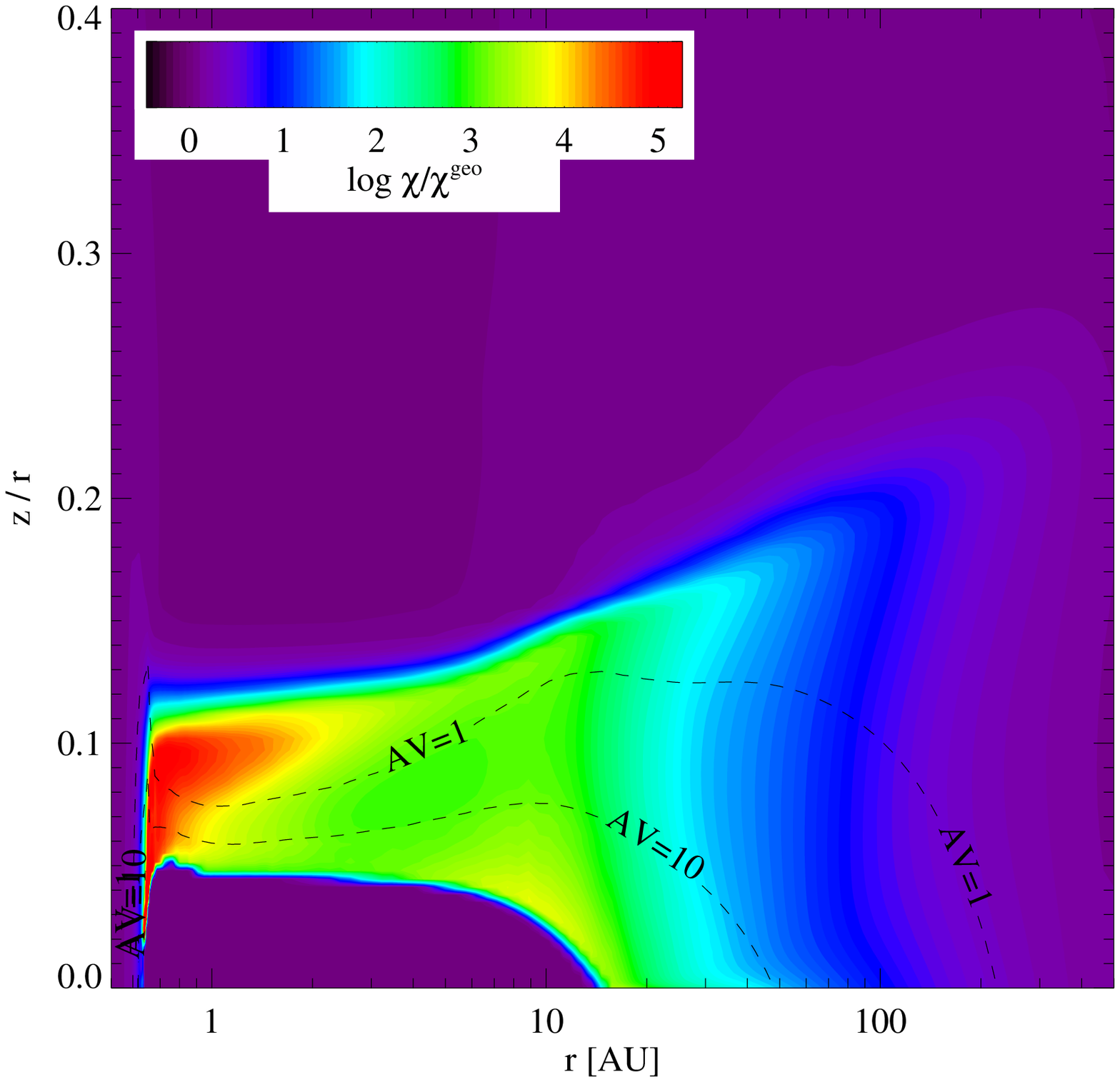} 
  \end{tabular}\\*[-3mm]  
  \caption{Comparison of the UV radiation field strengths $\chi$
    (Eq.~\ref{eq:chi}) between the simplified geometrical approach
    (l.h.s.), taking into account only vertical and radial dust
    extinction (see Eq.\,\ref{eq:twostream}), and the results of the
    full 2D radiative transfer including scattering (middle) for the
    $M_{\rm disk}\!=\!10^{-2}M_\odot$ model. The two dashed contour
    lines show $A_V\!=\!1$ and $A_V\!=\!10$ as calculated from the
    minimum of the radial and vertical column densities. The
    r.h.s. figure shows the ratio $\chi/\chi^{\rm geo}$, indicating
    that the regions mostly affected by scattering are situated
    roughly between $A_V\!=\!1$ and $A_V\!=\!10$.}
  \label{fig:chis}
\end{figure*}

\subsection{Photo-reactions}

Photon induced reaction rates can generally be written as
\begin{equation}
  R^{\rm ph} \;=\; 4\pi \int\!\sigma(\nu)\,\frac{J_\nu}{h\nu} \,d\nu
       \;=\; \frac{1}{h} \int\!\sigma(\lambda)\,\lambda u_\lambda \,d\lambda
  \label{eq:Rphoto}
\end{equation}  
where $\sigma(\nu)$ is the photo cross-section of the reaction, $\nu$
the frequency, $\lambda$ the wavelength, $h$ the Planck constant, and
$u_\lambda\!=\!\frac{4\pi}{c}J_\lambda$ the spectral photon energy
density $[\rm erg\,cm^{-4}]$, respectively. The proper calculation of
the photo-rates $R^{\rm ph}\,\rm[s^{-1}]$ according to
Eq.\,(\ref{eq:Rphoto}) would require the calculation of a detailed
(\ie UV-line resolved) radiative transfer including molecular
opacities to account for self-shielding effects (to get $J_{\nu}$) as
well as detailed knowledge about the wavelength-dependent cross
section $\sigma(\lambda)$, which is not always available.

{In this paper, we will apply the {\sc Umist\,2006} photo reaction
rates in combination with molecular self-shielding factors from the
literature instead.  The application of detailed molecular UV cross
sections in the calculated UV radiation field will be addressed in a
future paper.}

In the {\sc Umist} database, photo-rates are given as
\begin{equation}
  R^{\rm ph}_{\rm\scriptscriptstyle UMIST} \;=\; 2\,
  \chi_0\;\alpha\,\exp(-\,\gamma\,A_V^{\rm\scriptscriptstyle UMIST}) \ ,
  \label{eq:RUMIST}
\end{equation}
where $\chi_0$ is the unattenuated strength of the UV radiation field
with respect to a standard interstellar radiation field, $\alpha$ the
photo-rate in this standard ISM radiation field, and
$A_V^{\rm\scriptscriptstyle UMIST}$ the extinction at visual
wavelengths toward the UV light source. The photo-rates are derived
for semi-infinite slab geometry, that is radiation is coming only from
$2\pi$; this explains the factor 2 in Eq.\,(\ref{eq:RUMIST}).  In
arbitrary radiation fields, for other than 1D slab geometries, and for
dust properties different from the ones used in UMIST, it is not
obvious how to apply Eq.\,(\ref{eq:RUMIST}).
In the following, we will therefore carefully explain
our assumptions for the application of Eq.\,(\ref{eq:RUMIST}) to
protoplanetary disks.

\citet{Roellig2007} relate $\chi\!=\!1$ to a ``unit Draine
field'' and we will follow this idea in {\sc ProDiMo}. From the original
work by \citet{Draine1978}, \citet{Draine1996} deduced
\begin{equation}
   \lambda u_\lambda^{\rm Draine} = 1.71 \cdot 
         4\times10^{-14}
         \;\frac{31.016\lambda_3^2 - 49.913\lambda_3 + 19.897}{\lambda_3^5}
\end{equation}
for the standard ISM UV radiation field $[\rm erg\,cm^{-3}]$ where
$\lambda_3\!=\!\lambda/100{\rm\,nm}$. We apply an integral definition
of $\chi$ as
\begin{equation}
  \chi = \int_{91.2\,{\rm nm}}^{205\,{\rm nm}} 
         \lambda u_\lambda\,d\lambda \;\;\Bigg{/}\;
         \int_{91.2\,{\rm nm}}^{205\,{\rm nm}} 
         \lambda u_\lambda^{\rm Draine}\,d\lambda \ , 
  \label{eq:chi}
\end{equation}
The wavelength interval boundaries have been chosen to ensure coverage
of the most important photo-ionization and photo-dissociation processes
\citep{vanDishoeck2006}.
Numerical integration yields $F_{\rm
Draine}\!=\!\frac{1}{h}\int_{91.2\,{\rm nm}}^{205\,{\rm nm}} \lambda
u_\lambda^{\rm Draine}\,d\lambda
\!=\!1.921\times10^{+8}\rm\;cm^{-2}\,s^{-1}$.  Adopting the
wavelength boundaries 91.2\,nm and 205\,nm for the definition of our
spectral band 1, we can directly calculate $\chi$ from our banded
radiative transfer method, including scattering, by
\begin{equation}
  \chi = \frac{4\pi}{h\bar{\nu}_1} \,J_1 \Delta\nu_1 \,\Big{/}\, 
         F_{\rm Draine} \ ,
  \label{eq:chicalc}
\end{equation}
where we put $\bar{\nu}_1=\sqrt{\nu_0\nu_1}$. The unshielded
ISM photo-rate $\alpha$ is assumed to be given by
\begin{equation}
  \alpha \;=\; \frac{1}{2h} \int\!\sigma(\lambda)\,\lambda
  u_\lambda^{\rm Draine} \,d\lambda
  \label{eq:alpha}
\end{equation}  
and the coefficient $\gamma$ in Eq.\,(\ref{eq:RUMIST}) can be
identified as an effective, frequency-averaged opacity coefficient
which contains implicit information about the frequency-range of the
cross section $\sigma(\nu)$, the {\sc Umist} dust opacity and the
shape of the ISM radiation field assumed.  Neglecting gas extinction
and assuming constant dust properties along the line of sight, the UV
optical depth is
\begin{equation}
  \tau_{\rm UV} \,= \hat{\kappa}^{\rm ext}_1\,N_\HH
  \label{eq:tauUV}
\end{equation} 
where $N_\HH$ is the hydrogen nuclei column density toward the UV
light source and $\hat{\kappa}^{\rm ext}_1$ the dust extinction coefficient
per H-nucleus, averaged over spectral band 1. The ratio $A_V/\tau_{\rm
UV}$ depends on the frequency-dependence of the dust
opacities as
\begin{equation}
  A_V/\tau_{\rm UV} = 2.5 \log e 
   \cdot\kappa^{\rm ext}_{550\rm nm}/\kappa^{\rm ext}_1 \ .
  \label{eq:AV}
\end{equation}
However, in Eq.\,(\ref{eq:RUMIST}) we must not use $A_V$ as
calculated from our choice of dust properties! Instead, we have to use
the $A_V^{\rm\scriptscriptstyle UMIST}$ depth scale as used for the
compilation of the {\sc Umist} database. If we would use our $A_V$
(from dust properties in the disk), the use of $\gamma$ -- containing
{\sc Umist} dust properties -- would internally scale it to a
$\tau_{\rm UV}$ that is wrong\footnote{For species that can be
ionized with visual light like H$^-$, it might be actually better to
use our $A_V$ scale. But most photo-reactions occur in the UV and
$A_V$ is just used as an auxiliary variable.}. Thus, to obtain from
our UV optical depth the proper $A_V^{\rm\scriptscriptstyle UMIST}$,
we need the ratio $A_V^{\rm\scriptscriptstyle UMIST}/\tau_{\rm UV}$.
Since the exact {\sc Umist}-ratio is not known, we calculate it
according to Eq.\,(\ref{eq:AV}) for standard ``astronomical silicate''
grains \citep{Draine1984} with a size distribution
$f(a)\!\propto\!a^{-3.5}$ between 0.005\,$\mu$m and 0.25\,$\mu$m
\begin{equation}
  A_V^{\rm\scriptscriptstyle UMIST} = 0.216\;\tau_{\rm UV} \ ,
  \label{eq:AV2}
\end{equation}
whereas for larger disk dust, a value around 1 is more typical.
 
Another complicated problem is how to apply Eq.\,(\ref{eq:RUMIST}) in
disk geometry. For this purpose we introduce a geometric mean
intensity as it would be present, at least approximately, if only
extinction but no scattering would occur
\begin{equation}
  4\pi J_1^{\rm geo} = \Omega^\star I_1^\star e^{-\tau_{\rm UV}^{\rm rad}} 
         \;+\; \Omega^{\rm ISM} I_1^{\rm ISM} e^{-\tau_{\rm UV}^{\rm
         ver}} \ .
  \label{eq:twostream}
\end{equation}
$I_1^\star$ and $I_1^{\rm ISM}$ are the incident band-mean stellar and
interstellar intensities (see Sect.\,\ref{sec:irradiation}), and
$\tau_{\rm UV}^{\rm rad}$ and $\tau_{\rm UV}^{\rm ver}$ are the radial
(toward the star) and vertical (upwards) UV optical depth,
respectively. $\Omega^\star$ is the solid angle occupied by the star
and $\Omega^{\rm ISM}\!=\!4\pi-\Omega^\star$ the remainder of the full
solid angle. Switching to corresponding $\,\chi$ variables, we find
\begin{equation}
  \chi^{\rm geo} = 
  \frac{\Omega^\star}{4\pi}    \,\chi_0^\star e^{-\tau_{\rm UV}^{\rm rad}}
 \;+\;
  \frac{\Omega^{\rm ISM}}{4\pi}\,\chi_0^{\rm ISM} e^{-\tau_{\rm UV}^{\rm ver}}
\end{equation}
where $\chi^\star_0$ is calculated with $J_1\!=\!I_1^\star$ from
Eq.\,(\ref{eq:chicalc}), and $\chi_0^{\rm ISM}$ with $J_1\!=\!I_1^{\rm
ISM}$. This decomposition into two slab geometries allows us to apply
Eq.\,(\ref{eq:RUMIST}) and calculate the photo-rates as
\begin{equation}
  R^{\rm ph} = \frac{\chi}{\chi^{\rm geo}} \,\Bigg(
    \frac{\Omega^\star}{4\pi} 
    \chi_0^\star\,\alpha^\star e^{-\gamma\,
      A_{V,\scriptstyle\rm\,rad}^{\rm\scriptscriptstyle UMIST}}
  \;+\;\frac{\Omega^{\rm ISM}}{2\pi} 
    \chi_0^{\rm ISM}\,\alpha\, e^{-\gamma\,
      A_{V,\scriptstyle\rm\,ver}^{\rm\scriptscriptstyle UMIST}}
  \Bigg)
  \label{eq:photoR}
\end{equation}
{This approach to calculate the photo-rates according to
Eq.\,(\ref{eq:photoR}) can be extended for molecular self-shielding
factors (see Sect.~\ref{sec:specialUV}) and states a compromise
between the usual two-stream approximation and a proper treatment of
UV line-resolved radiative transfer according to
Eq.\,(\ref{eq:Rphoto}). The factor $\,\chi/\chi^{\rm geo}$ corrects
for the mayor shortcomings of the two-stream approximation, \ie
the effects of scattering and the assumptions about the geometry of
the radiation field made in Eq.\,(\ref{eq:twostream}).
Figure~\ref{fig:chis} shows that the enhancements $\,\chi/\chi^{\rm
geo}$ are} very close to 1 in the upper, directly irradiated layers of
the disk, but may be as large as $10^5$ at the inner rim and in the
warm intermediate layer of the disk, and about $10^3$ in the outer
midplane due to scattering. $\alpha^\star$ is the unshielded
photo-rate for $4\pi$-irradiation with star light, divided by
$\chi^\star_0$. It can be calculated from $\sigma(\lambda)$ if known,
or is assumed to be identical to $2\alpha$ otherwise.

\subsection{Special UV photo reactions}
\label{sec:specialUV}
For the photo-ionization of neutral carbon,
$\alpha\!=\!3\times10^{-10}\rm s^{-1}$ is taken from the {\sc Umist}
database, whereas $\alpha^\star$ is calculated according to the
frequency-dependent stellar irradiation and the bound-free cross
section of \citep{Osterbrock1989}. Molecular shielding by H$_2$ and
self-shielding is taken into account via the following factors from
\citet{Kamp2000}
\begin{eqnarray}
  s_{\rm C,C}   &=& \exp(-\sigma_{\!\rm C}^{\rm bf} N_{\rm C}) \\
  s_{\rm C,H_2} &=& \exp\left(-0.9\,\Tg^{0.27}
            \Big(\frac{N_{\rm H_2}}{10^{22}\rm cm^{-2}}\Big)^{0.45}\right)
\end{eqnarray} 
\begin{equation}
  R^{\rm ph}_{\rm C} = 
    s_{\rm C,C}\,s_{\rm C,H_2}\;\chi_0\,\alpha\,\exp(-\tau_{\rm UV}) \ ,
  \label{eq:photoRC}
\end{equation}
\ie we refrain from an indirect formulation with $A_V$ in cases we
have the cross sections at hand. The approximation of H$_2$ shielding
for the C ionization is strictly valid at low temperatures only
($T\!<\!300\,$K); for higher temperatures the factor 0.9 should be
dropped.  $N_{\rm C}$ and $N_{H_2}$ are the neutral carbon and
molecular hydrogen column densities toward the UV light source,
respectively, and $\sigma_{\!\rm C}^{\rm bf}\!=\!1.1\times10^{-17}\rm
cm^2$ the FUV-averaged cross section.  The neutral carbon ionization
rates due to radial and vertical UV irradiation are calculated
separately according to Eq.\,(\ref{eq:photoRC}), then multiplied by
the respective solid angles and added together, and then corrected for
scattering as in Eq.\,(\ref{eq:photoR}).

For the photo-dissociation rate of molecular hydrogen, the same
procedure applies with a H$_2$ self-shielding factor taken from
\citet[][see their Eq.37]{Draine1996}. We assume
$\alpha\!=\!4.2\times10^{-11}\rm s^{-1}$ \citep{Draine1996}.
\begin{equation}
  s_{\rm H_2,H_2} = 
         \frac{0.965}{(1+x/b_{\rm H_2})^2} 
       + \frac{0.035}{c\,\exp(8.5\times 10^{-4}\,c)}
\end{equation}
with $x\!=\!N_{H_2}/5\times 10^{14}\rm\,cm^{-2}$, $b_{\rm H_2}$ the
H$_2$ UV line broadening parameter in [km/s] and $c\!=\!(1+x)^{1/2}$.
The line broadening parameter $b$ is defined as $\rm{FWHM}/(4 \ln
2)^{1/2}$; it contains the sum of thermal and turbulent velocities $b
= (\frac{2 k T}{m} + \Delta v^2)^{1/2}$.  Observations of line width in
protoplanetary disks show that the turbulent velocities are below
0.1~km/s \citep[e.g.][]{Guilloteau1998,Simon2000}.

The CO photo-dissociation rate is calculated from detailed band
opacities in a similar fashion, taking into account the shielding by
molecular hydrogen and the self-shielding.
\begin{equation}
  s_{\rm CO,H_2} = \exp\left(-5\cdot 10^{\displaystyle x}\right) 
\end{equation}
with $x\!=\!9.555\times 10^{-4}\left(\log_{10}\!N_{H_2}\right)^{2.684}
- 3.976$.  The CO photodissociation rate for each band is interpolated
from pre-tabulated rates using the CO column density, gas temperature
and line broadening parameter as input \citep{Kamp2000,Bertoldi1996}. 


\subsection{H$_2$ formation on grains}
\label{sec:H2form}
The formation of H$_2$ on grain surfaces $\rm H + H + {\rm grain} \to
H_2 + {\rm grain}$ is taken into account according to \citep{Cazaux2002}
\begin{equation}
  R_{\rm H_2} = \frac{1}{2}\,v^{\rm th}_{\rm H}(\Tg)
                \,n_d\,4\pi\langle a^2\rangle
                \,\alpha_H\,\varepsilon(\Td) 
  \label{eq:H2form}
\end{equation}
with latest updates for the temperature-dependent efficiency
$\varepsilon(\Td)$ from S.~Cazaux (2008, priv.comm.).  $v^{\rm
th}_{\rm H}\!=\!(k\Tg/(2\pi\,m_{\rm H}))^{1/2}$ is the thermal
relative velocity of the hydrogen atom, $n_d\,4\pi\langle
a^2\rangle$ is the total surface of the dust component per volume, and
$\alpha_H\!\approx\!0.223$ is the sticking coefficient, which results
in $\frac{1}{2}\,v^{\rm th}_{\rm H}(100\,{\rm K})\,4\pi\langle
a^2\rangle\,(n_d/\nH)\,\alpha_H\!=\!3\times 10^{-17}\rm cm^3/s$ for
standard ISM grain parameter $\rho_d/\rho\!=\!0.01$,
$\amin\!=\!0.005\,\mu$m, $\amax\!=\!0.25\,\mu$m, $\apow\!=\!3.5$ and
$\rho_{\rm gr}\!=\!2.5\,$g/cm$^{-3}$. The rate coefficient $R_{\rm
H_2}$ still needs to be multiplied by the neutral hydrogen particle
density $n_{\rm H}$ to get the H$_2$ formation rate $[\rm
cm^{-3}s^{-1}]$.

\subsection{Chemistry of excited H$_2$}
\label{sec:excH2}
13 reactions for vibrationally excited molecular hydrogen H$_2^\star$
are taken into account as described in \citep{Tielens1985}.
The FUV pumping rate ${\rm H_2} + h\nu \to {\rm H_2}^\star + h\nu'$ is
assumed to be 10 times the H$_2$ photo-dissociation rate. Two additional
reactions are added for the collisional excitation by 
H and H$_2$ as inverse of the de-excitation reactions
\begin{eqnarray}
 & {\rm H_2 + H  \to H_2^\star + H} :    & \quad  R=C_{ul}^{\rm H}(\Tg)     
                                        \,\exp(-\Delta E/k\Tg)\\
 & {\rm H_2 + H_2 \to H_2^\star + H_2} : & \quad  R=C_{ul}^{\rm  H_2}(\Tg) 
                                        \,\exp(-\Delta E/k\Tg) \ ,
\end{eqnarray}
where the energy of the pseudo vibrational level $\Delta E=2.6\,$eV as
well as the collisional de-excitation rate coefficients $C_{ul}^{\rm
H}$ and $C_{ul}^{\rm H_2} \;\rm[cm^{-3}s^{-1}]$ are given in
\citep{Tielens1985}.

\subsection{Ice formation and evaporation}
\label{sec:ice}

The formation of ice mantles on dust grains plays an important
role for the chemistry in the dark and shielded midplane.  At
the moment, five ices are considered: CO\#, CO$_2$\#, H$_2$O\#,
CH$_4$\# and NH$_3$\# which are treated as additional species in the
chemistry (Sects.~\ref{sec:chem} and \ref{sec:kinCE}).  Apart from
the adsorption and desorption reactions of these species and the H$_2$
formation on grains (Sect.~\ref{sec:H2form}) no other surface reactions are
currently taken into account. In particular, we do not form water on
grain surfaces.

Considering collisional adsorption, and thermal, cosmic-ray and
photo-desorption, the total formation rate of ice species $i$ is
\begin{equation}
  \frac{d n_{i\#}}{dt} = n_i R^{\rm ads}_i \,-\, n_{i\#}^{\rm desorb}
            \left(R^{\rm des, th}_i\!
                 +R^{\rm des, ph}_i\!
                 +R^{\rm des, cr}_i\right)
\end{equation}
where $n_{i\#}$ is the density of ice units $i$ and 
$n_{i\#}^{\rm desorb}$ the fraction of $n_{i\#}$ located
in the uppermost active surface layers of the ice mantle.

\subsubsection{Adsorption}

A gas species will adsorb on grain surfaces upon collision.  The
adsorption rate [s$^{-1}$] is the product of the sticking coefficient
$\alpha$, the total grain surface area per volume $4\pi\langle
a^2\rangle n_d$ and the thermal velocity $v^{\rm
th}_i\!=\!(k\Tg/(2\pi m_i))^{1/2}$
\begin{equation}
  R^{\rm ads}_i = 4\pi\langle a^2\rangle n_d\,\alpha\,v^{\rm th}_i
\end{equation}
where $m_i$ is the mass of gas species $i$.  We assume unit sticking
coefficient ($\alpha\!=\!1$) for all species heavier than Helium
\citep{Burke1983}. 

\subsubsection{Desorption}

A chemical species with internal energy greater than the energy that
binds it to a grain surface will desorb. Desorption mechanisms depend
on the source of the internal energy. 

\smallskip
\noindent{\sl 1. Thermal desorption:}\ \ An ice species $i$ at the
surface of a grain at temperature $T_d$ has probability to desorb
\begin{equation}
  R^{\rm des, th}_i = \nu^{\rm osc}_i 
                      \exp\left(-\frac{E^{\rm ads}_i}{kT_d}\right) \ ,
  \label{eq:Rdesth}
\end{equation}
where $\nu^{\rm osc}_i\!=\!(2\,n_{\rm surf}\,k E^{\rm ads}_i/(\pi^2
m_i))^{1/2}$ is the vibrational frequency of the species in the
surface potential well of ice species $i$, $n_{\rm surf}\!=\!1.5\times
10^{15}$\,cm$^{-2}$ is the surface density of adsorption sites and
$E^{\rm ads}_i$ is the adsorption binding energy. The adopted values
are provided in Table~\ref{tab:adsorption_energy}. Following
\citet{Aikawa1996}, the number density of ice units at the active
surface $n_{i\#}^{\rm desorb}$ is given by
\begin{equation}
  n_{i\#}^{\rm desorb} = \left(\begin{array}{ll}
    \displaystyle  n_{i\#}             
        & \quad,\ \ n^{\rm ice}_{\rm tot}<n_{\rm act} \\
    \displaystyle  n_{\rm act} \frac{n_{i\#}}{n^{\rm ice}_{\rm tot}} 
        & \quad,\ \ n^{\rm ice}_{\rm tot}\ge n_{\rm act} 
  \end{array}\right.
\end{equation}
where $n_{\rm act}\!=\!4\pi\langle a^2\rangle n_d\,n_{\rm surf}N_{\rm Lay}$
is the number of active surface places in the ice mantle per volume and 
$N_{\rm Lay}$ is the number of surface layers to be considered
as ``active''. We assume $N_{\rm Lay}\!=\!2$ in accordance with
\citep{Aikawa1996}. $n^{\rm ice}_{\rm tot}\!=\!\sum_j n_{j\#}$ is the
number density of ices.

\smallskip
\noindent{\sl 2. Photo-desorption:}\ \ Absorption of a UV photon by a
surface species can increase the species internal energy enough to
induce desorption. The photo-desorption rate of species $i$ is given by
\begin{equation}
  R^{\rm des, ph}_i = \pi\langle a^2\rangle \frac{n_d}{n_{\rm act}}
                      \,Y_i\,\chi F_{\rm Draine}
\end{equation}
where $Y_i$ is the photo-desorption yield (see
Table~\ref{tab:adsorption_energy}), $\chi F_{\rm Draine}$ is a
flux-like measure of the local UV energy density [photons/cm$^2$/s]
computed from continuum radiative transfer (Eqs.\,\ref{eq:chi},
\ref{eq:chicalc}).  Photo-desorption can enhance gas-phase water
abundances by orders of magnitude in outer region of disks
\citep{Willacy2000ApJ...544..903W, Dominik2005ApJ...635L..85D,
Oberg2008arXiv0812.1918O}.

\begin{table}
\centering
\caption{Adsorption energies and photo-desorption yields.}
\label{tab:adsorption_energy}
\begin{tabular}{lcc}
\\[-4.5ex]
\hline
species & adsorption energy    & photo-desorption yield \\
        & $E^{\rm ads}_i$ [K]  & $Y_i$ [per UV photon] \\
\hline
\hline
&&\\[-2.2ex]
CO\#      &    960 $^a$   & 2.7 $\times$ 10$^{-3}$ $^c$\\
CO$_2$\#  &   2000 $^a$   & 1.0 $\times$ 10$^{-3}$ $^e$\\
H$_2$O\#  &   4800 $^b$   & 1.3 $\times$ 10$^{-3}$ $^d$\\
CH$_4$\#  &   1100 $^a$   & 1.0 $\times$ 10$^{-3}$ $^e$\\  
NH$_3$\#  &    880 $^a$   & 1.0 $\times$ 10$^{-3}$ $^e$\\
\hline
\end{tabular}\\[0.5mm]
\hspace*{2mm}\begin{minipage}{8cm}
\footnotesize
$a$: \citet{Aikawa1996};
$b$: \citet{Sandford1990};
$c$: \citet{Oberg2007ApJ...662L..23O}; 
$d$: \citet{Westley1995Natur.373..405W}; 
$e$: assumed values. 
\end{minipage}
\end{table}

\smallskip
\noindent{\sl 3. Cosmic-ray induced desorption:}\ \ Cosmic-rays
hitting a grain can locally heat the surface and trigger
desorption. Cosmic-rays can penetrate deep into obscured regions,
maintaining a minimum amount of species in the gas-phase. Cosmic-ray
fluxes in disks may be higher than in molecular clouds because of the
stellar energetic particles in addition to the galactic
component. X-ray photons can also penetrate deep inside the disk and
locally heat a dust grain but X-ray induced desorption is not included
in the code yet. We adopt for the cosmic-ray desorption the formalism
of \citet{Hasegawa1993MNRAS.261...83H}.
\begin{equation}
  R^{\rm des, cr}_i = f(70K)\,R^{\rm des, th}_i(70K)\,
                      \frac{\zeta_{\rm CR}}{5\times 10^{-17}{\rm s}^{-1}}
\end{equation}
where $\zeta_{\rm CR}$ is the cosmic ray ionization rate of H$_2$,
$f(70K)\!=\!3.16 \times 10^{-19}$ the 'duty-cycle' of the grain
at 70\,K and $R^{\rm des, th}_i(70\,\rm K)$ the thermal desorption rate
for species $i$ at temperature $\Td\!=\!70$\,K. The adopted
value for $f(70K)$ is strictly valid only for 0.1 $\mu$m grains
in dense molecular clouds.

\subsection{Kinetic chemical equilibrium}
\label{sec:kinCE}

Assuming kinetic chemical equilibrium in the gas phase, and between gas
and ice species, we have $\frac{d n_i}{dt}\!=\!0$ in
Eq.\,(\ref{eq:chem}) and obtain $i\!=\!1\,...\,\Nsp$ non-linear equations
for the unknown particle densities $n_j$ $(j\!=\!1\,...\,\Nsp)$
\begin{equation}
  F_{\!i}(n_j) \,=\,0 \ .
  \label{eq:chemEq}
\end{equation}
It is noteworthy that the electron density $n_{\rm e}$ is not among
the unknowns, but is replaced by the constraint of charge conservation
\begin{equation}
  n_{\rm e} = \sum_j n_j \,z_j
\end{equation}
where $z_j$ is the charge of species $j$ in units of the elementary
charge. The explicit dependency of $n_{\rm e}$ on the particle
densities $n_j$ causes additional entries in the
chemical Jacobian $d F_i/d n_j = \partial
F_j/\partial n_j + \partial F_i/\partial n_{\rm e} \cdot \partial n_{\rm
e}/\partial n_j$.


\subsection{Element conservation}

The system of Eqs.~(\ref{eq:chem}) is degenerate because every
individual chemical reaction obeys several element conservation
constraints, and therefore, certain linear combinations of $F_{\!j}$
can be found which cancel out, making the equation system
under-determined. Only if the element conservation is implemented in
addition, the system (Eqs.~\ref{eq:chem}) becomes well-defined.

Considering the total hydrogen nuclei density $\nH$ as given, the
conservation of element $k$ can be written as
\begin{equation}
  \nH\,\epsilon_k - \sum_i n_i\,\nu_{i,k} \,=\, 0\ ,
  \label{eq:ElCons1}
\end{equation}
resulting in $N_{\rm el}$ auxiliary conditions.  $\epsilon_k$ is the
elemental abundance of element $k$ normalized to hydrogen and
$\nu_{i,k}$ are the stoichiometric coefficient of species $i$ with respect
to element $k$.

Alternatively, the gas pressure $p$ may be considered as the given
quantity and the relative element conservation can be expressed by
\begin{equation}
  \hat\epsilon_k \Big(n_{\rm e}\,m_{\rm e}+\sum_i n_i m_i\Big)
          - \sum_i n_i\,\nu_{i,k}\,m_k \,=\,0\ ,
  \label{eq:ElCons2}
\end{equation}
where $\hat\epsilon_k=\epsilon_k m_k/(\sum_{k'} \epsilon_{k'} m_{k'})$
is the relative mass fraction of element $k$, $m_i$ the mass of a gas
particle of kind $i$ and $m_k$ the mass of element $k$.  Since summing
up all Eqs.\,(\ref{eq:ElCons2}) for $k\!=\!1\,...\,N_{\rm el}$ results
in $\rho\!=\!\rho$, one of these equations is redundant and can be
replaced by the constraint of given pressure
\begin{equation}
  p - \sum_i n_i \left( 1 + z_i\right) k\Tg \,=\,0\ ,
  \label{eq:pgas}
\end{equation}
where $k$ the Boltzmann constant. The element conservation is
implemented by replacing $N_{\rm el}$ selected components of $F_{\!j}$
in Eq.\,(\ref{eq:chemEq}) by these auxiliary conditions, either
according to Eq.\,(\ref{eq:ElCons1}) or according to
Eqs.\,(\ref{eq:ElCons2}) and (\ref{eq:pgas}), after suitable
normalization. For this purpose, we choose for every element $k$ the
index $j$ that belongs to the most abundant species containing this
element. Eq.\,(\ref{eq:pgas}) overwrites the entry for the most
abundant H-containing species.

The global iteration, which solves the hydrostatic disk structure
consistently with the chemistry and heating \& cooling balance (see
Fig.~\ref{fig:iteration}), is found to converge only if the chemistry
is solved for constant pressure $p$. Since the vertical hydrostatic
condition (Eq.\,\ref{eq:motion2}) is a pressure constraint, it is
essential to ensure that the chemistry solver, coupled to the
$\Tg$-determination via heating\plus cooling balance, is not allowed
to change $p$ as it would be the case if $\nH$ was fixed. At given
pressure $p$, $\Tg$ may be found to increase during the course of the
iteration, but only if simultaneously $\nH$ drops, thereby conserving
the $p$-structure within one global iteration step.

\begin{table}
\centering
\caption{Assumed element abundances in (gas\,+\,ice)}
\label{tab:Abundances}
\begin{tabular}{c|cc}
\\[-4.5ex]
\hline 
element & $12+\log\epsilon$ & mass fraction $\hat\epsilon$\\
\hline 
\hline 
&&\\[-2.2ex]
H   &  12.00 & 7.66$\times 10^{-1}$\\
He  &  10.88 & 2.28$\times 10^{-1}$\\
C   &   8.11 & 1.19$\times 10^{-3}$\\
N   &   7.33 & 2.28$\times 10^{-4}$\\
O   &   8.46 & 3.52$\times 10^{-3}$\\
Mg  &   6.62 & 7.76$\times 10^{-5}$\\
Si  &   6.90 & 1.70$\times 10^{-4}$\\
S   &   6.28 & 4.64$\times 10^{-5}$\\
Fe  &   6.63 & 1.82$\times 10^{-4}$\\
\hline 
\end{tabular}\\[1mm]
\footnotesize
This choice of element abundances implies $\rho=1.315{\rm\,amu}\cdot\nH$.\\
\end{table}

\subsection{Numerical solution of chemistry}

The non-linear equation system (\ref{eq:chemEq}), expressing the
kinetic chemical equilibrium including element conservation, is
usually solved by means of a self-developed, globally convergent
Newton-Raphson method.  A quick and reliable numerical solution of the
Eqs.\,(\ref{eq:chemEq}) is crucial for the computational time
consumption, stability, and global convergence of our model. Our
numerical experience shows that a careful storage of converged results
(particle densities) is the key to increase stability and
performance. These particle densities are used as initial guesses for
the next time the Newton-Raphson method is invoked, either in form of
a downward-outward sweep through the grid (first iteration), or from
the last results of the same point (following iterations).  

In cases, where the solution by the Newton-Raphson method fails, we
fall back to the time-dependent case and solve Eqs.\,(\ref{eq:chem})
by means of the ODE solver {\sc Limex} \citep{Deuflhard1987}
for $10^7\,$yrs, which is much slower but in practice gives the same
results as the Newton-Raphson method.


\section{Gas thermal balance}
\label{sec:heatcool}

The net gain of thermal kinetic energy is written as
\begin{equation}
  \frac{de}{dt} =  
    \sum\limits_k \Gamma_k(\Tg,n_{\rm sp}) 
  - \sum\limits_k \Lambda_k(\Tg,n_{\rm sp})
  \label{eq:energ}
\end{equation}
where $\Gamma_k$ and $\Lambda_k$ are the various heating and cooling
rates $\rm [erg\,cm^{-3}\,s^{-1}]$ which are detailed in the
forthcoming sub-sections. Restricting ourselves to the case of thermal
energy balance, we assume $de/dt\!=\!0$ in the following and
Eq.\,(\ref{eq:energ}) states an implicit equation for the unknown
kinetic gas temperature $\Tg$. Since the heating and cooling rates
depend not only on $\Tg$, but also on the particle densities $n_{\rm
sp}$, which themselves depend on $\Tg$, an iterative process is required
during which $\Tg$ is varied and the the chemistry is re-solved until
$\Tg$ satisfies Eq.~(\ref{eq:energ}).

\subsection{Non-LTE treatment of atoms, ions and molecules}
\label{sec:NLTE}

The most basic interaction between matter and radiation is the
absorption and emission of line photons by a gas
particle, which can be an atom, ion or molecule.  We consider a
$N$-level system with bound-bound transitions only and calculate the
level populations $n_j\,\rm[cm^{-3}]$ by means of the statistical
equations
\begin{equation} 
  n_i \sum\limits_{j\ne i} R_{ij} = \sum\limits_{j\ne i} n_j R_{ji} 
  \label{eq:NLTE} \quad,
\end{equation} 
which are solved together with the equation for the conservation of the
total particle density of the considered species 
$\sum_i n_i\!=\!n_{\rm sp}$. The rate coefficients are given by 
\citep{Mihalas1978}:
\begin{eqnarray} 
  R_{ul} &=& A_{ul}
          +  B_{ul}\Jquer_{ul} + C_{ul} \label{eq:Rul} \nonumber\\ 
  R_{lu} &=& B_{lu}\Jquer_{ul} + C_{lu} 
  \label{eq:Rlu} \quad, 
\end{eqnarray} 
where $u$ and $l$ label an upper and lower level, respectively.
$A_{ul}$, $B_{ul}$, $B_{lu}$, $C_{ul}$ and $C_{lu}$ are the Einstein
coefficients for spontaneous emission, absorption, stimulated emission
and the rate coefficients for collisional (de-)excitation,
respectively.  Additionally we have the Einstein relations
$B_{ul}/A_{ul}\!=\!c^2/(2h\nu_{ul}^3)$, $B_{lu}/B_{ul}\!=\!g_u/g_l$
and the detailed balance relation $C_{lu}/C_{ul}\!=\!g_u/g_l \cdot
\exp(-\Delta E_{ul}/k\Tg)$, where $\nu_{ul}$, $g_u$, $g_l$ and $\Delta
E_{ul}$ are the line center frequency, the statistical weights of the
upper and lower level and the energy difference, respectively.
The line integrated mean intensity is given by
\begin{equation}
  \Jquer_{ul} = \frac{1}{4\pi}
  \iint\!\phi_{ul}(\nu,\nn)\,I_{\nu}(\nn)\,d\nu\,d\Omega
  \label{eq:Jquer}
\end{equation} 
where $\phi_{ul}(\nu,\nn)$ is the line profile function in direction $\nn$.

\subsubsection{Escape probability treatment}

\begin{figure}
  \centering
  \includegraphics[height=2.9cm,width=4.5cm]{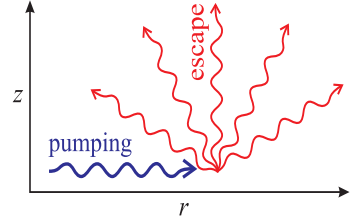}
  \caption{Different pumping and escape probabilities 
           according to the predominantly radial irradiation and
           the predominantly vertical escape.}
  \label{fig:escape}
  \vspace*{-1mm}
\end{figure}

The spectral intensity $I_{\nu}$ in Eq.\,(\ref{eq:Jquer}) is
affected by line absorption and emission. Assuming that the line
source function (Eq.\,\ref{eq:lsource}) varies slowly in a local
environment where the line optical depths (Eq.\,\ref{eq:tauL})
build up rapidly, we can approximate for a static, plane-parallel medium
\begin{equation}
  I_{\nu}(\mu) \,\approx\, 
    \Icont_{\nu_{ul}}(\mu)\exp\Big(\frac{\phi_{ul}(\nu)\,\tauL_{ul}}{\mu}\Big) 
   +\SL_{ul}\left(1-\exp\Big(\frac{\phi_{ul}(\nu)\,\tauL_{ul}}{\mu}\Big)\right)
   \label{eq:Icont}
\end{equation}
where $\Icont_{\nu_{ul}}(\mu)$ is the continuous
background intensity which propagates backward along the ray
in direction $\mu$. The direction $\mu$=1 points ``outward''
($\mu\!=\!\cos\theta$). The line source function and the perpendicular line
optical depth are given by \citep{Mihalas1978}
\begin{eqnarray}
  \SL_{ul} &=& \frac{2h\nu_{ul}^3}{c^2}
             \left( \frac{g_u n_l}{g_l n_u} - 1 \right)^{-1} 
  \label{eq:lsource}\\
  \tauL_{ul} &=& \frac{A_{ul}\,c^3}{8\pi\nu_{ul}^3\Delta {\rm v}_D}
     \int_z^{z_{\rm max}}\!\!\Big(n_l(z')\frac{g_u}{g_l}-n_u(z')\Big)\,dz'
  \label{eq:tauL}
\end{eqnarray}
where $\Delta {\rm v}_D$ is the (turbulent + thermal) velocity Doppler
width of the line, assumed to be constant along the line of sight in
Eq.\,(\ref{eq:tauL}).  Equations (\ref{eq:Jquer}) and (\ref{eq:Icont})
can be combined to find
\begin{eqnarray}
  \Jquer_{ul} &=& \frac{1}{2} \int_{-1}^1 
      \pe_{ul}(\mu)\,\Icont_{\nu_{ul}}(\mu)
    + \left(1-\pe_{ul}(\mu)\right) \SL_{ul} \;d\mu \\
    &\approx& \Pp_{ul}\,\Jcont_{\nu_{ul}} 
  \;+\; \left(1-\Pe_{ul}\right)\,\SL_{ul}  
  \label{eq:Jquerapprox}
\end{eqnarray}
where the direction-dependent and the mean escape probabilities 
are found to be
\begin{eqnarray}
  \pe_{ul}(\mu) &=& \int_{-\infty}^{+\infty} \phi(x)\,
               \exp\big(-\frac{\tauL_{ul}\,\phi(x)}{\mu}\big)\,dx \\
  \Pe_{ul} &=& \frac{1}{2}\int_{-1}^{\,1} \pe_{ul}(\mu)\;d\mu
\end{eqnarray}
with dimensionless line profile function
$\phi(x)\!=\!\exp(-x^2)/\!\sqrt{\pi}$,\linebreak
$x\!=\!(\nu-\nu_{ul})/\Delta\nu_D$ and frequency width
$\Delta\nu_D\!=\!\nu_{ul}\Delta {\rm v}_D/c$.
Using Eq.\,(\ref{eq:Jquerapprox}), it is straightforward to show that
the unknown line source function $\SL_{ul}$ can be eliminated, and the
leading term $n_u A_{ul}$ cancels out, when considering the net rate
$n_u A_{ul} + (n_u B_{ul} - n_l B_{lu})\,\Jquer_{ul} = n_u
A_{ul}\Pe_{ul} + (n_u B_{ul} - n_l
B_{lu})\Pp_{ul}\Jcont_{\nu_{ul}}$. Thus, we can solve the statistical
rate Eqs.\,(\ref{eq:NLTE}) with modified rate coefficients
\begin{eqnarray} 
  \tilde{R}_{ul} &=& A_{ul}\Pe_{\!ul} 
          +  B_{ul}\Pp_{\!ul}\Jcont_{\nu_{ul}} + C_{ul} \nonumber\\ 
  \tilde{R}_{lu} &=& B_{lu}\Pp_{\!ul}\Jcont_{\nu_{ul}} + C_{lu} 
  \quad, 
  \label{eq:Rmod}
\end{eqnarray} 
which is known as escape probability formalism
\citep{Avrett1965,Mihalas1978}. $\Pe_{\!ul}$ is the mean probability
for line photons emitted from the current position to escape the local
environment and $\Pp_{\!ul}$ the mean probability for continuum
photons to arrive at the current position.  $\Jcont_{\nu_{ul}}$ is the
continuous mean intensity at line center frequency
$\nu_{ul}$, as would be present if no line transfer effects took
place. In semi-infinite slab symmetry, all directions $\mu\!<\!0$ have
infinite line optical depth and can be discarded from the calculation
of the escape probabilities
\begin{eqnarray}
  \Pe_{ul}(\tauL_{ul}) &=& \frac{1}{2}\int_{-\infty}^{+\infty}\!\!\!\!
               \phi(x)\!\!\int_0^1\!\!
               \exp\big(-\frac{\tauL_{ul}\phi(x)}{\mu}\big)\;d\mu\,dx \\
             &=& \frac{1}{2}\int_{-\infty}^{+\infty}\!\!\!\!
	       \phi(x)\,E_2\big(\tauL_{ul}\phi(x)\big)\,dx
  \label{eq:pesc}
\end{eqnarray}
This function is numerically fitted as\\[2mm]
\noindent\resizebox{\columnwidth}{!}{$
  \Pe_{ul}(\tau) = \left\{ \begin{array}{ll}
   0.5                                        & , \tau\!\leq\!0\\[1mm]
   0.5 + (0.1995\ln\tau\!-\!0.2484)\tau  
       -0.04594\tau^2                         & , 0\!<\!\tau\!\leq\!0.6\\[1mm]
   \displaystyle\frac{1 - \exp(-1.422\tau)}
     {3.324\tau+0.2852\tau^2}                 & , 0.6\!<\!\tau\!\leq\!9\\[4mm]
   \displaystyle\frac{0.1999}{\tau}
     \big(\ln(0.4799\tau)\big)^{-0.4195}      & , 9\!<\!\tau
\end{array}\right.$}\\[2mm]

\noindent Considering the pumping probability as defined by
Eq.\,(\ref{eq:Jquerapprox}), it is noteworthy that
$\Pp_{ul}\!\approx\!\Pe_{ul}$ is only valid in an almost iso\-tropic
background radiation field. In disk symmetry, much of the pumping is
due to direct star light (see Fig.\,\ref{fig:chis}) which has a very
pointed character. In the optically thick midplane, the continuous
radiation field is almost isotropic, but here the pumping is
pointless, because the radiation is thermalized and the collisional
processes dominate. Considering near to far IR wavelengths at a
certain height above the midplane, the irradiation from underneath
plays a role, but these directions are just the opposite of what is
considered in Eq.\,(\ref{eq:pesc}), and so using
$\Pp_{ul}\!\approx\!\Pe_{ul}$ would be strongly misleading. Thus, we
approximate
\begin{eqnarray}
  \Pp_{ul}(\tau^{\rm rad}_{ul}) &=& \int_{-\infty}^{+\infty}\!\!\!\!
               \phi(x)\,\exp\big(-\tau^{\rm rad}_{ul}\phi(x)\big)\;dx 
  \label{eq:pump}
\end{eqnarray}
with $\tau^{\rm rad}_{ul}$ now being the radially inward line
optical depth. This function is numerically fitted as\\[2mm]
\noindent\resizebox{\columnwidth}{!}{$
  \Pp_{ul}(\tau) = \left\{ \begin{array}{ll}
     1                                         & , \tau\!\leq\!0\\[1mm]
     1 - 0.3989\tau 
       + 0.09189\tau^2
       - 0.01497\tau^3                         & , 0\!<\!\tau\!\leq\!0.9\\[1mm]
  \displaystyle\frac{1-\exp(-0.6437\tau)}
   {0.6295\tau + 0.07008\tau^2}                & , 0.9\!<\!\tau\!\leq\!9\\[4mm]
  \displaystyle\frac{0.8204}{\tau}
   \big(\ln(0.3367\tau)\big)^{-0.4306}         & , 9\!<\!\tau
\end{array}\right.$}\\[2mm]

\begin{figure}
\centering
\includegraphics[height=6.5cm,width=8cm]{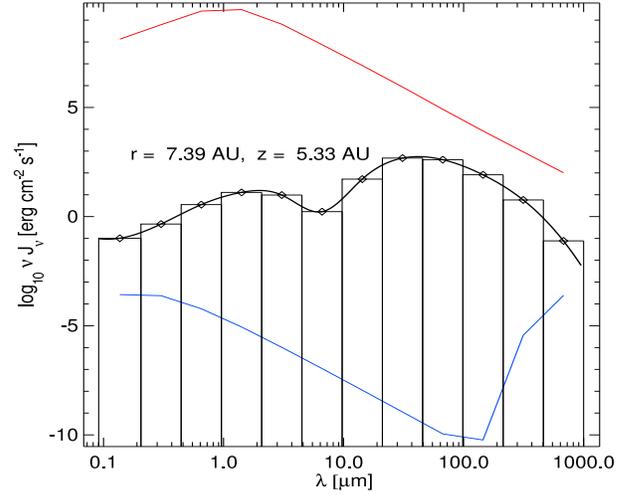}
\caption{Continuum mean intensities as input for non-LTE modelling.
         The calculated band-mean mean intensities are shown for one
         particular point $(r,z)$ in a model (12 black dots) and a
         cubic spline interpolation through these points (black
         line). The vertical lines indicate the interval boundaries of
         the 12 spectral bands.  The red line shows the band-mean
         incident stellar intensities $\nu I_\nu^\star$ and the blue
         line shows the incident interstellar intensities $\nu
         I_\nu^{\rm ISM}$. The radiation field has two major
         components, the dust attenuated UV -- near IR part,
         originating mainly from the star, and the self-generated mid
         -- far IR part, originating from thermal dust emission in the
         disk.}
\label{fig:Jspline}
\end{figure}

\subsubsection{Background radiation field}

The continuum background mean intensities $\Jcont_{\nu_{ul}}$ have an
important impact on the gas energy balance. For example, in strong
continuum radiation fields, the reverse process to line cooling,
namely line absorption followed by collisional de-excitation,
dominates.  $\Jcont_{\nu_{ul}}$ is identified to be just given by the
mean intensities calculated from the dust continuum radiative transfer
(see Sect.~\ref{sec:RT}). In order to obtain the required
monochromatic mean continuum intensities at the line center positions,
we apply a cubic spline interpolation to the calculated local
continuum $J_\nu^{\rm cont}(r,z)$ in frequency space as depicted in
Fig.~\ref{fig:Jspline}.

\subsubsection{Solving the statistical equations}

Equations (\ref{eq:NLTE}, \ref{eq:tauL}, \ref{eq:Rmod}) form a system
of coupled equations for the unknown population numbers $n_i$. Since
the line optical depths (Eq.\,\ref{eq:tauL}) depend partly on the
local $n_i$, these equations must be solved iteratively. We apply a
fully implicit integration scheme for the numerical solution of
(Eq.\,\ref{eq:tauL}) where the line optical depth increment along the
last downward integration step, between the previous and the current
grid point, is given by the local populations, \ie
\begin{equation}
  \Delta\tauL_{i} = \frac{A_{ul}\,c^3}{8\pi\nu_{ul}^3\Delta {\rm v}_D}
     \frac{n_l\frac{g_u}{g_l}-n_u}{\nH} 
     \big(N_\HH(z_i)-N_\HH(z_{i+1})\big) \ .
\end{equation}
where $N_\HH(z_i)$ is the vertical hydrogen
column density at grid point $z_i$, and $n_u$, $n_l$ and $\nH$ refer
to the current grid point $i$. A simple $\Lambda$-type iteration
scheme is found to converge within typically 0 to 20 iterations.
The outward radial line optical depth increments
$\Delta\tau_j^{\rm rad}$ between $r_{j-1}$ and $r_j$ are calculated 
in a similar fully implicit fashion.

\subsubsection{Calculation of the heating/cooling rate}

Once the statistical equations (Eqs.\,\ref{eq:NLTE}) have been solved,
the radiative heating and cooling rates can be determined. There are
two valid approaches. For the net cooling rate, one can either
calculate the net creation rate of photon energy (radiative approach),
or one can calculate the total destruction rate of thermal energy
(collisional approach).
\begin{eqnarray}
  \Gamma_{\rm rad} &=& \sum\limits_{u>l} n_l \Delta E_{ul} 
                       \Pp_{ul}B_{lu}\Jcont_{\nu_{ul}}\\
  \Lambda_{\rm rad}&=&\sum\limits_{u>l} n_u \Delta E_{ul} 
            \,\big(\Pe_{ul}A_{ul} + \Pp_{ul}B_{ul}\Jcont_{\nu_{ul}}\big)\\
  \Gamma_{\rm col} &=& \sum\limits_{u>l} n_u C_{ul} \Delta E_{ul}\\
  \Lambda_{\rm col}&=& \sum\limits_{u>l} n_l C_{lu} \Delta E_{ul}
\end{eqnarray}
Both approaches must yield the same net result $\Gamma_{\rm
rad}\!-\!\Lambda_{\rm rad}\!=\!\Gamma_{\rm coll}\!-\!\Lambda_{\rm
coll}$, which can be used to check the quality of the numerical
solution. In practise, one pair of these heating/cooling rates is
often huge in comparison to the other pair, \eg $\{\Gamma_{\rm
rad},\Lambda_{\rm rad}\} \gg \{\Gamma_{\rm col},\Lambda_{\rm col}\}$
in a thin gas with radiatively controlled populations, and
$\{\Gamma_{\rm rad},\Lambda_{\rm rad}\} \ll \{\Gamma_{\rm
col},\Lambda_{\rm col}\}$ in a dense gas with close to LTE
populations. Thus, it is numerically favorable to choose
\begin{eqnarray}
  \Gamma  &=& \left\{\begin{array}{lccl}
                \Gamma_{\rm col} &&& , \Gamma_{\rm rad}>   \Gamma_{\rm col}\\
                \Gamma_{\rm rad} &&& , \Gamma_{\rm rad}\le \Gamma_{\rm col}
              \end{array}\right.\\
  \Lambda &=& \left\{\begin{array}{lccl}
                \Lambda_{\rm col}&&& , \Gamma_{\rm rad}>   \Gamma_{\rm col}\\
                \Lambda_{\rm rad}&&& , \Gamma_{\rm rad}\le \Gamma_{\rm col}
              \end{array}\right.
\end{eqnarray}

\begin{table}
\caption{non-LTE model atoms, ions and molecules}
\label{tab:nonLTE}
\begin{tabular}{c|cc|c|c}
\\[-4.5ex]
\hline
species & \hspace*{-1mm}\#levels\hspace*{-1mm} 
        & \hspace*{-3mm}\#lines\hspace*{-2mm} 
        & coll.\,partners & reference\\
\hline
\hline
O\,I               &   3 &   3 & p-H$_2$,o-H$_2$,H,H$^+$,e$^-$    
                   & \hspace*{-1mm}$\Lambda$-database\hspace*{-1mm}\\
C\,I               &   2 &   1 & 
      \hspace*{-1mm}p-H$_2$,o-H$_2$,H,H$^+$,He,e$^-$\hspace*{-2mm} 
                   & \hspace*{-1mm}$\Lambda$-database\hspace*{-1mm}\\
C\,II              &   3 &   3 & H$_2$,H,e$^-$        
                   & \hspace*{-1mm}$\Lambda$-database\hspace*{-1mm}\\
Mg\,II             &   8 &  12 & e$^-$                & {\sc Chianti}\\
Fe\,II             &  80 & 477 & e$^-$                & {\sc Chianti}\\
Si\,II             &  15 &  35 & e$^-$                & {\sc Chianti}\\
S\,II              &   5 &   9 & e$^-$                & {\sc Chianti}\\
\hspace*{-2mm}CO rot. \& ro-vib.\hspace*{-1mm} 
                   & 110 & 243 & p-H$_2$,o-H$_2$,H,He,e$^-$ & see text\\
\hspace*{-2mm}o-H$_2$ ro-vib.\hspace*{-1mm}    
                   &  80 & 803 & p-H$_2$,o-H$_2$,H,He       & see text\\
\hspace*{-2mm}p-H$_2$ ro-vib.\hspace*{-1mm}    
                   &  80 & 736 & p-H$_2$,o-H$_2$,H,He       & see text\\
o-H$_2$O rot.      &  45 & 158 & H$_2$                
                   & \hspace*{-1mm}$\Lambda$-database\hspace*{-1mm}\\
p-H$_2$O rot.      &  45 & 157 & H$_2$                
                   & \hspace*{-1mm}$\Lambda$-database\hspace*{-1mm}\\
\hline
\end{tabular}
\end{table}

\subsubsection{Atomic and molecular data}

The atomic and molecular data for O{\sc i}, C{\sc i}, C{\sc ii} and
H$_2$O (energy levels $E_i$, statistical weights $g_i$, Einstein
coefficients $A_{ul}$, and collision rates $C_{ul}$ are taken from
Leiden's {\sc Lambda}-database \citep{Lambda2005}, see
Table~\ref{tab:nonLTE}. In addition to these low-temperature coolants,
we have included several ions as high-temperature coolants from the
{\sc Chianti}-database \citep{Chianti1997}: Mg\,{\sc ii}, Fe\,{\sc
ii}, Si\,{\sc ii} and S\,{\sc ii}, taking into account all energy
levels up to about 60\,000\,cm$^{-1}$. This database has collisional
data for free electrons only, but since we consider only ions of
abundant elements here, the electron concentration is always rather
high wherever these ions are abundant. Since the electron collisional rates
are typically $10^4$ times larger than
those of heavy particles, the thereby introduced error seems
acceptable.

For CO, we have merged level and radiative data ($E_i$, $g_i$ and
$A_{ul}$) of the rotational and ro-vibrational states
($v\!=\!0,1,2,3,4$) from the {\sc Hitran} database \citep{Hitran2005}
with collisional data among the rotational levels from the {\sc
Lambda} database. For the vibrational collisions we use the
$C_{1\to0}$ data for H and H$_2$ de-exciting collisions from
\citep{Neufeld1994} and for He collisions from \citep{Millikan1964}.
The de-exciting rate coefficients for other than $1\!\to\!0$
vibrational transitions are estimated according to the formula
provided by \citep{Elitzur1983}
\begin{equation}
  C_{v'\to v} = (v-v')\,C_{1\to0} \exp\left(-\frac{(v-v'-1)\,1.5\,\theta/\Tg}
                                                 {1+1.5\,\theta/\Tg}\right) 
\end{equation}
where $\theta\!=\!\hbar\omega/k$ is the difference between the first
vibrationally excited and the ground state energy. For detailed
ro-vibrational modelling, these total vibrational collisional rates
still need to be spread over the rotational sub-states.  For
simplicity, however, we assume $C_{v'J'\to vJ}\approx C_{v'\to
v}$ for every rotational state $J$.

For \hh, the level and radiative data (quadrupole transitions) 
are taken from \citet{Wolniewicz1998ApJS..115..293W}. We
include calculated collisional excitations by H
\citep{Wrathmall2007MNRAS.382..133W}, ortho- and para-\hh, and Helium
\citep{LeBourlot1999MNRAS.305..802L}.
The H$_2$ and H$_2$O ortho to para abundance ratios are assumed to be 
at thermal equilibrium according to the gas temperature.

\subsection{Specific heating processes}

Below, we list further heating processes that are not covered by
Sect.~\ref{sec:NLTE}.
Photoelectric heating, cosmic ray ionization, carbon photo-ionization
and H$_2$ photo-dissociation are still radiative processes, while
other heating mechanisms are of chemical nature, such as H$_2$
formation heating, or of dynamical nature, such as viscous heating.

\subsubsection{Photo-electric heating}

UV photons impinging on dust grains can eject electrons with
super-thermal velocities which then thermalize through collisions with
the gas. The efficiency of this process decreases strongly with grain
charge (positively charged grains are less efficient heaters). The
grain charge is set by the balance of incoming UV flux that ejects
electrons and collisional recombination. The collision rate for
recombination scales with electron density, thermal velocity and the
ratio between potential and thermal energy ($\Phi\!=\!eU/k\Tg$, with
$U$ being the grain potential). Thus the grain charge can be
parameterized by a 'so-called' grain charge parameter \citep{Bakes1994}
\begin{equation}
  x = \sqrt{\Tg}\,\frac{\chi}{n_{\rm e}} \ .
\end{equation}  
The probability of electron ejection after photon absorption (yield),
is generally taken from experimental data on bulk material with large
flat surfaces, and then applied to (smaller) astrophysical dust grains
to compute the photoelectric heating rates. The heating process is
thought to be less effective for micron-sized grains compared to small
ISM dust grains. The reason is that a photo-electron can more easily
be trapped within the matrix of a large grain, thus lowering the
photoelectric yield. Experimental data on realistic astrophysical dust
grains is sparse and only recently \citep{Abbas2006} carried out
experiments with sub-micron to micron sized individual dust
grains. They measure yields that are larger than those of bulk flat
surfaces and they find an increasing yield with increasing grain
size. However, the underlying physics of these experiments are not yet
well understood.

\citet{Kamp2000} provide a formula to approximate the photoelectric
heating rate for large graphite and silicate grains using the
photoelectric yields of bulk material from \citet{Feuerbacher1972}.
For silicate grains, the photoelectric heating rate $\Gamma_{\rm PE}$
and the efficiency $\epsilon$ are
\begin{eqnarray}
  \Gamma_{\rm PE} &=& 2.5\times 10^{-4} \,\sigma^{\rm abs}\nH\,\epsilon\,\chi
                      \\
  \epsilon &=& \frac{0.06}{1+1.8\times10^{-3}\,x^{\,0.91}}
              +\frac{y\,(10^{-4} \Tg)^{1.2}}{1+0.01 x} \\[2mm]
         y &=& \left\{ \begin{array}{ll}
     0.7   \hspace*{3mm}   & , x \leq 10^{-4}\\[1mm]
     0.36                  & , 10^{-4} < x \leq 1\\[1mm]
     0.15                  & , x > 1
        \end{array}\right.
\end{eqnarray}
valid for electron particle densities $10^{-5}{\rm cm^{-3}}\!<\!n_{\rm
e}\!<\!10^5{\rm cm^{-3}}$, gas temperatures $10\,{\rm
K}\!<\!\Tg\!<10000\,$K, and strength of FUV radiation field
$10^{-5}\!<\!\chi\!<\!10^5$. Here, $\sigma^{\rm abs}$ is the grain absorption
cross section per H-nucleus ($\sigma^{\rm abs}\nH\!=\!\kappa_1^{\rm abs}$,
see Eqs.\,\ref{eq:kband} and \ref{eq:kappa_dust}).


\subsubsection{PAH heating}

Very small dust grains such as polycyclic aromatic hydrocarbons (PAHs)
are an extremely efficient heating source for the gas. The
photoelectric heating rate can be written separately from the rest of
the grain size distribution using only the first term of the \citep{Bakes1994}
efficiency formulation
\begin{equation}
  \Gamma_{\rm PE} = f_{\rm PAH}\,10^{-24}\nH\,\epsilon\,\chi 
\end{equation}
where the efficiency $\epsilon$ is given by
\begin{equation}
  \epsilon = \frac{0.0487}{1 + 4\times 10^{-3}x^{\,0.73}}
\end{equation}
In the ISM, the abundance of PAHs is $f_{\rm PAH}\!=\!1$. For disks, this
value can be scaled according to the observed strength of the PAH bands.

\subsubsection{Carbon photo-ionization}

Ionization of carbon releases electrons with energies around 1~eV
\citep{Black1987}. Subsequent collisions heat the gas as
\begin{equation}
  \Gamma_{\rm C} = 1.602\times 10^{-12}\,R^{\rm ph}_{\rm C}\,n_{\rm C}
\end{equation}
where the photo-ionization rate $R^{\rm ph}_{\rm C}$ is
given by Eq.(\ref{eq:photoRC}).

\subsubsection{H$_2$ photo-dissociation heating}

Photo-dissociation of molecular hydrogen occurs via {\rm UV line
absorption} into an electronically excited state followed by
spontaneous decay into an unbound state of the two hydrogen atoms. The
kinetic energy of such H-atoms is typically 0.4~eV
\citep{Stephens1973}, leading to an approximate heating rate of
\begin{equation}
  \Gamma_{\rm ph}^{\rm H_2} = 6.4 \times 10^{-13}\,R_{\rm ph}^{\rm H_2} 
                              \,n_{\rm H_2}
\end{equation}
Here, $R_{\rm ph}^{\rm H_2}$ is the H$_2$ photo-dissociation rate
given in Sect.~\ref{sec:specialUV} including dust and H$_2$
self-shielding.

\subsubsection{cosmic ray heating}

Cosmic rays have a typical attenuation depth of 96\,g\,cm$^{-2}$ and
thus reach much deeper than stellar FUV photons \citep[$\sim
10^{-3}$g\,cm$^{-2}$, see][for an overview]{Bergin2007}. They
ionize atomic and molecular hydrogen and this inputs approximately 3.5
and 8~eV into the gas for H and H$_2$, respectively
\citep{Jonkheid2004}. The heating rate can then be written as
\begin{equation}
  \Gamma_{\rm CR} = \zeta_{\rm CR} \left( 5.5 \times 10^{-12} n_{\rm H} 
       + 2.5 \times 10^{-11} n_{\rm H_2} \right)
\end{equation}
where $\zeta_{\rm CR}$ is the primary cosmic ray ionization rate.

\subsubsection{H$_2$ formation heating}

The formation of H$_2$ on dust surfaces releases the
binding energy of 4.48~eV. Due to the lack of laboratory data, we
follow the approach by \citet{Black1976} and assume that this
energy is equally distributed over translation, vibration and
rotation. Hence, about 1.5~eV per reaction is liberated as heat 
\begin{equation}
\Gamma_{\rm form}^{\rm H_2} = 2.39 \times 10^{-12}\,R_{\rm H_2}\,n_{\rm H}
\end{equation}
where the H$_2$ formation rate $R_{\rm H_2}$ is given
in Sect.~\ref{sec:H2form}. 

\subsubsection{Heating by collisional de-excitation of H$_2^\star$}

The fluorescent excitation of H$_2$ by UV photons $\rm H_2 + h\nu \to
H_2^{\star\star} \to \rm H_2^\star + h\nu'$ produces vibrationally
excited molecular hydrogen H$_2^\star$ \citep{Tielens1985}, and the
vibrational excitation energy can be converted into thermal energy by
collisions. The heating rate is
\begin{eqnarray}
\Gamma_{\rm coll}^{\rm H_2^\star} &=& \Delta E\, 
     R^{\rm coll}_{\rm H_2\to H_2^\star} 
     \left( n_{\rm H_2^\star} - n_{\rm H_2} 
            \exp\Big(-\frac{\Delta E}{k\Tg}\Big)\right) 
  \label{eq:QexcH2}\\
  R^{\rm coll}_{\rm H_2\to H_2^\star} &=&
          n_{\rm H}  \,C_{ul}^{\rm H}(\Tg)  
        + n_{\rm H_2}\,C_{ul}^{\rm H_2}(\Tg)
\end{eqnarray}
where the excitation energy of the pseudo vibration level 
$\Delta E$ and the collisional de-excitation rates $C_{ul}$ are given
in \citep{Tielens1985}, see also Sect.~\ref{sec:excH2}. The second
term in Eq.\,(\ref{eq:QexcH2}) corrects for collisional excitation.

\subsubsection{Viscous heating}

Due to high optical thickness, radiative heating cannot penetrate
efficiently to the midplane. These dense layers can instead also be
heated by local viscous dissipation \citep{Frank1992}
\begin{equation}
\Gamma_{\rm vis} = \frac{9}{4}\,\rho\,\nu_{\rm kin}\,\Omega_{\rm kep}^2
                  \ .
  \label{eq:Qvis}
\end{equation}
In the absence of a well-understood mechanism, angular momentum
transport is conceptualized using the kinematic -- or turbulent --
viscosity $\nu_{\rm kin}$ often parameterized as an $\alpha$-viscosity
\citep{Shakura1973}
\begin{equation}
  \nu_{\rm kin} = \alpha\,c_T\,H_g\ ,
\end{equation}
where $\alpha$ is a dimensionless scaling factor, $c_T^2\!=\!p/\rho$
is the isothermal sound speed, $H_g\!=\!c_T/\Omega_{\rm kep}$ is the
gas scale height, and $\Omega_{\rm kep}\!=\!v_\phi/r$ is the
Keplerian angular velocity (see Eq.\,\ref{eq:vKepler}).  For
$r\!\la\!2$\,AU, the viscous heating is known to be capable of
dominating the energy balance in the midplane
\citep{Dalessio1998}\footnote{Without further adjustments, the
viscous heating rate according to Eq.\,(\ref{eq:Qvis}) scales as
$\Gamma_{\rm vis}\!\propto\!p$ at given radius $r$. Since all known
cooling rates scale as $\Lambda\!\propto\!\rho^2$ in the low density
limit, there is always a critical height $z$ above which the viscous
heating would dominate the energy balance and lead to ever increasing
$\Tg$ (well above 20000\,K) with increasing height $z$. We consider
this behavior as an artefact of the concept of viscous heating and/or
$\alpha$-viscosity.}.

\subsection{Specific cooling processes}

Most cooling processes are radiative in nature and covered in
Sect.~\ref{sec:NLTE}. However, two prominent high temperature cooling
processes are treated in a simpler approximative fashion:
Lyman-$\alpha$ and O{\sc i}-630nm cooling.

\subsubsection{Ly-$\alpha$ cooling}

Cooling through the Lyman-$\alpha$ line becomes efficient at
temperatures of a few 1000~K \citep{Sternberg1989}. Given the
densities of atomic hydrogen $n_{\rm H}$ and electrons $n_{\rm e}$,
the cooling rate can be written as
\begin{equation}
  \Lambda_{\rm Ly-\alpha} = 7.3 \times 10^{-19} n_{\rm H}\,n_{\rm e} 
              \exp\left( -118\,400/\Tg\right) \ .
\end{equation}

\subsubsection{O{\sc i}-630nm cooling}

Line emission from the meta-stable $^1$D level of neutral oxygen
efficiently cools the gas at temperatures in excess of a few
1000~K. With $n_{\rm O}$ denoting the neutral oxygen particle density 
the cooling rate is \citep{Sternberg1989}
\begin{equation}
  \Lambda_{\rm OI\,630nm} = 1.8 \times 10^{-24} n_{\rm O}\,n_{\rm e} 
          \exp\left( -22\,800/\Tg\right) \ .
\end{equation}

\subsection{Miscellaneous heating/cooling processes}

We list below two additional processes that can cause either heating
or cooling of the gas.

\subsubsection{Thermal accommodation}

Following \citep{Burke1983}, the energy exchange rate by
inelastic collisions between grains and gas particles is
\begin{equation}
\Gamma_{\rm g-g} = 4 \times 10^{-12}\,\pi\langle a^2\rangle\,n_d\,\nH 
                   \,\alpha_{\rm T}\,\sqrt{T_{\rm g}}\,(\Td-\Tg)
\end{equation}
For gas temperatures $\Tg$ higher than dust temperatures $\Td$,
this rate turns into a cooling rate $\Lambda_{\rm g-g}$. The thermal
accommodation coefficient $\alpha_{\rm T}$ is set to the typical value
for silicate and graphite dust of $0.3$ \citep{Burke1983}.

\subsubsection{free-free heating/cooling}
Free-free transitions directly convert photon energy into thermal
energy (ff-heating) or vice versa (ff-cooling) during electron
encounters. The heating rate $\Gamma_{\rm ff}$ and cooling
rate $\Lambda_{\rm ff}$ are given by
\begin{eqnarray}
\Gamma_{\rm ff}  &=& 4\pi \int \kappa^{\rm ff}_\nu J_\nu\,d\nu\\ 
\Lambda_{\rm ff} &=& 4\pi \int \kappa^{\rm ff}_\nu B_\nu(\Tg)\;d\nu  \\
\kappa^{\rm ff}_\nu &=& 
                  n_{\rm e}^2\,\sigma^{\rm ff}_\nu 
                + n_{\rm e}\,n_{\rm H}\,\sigma^{\rm H^-ff}_\nu  
                + n_{\rm e}\,n_{\rm H_2}\,\sigma^{\rm H_2^-ff}_\nu  
                + n_{\rm e}\,n_{\rm He}\,\sigma^{\rm He^-ff}_\nu \ ,
\end{eqnarray}
where $\kappa^{\rm ff}_\nu$ is the free-free gas opacity [cm$^{-1}$]. The
free-free cross-sections $\sigma^{\rm ff}_\nu$ [cm$^5$] for
bremsstrahlung of singly ionized gases are taken from \citep{Hummer1988},
for H$^-$ff from \citep{Stilley1970}, 
for H$_2^-$ff from \citep{Somerville1964}, and 
for He$^-$ff from \citep{John1994}.


\section{Sound Speeds} 

After the chemistry (see Sect.~\ref{sec:chem}) and the thermal gas
energy balance (see Sect.~\ref{sec:heatcool}) have been solved
throughout the disk volume, all particle densities $n_i$ and the
kinetic temperature of the gas $\Tg$ are known, and \ProDiMo can update
the isothermal sound speeds on the numerical grid $\cT2(r_j,z_k)$
as preparation for the next iteration of the hydrostatic disk
structure (see Sect.~\ref{sec:struc}).
\begin{eqnarray}
  \rho &=& n_{\rm e}\,m_{\rm e} + \sum_i n_i\,m_i \\
  p    &=& \Big( n_{\rm e} + \sum_i n_i\Big) \,k\Tg \\
  \cT2 &=& p/\rho
\end{eqnarray}



\begin{figure*}
  \centering
  \begin{tabular}{cc}
    gas in thermal balance & $\Tg\!=\!\Td$ assumed \\[-1mm]
    \hspace*{-5mm}\includegraphics[height=7.5cm,width=8.8cm]{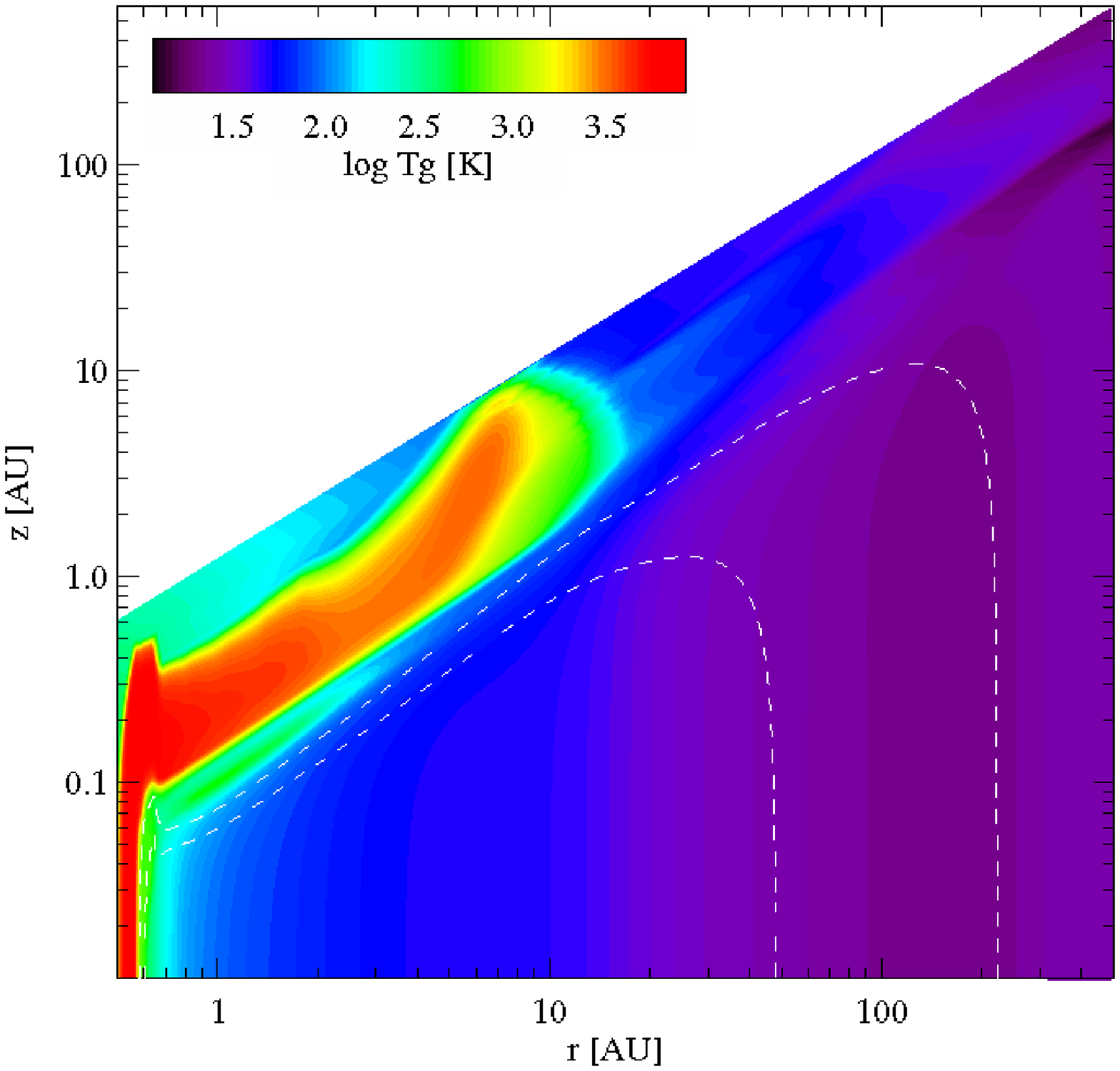} &
    \hspace*{-6mm}\includegraphics[height=7.5cm,width=8.8cm]{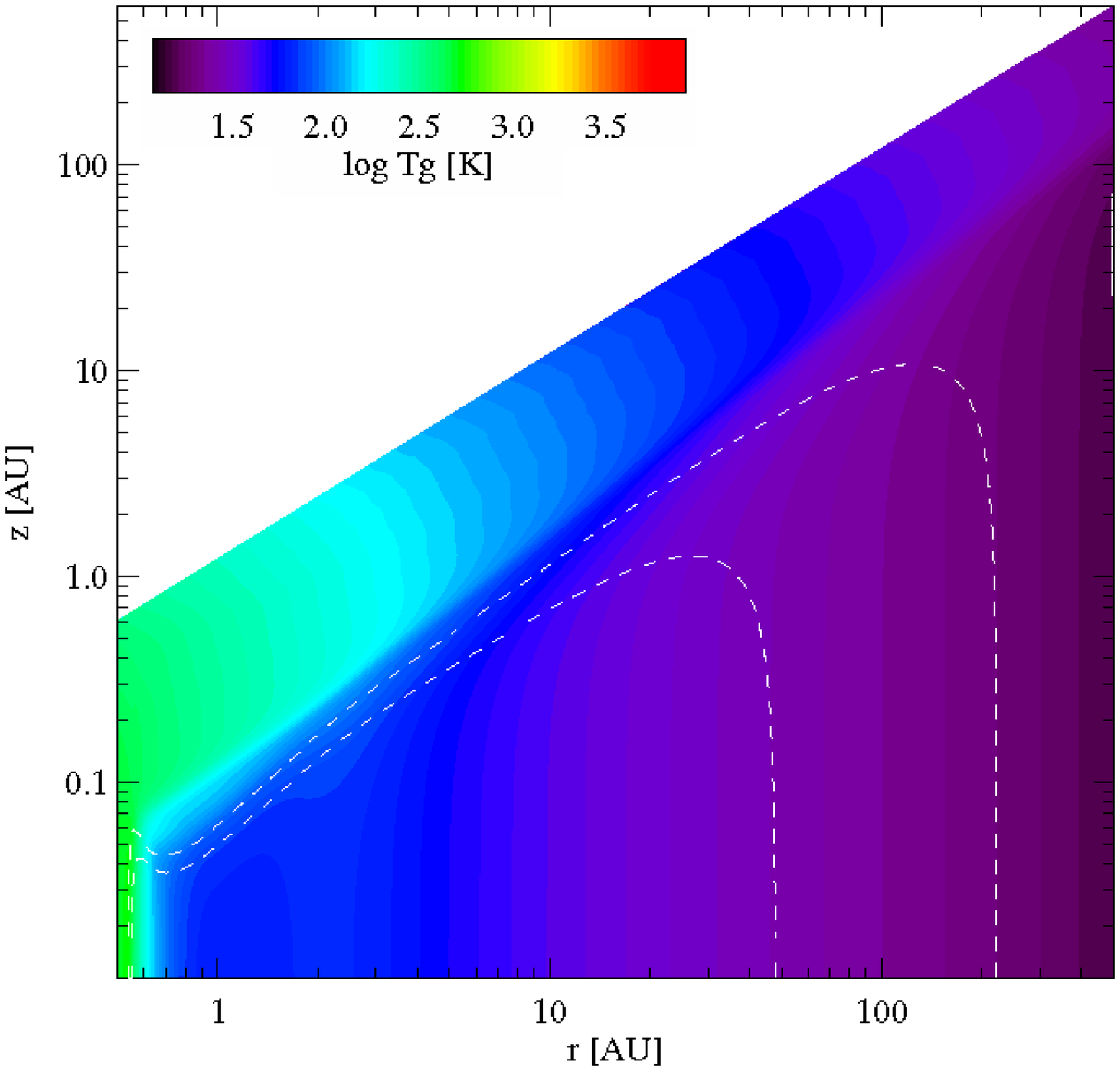} 
  \end{tabular}\\*[-3mm]  
  \caption{Gas temperature structure $\Tg(r,z)$ in a model for a T\,Tauri
           type disk with $M_{\rm disk}\!=\!0.01\,M_\odot$ (l.h.s.). See
	   further model parameter in Table~\ref{tab:Parameter}. On the
           r.h.s. we show the results for the same parameter,
           if $\Tg\!=\!\Td$ is assumed. The white dashed lines show
           $A_V\!=\!1$ and $A_V\!=\!10$.}
  \label{fig:Tgas}
\end{figure*}

\begin{figure*}
  \centering
  \begin{tabular}{cc}
    gas in thermal balance & $\Tg\!=\!\Td$ assumed \\[-1mm]
    \hspace*{-5mm}\includegraphics[height=7.5cm,width=8.8cm]{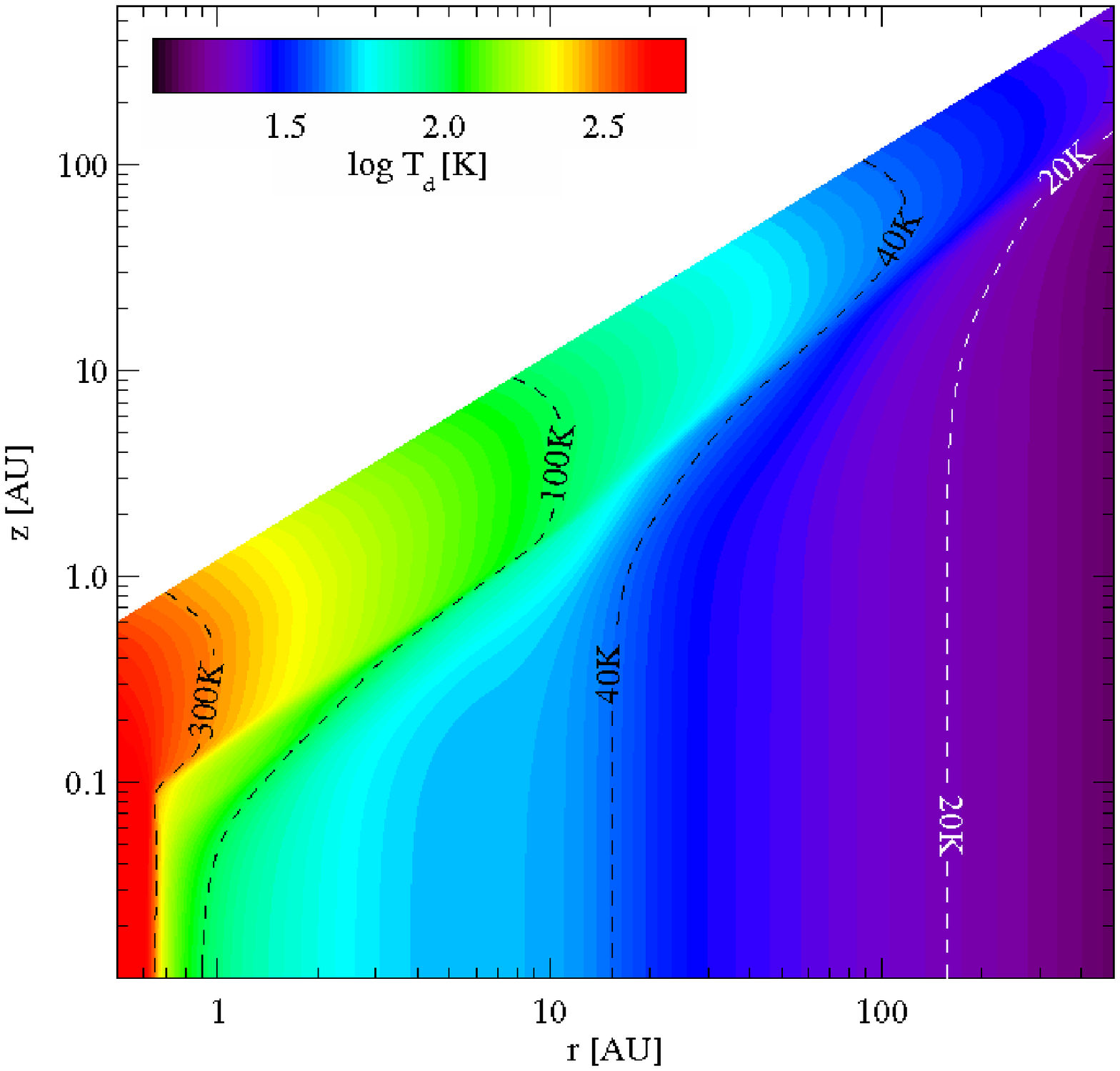} &
    \hspace*{-6mm}\includegraphics[height=7.5cm,width=8.8cm]{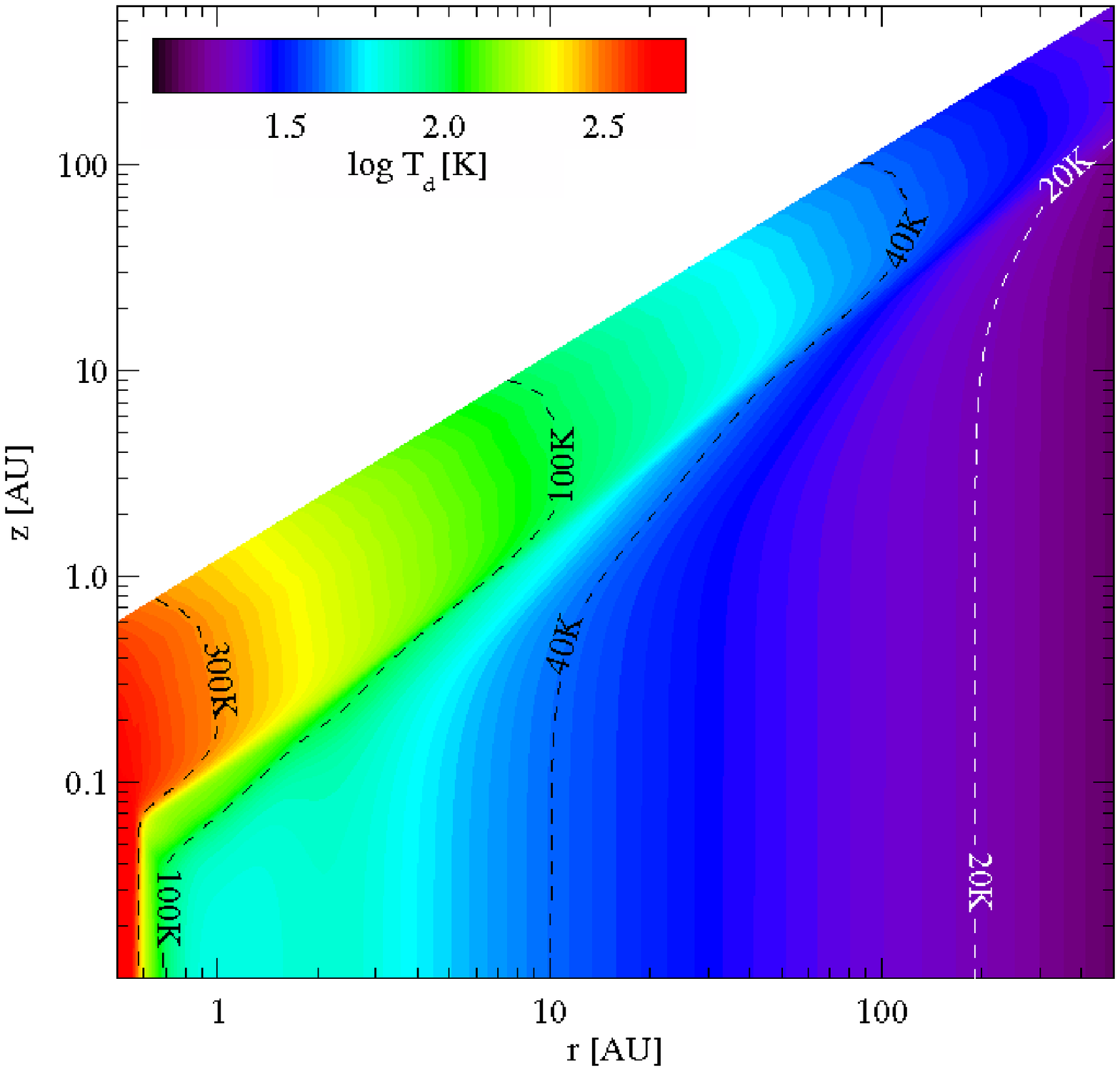} 
  \end{tabular}\\*[-3mm]  
  \caption{Dust temperature structure $\Td(r,z)$. The differences between
    the l.h.s. and the r.h.s. model are caused by the different 
    density structures which are a consistent result of the entire coupled
    physical problem.}
  \label{fig:Tdust}
\end{figure*}

\begin{table}
\centering
\caption{Parameter of the model depicted in Figs.~\ref{fig:Tgas} 
         to \ref{fig:Tcum}.}
\label{tab:Parameter}
\begin{tabular}{l|c|c}
\\[-4.5ex]
\hline
 quantity & symbol & value\\
\hline 
\hline 
stellar mass                      & $M_\star$          & $1\,M_\odot$\\
effective temperature             & $T_{\rm eff}$      & $5770\,$K\\
stellar luminosity                & $L_\star$          &$1\,L_\odot$\\
\hline
disk mass                         & $M_{\rm disk}$     & $0.01\,M_\odot$\\
inner disk radius                 & $\Rin$             & 0.5\,AU$^{\,(1)}$\\
outer disk radius                 & $\Rout$            & 500\,AU\\
radial column density power index & $\epsilon$         & 1.5\\
\hline
dust-to-gas mass ratio$^{\,(2)}$  & $\rho_d/\rho$      & 0.01\\
minimum dust particle radius      & $\amin$            & $0.1\,\mu$m\\
maximum dust particle radius      & $\amax$            & $10\,\mu$m\\
dust size distribution power index& $\apow$            & 2.5\\
dust material mass density        & $\rho_{\rm gr}$    & 2.5\,g\,cm$^{-3}$\\
\hline 
&&\\[-2.2ex]
strength of incident ISM UV       & $\chi^{\rm ISM}$   & 1\\
cosmic ray ionization rate of H$_2$  
                  & $\zeta_{\rm CR}$   
                  & $5\times 10^{-17}$~s$^{-1}$\\
abundance of PAHs relative to ISM & $f_{\rm PAH}$      & 0.12\\
$\alpha$ viscosity parameter      & $\alpha$           & 0\\
\hline
\end{tabular}\\[1mm]
\hspace*{3mm}\begin{minipage}{7cm}
\footnotesize
$(1)$: soft inner edge applied, see Sect.~\ref{sec:SoftEdge}\\
$(2)$: dust optical constants from \citet{Draine1984}
\end{minipage}
\end{table}


\section{Results}

We apply our \ProDiMo model to a typical passive protoplanetary disk
of mass $M_{\rm disk}\!=\!0.01\,M_\odot$ which extends from 0.5\,AU to
500\,AU. The central star is assumed to be a T\,Tauri-type ``young
sun'' with parameters $T_{\rm eff}\!=\!5770\,$K and
$L_\star\!=\!1\,L_\odot$, and to emit excess UV of predominantly
chromospheric origin as shown in Fig.~\ref{fig:Iinc}.  The stellar UV
excess creates an unshielded UV radiation strength of about
$\chi\!=\!2\times10^6$ at 1\,AU (see Eq.~\ref{eq:chi}). Further
parameter of our model are summarized in
Table~\ref{tab:Parameter}. Our selection of elements and chemical
species is outlined in Table~\ref{tab:Species}, and the applied
element abundances are listed in Table~\ref{tab:Abundances}.

The model uses a 150$\times$150 grid of points which are arranged
along radial and vertical rays which enables us to calculate the
respective column densities and line optical depths in a simple
way. The spatial resolution is much higher in the inner regions and
the grid points are also somewhat concentrated toward the
midplane. About half of the grid points are located inside of
2.25\,AU in this model to resolve the strong gradients in the
radiation field and in the thermal and chemical structure occuring
just inside of the inner rim.

\begin{figure*}
  \centering
  \begin{tabular}{cc}
    gas in thermal balance & $\Tg\!=\!\Td$ assumed \\[-1mm]
    \hspace*{-5mm}\includegraphics[height=7.5cm,width=8.8cm]{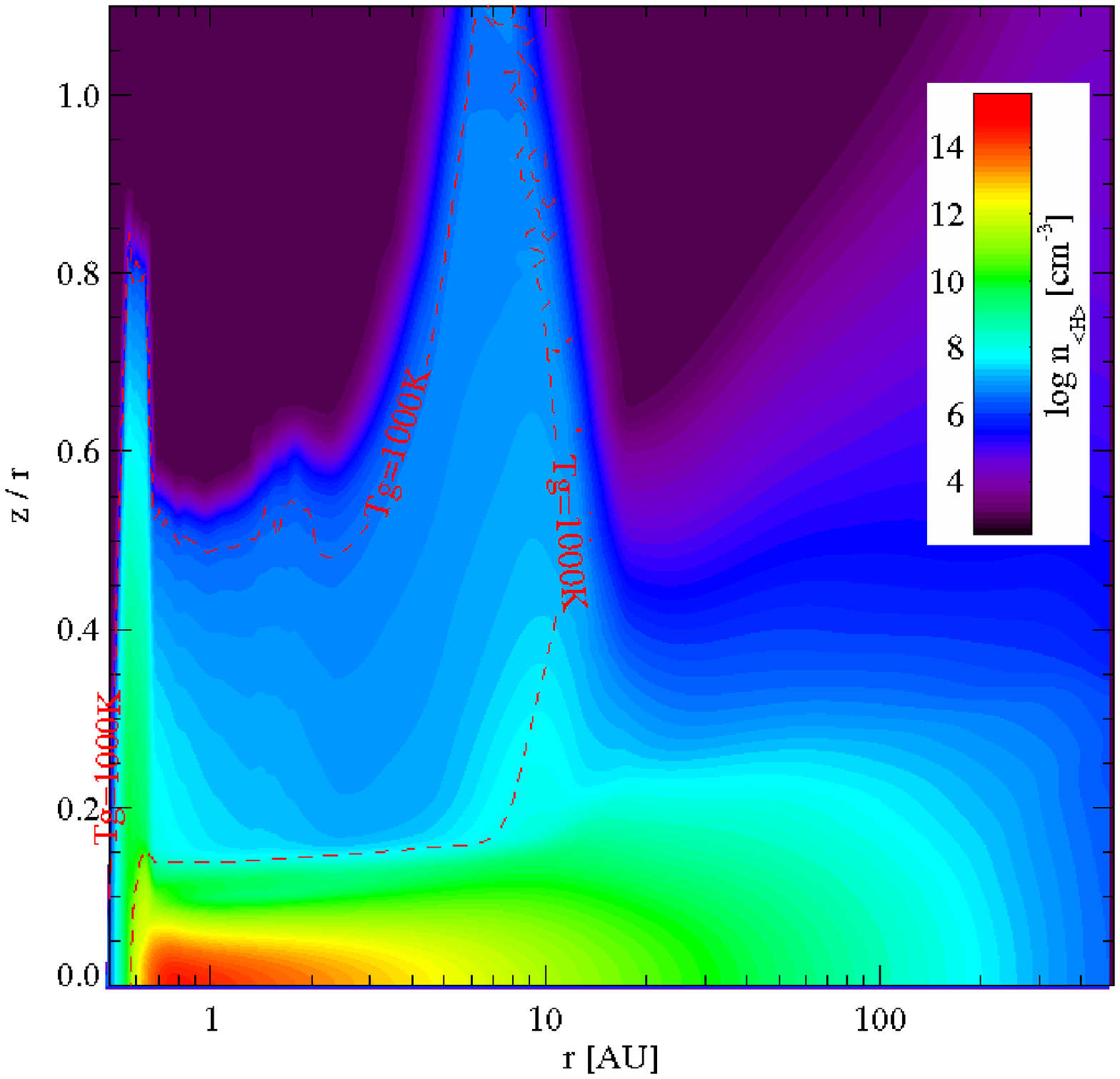} &
    \hspace*{-6mm}\includegraphics[height=7.5cm,width=8.8cm]{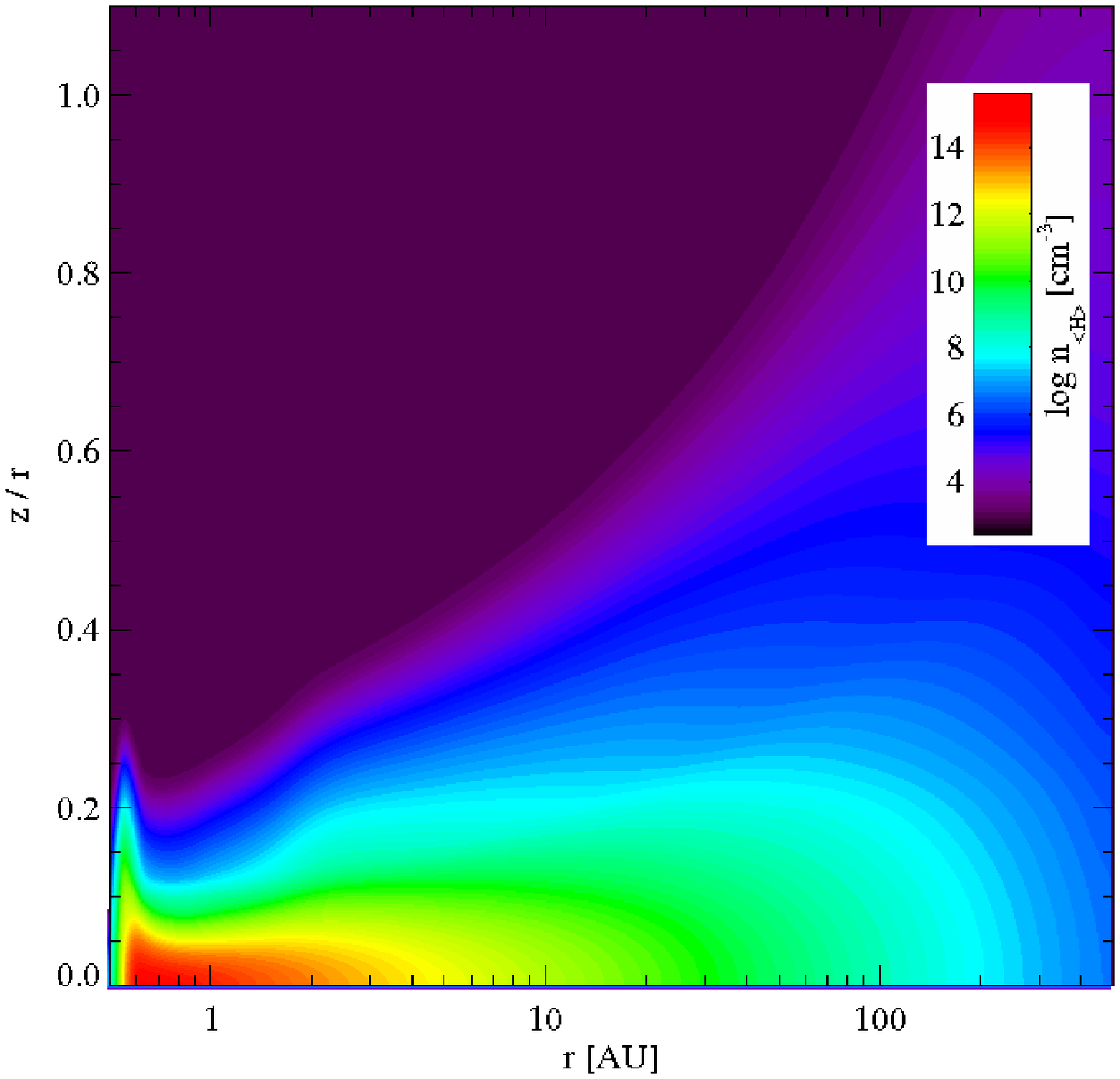} 
  \end{tabular}\\*[-3mm]  
  \caption{Density structure $\nH(r,z)$ as function of relative height
    $z/r$ and $\log r$. The red dashed line on the
    l.h.s. encircles hot regions $\Tg\!>\!1000\,$K.}
  \label{fig:dens}
\end{figure*}

\begin{figure*}
  \centering
  \begin{tabular}{ccc}
    gas in thermal balance & $\Tg\!=\!\Td$ assumed \\[-1mm]
    \hspace*{-5mm}\includegraphics[height=6.9cm,width=8.8cm]
                  {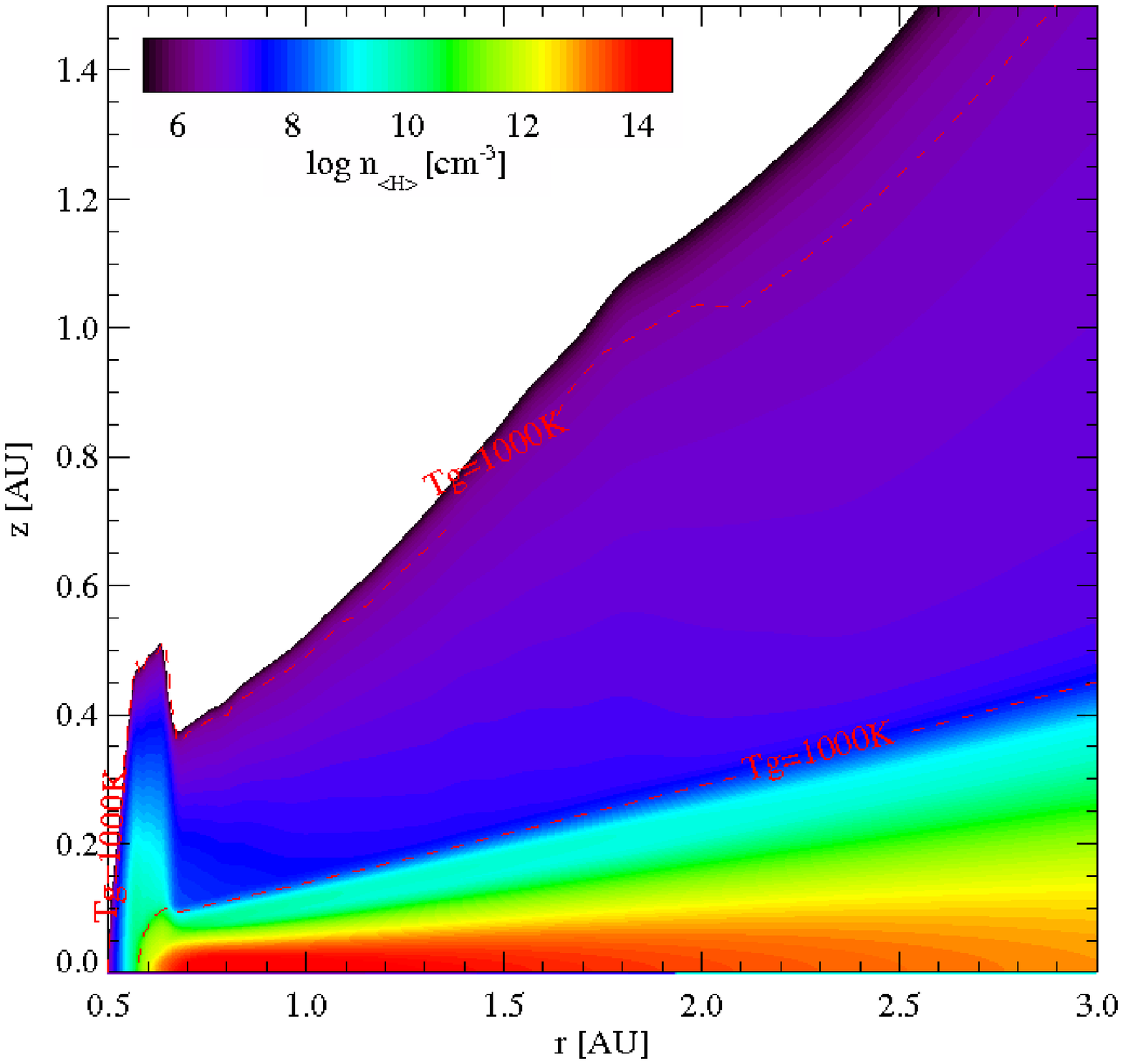} &
    \hspace*{-6mm}\includegraphics[height=6.9cm,width=8.8cm]
                  {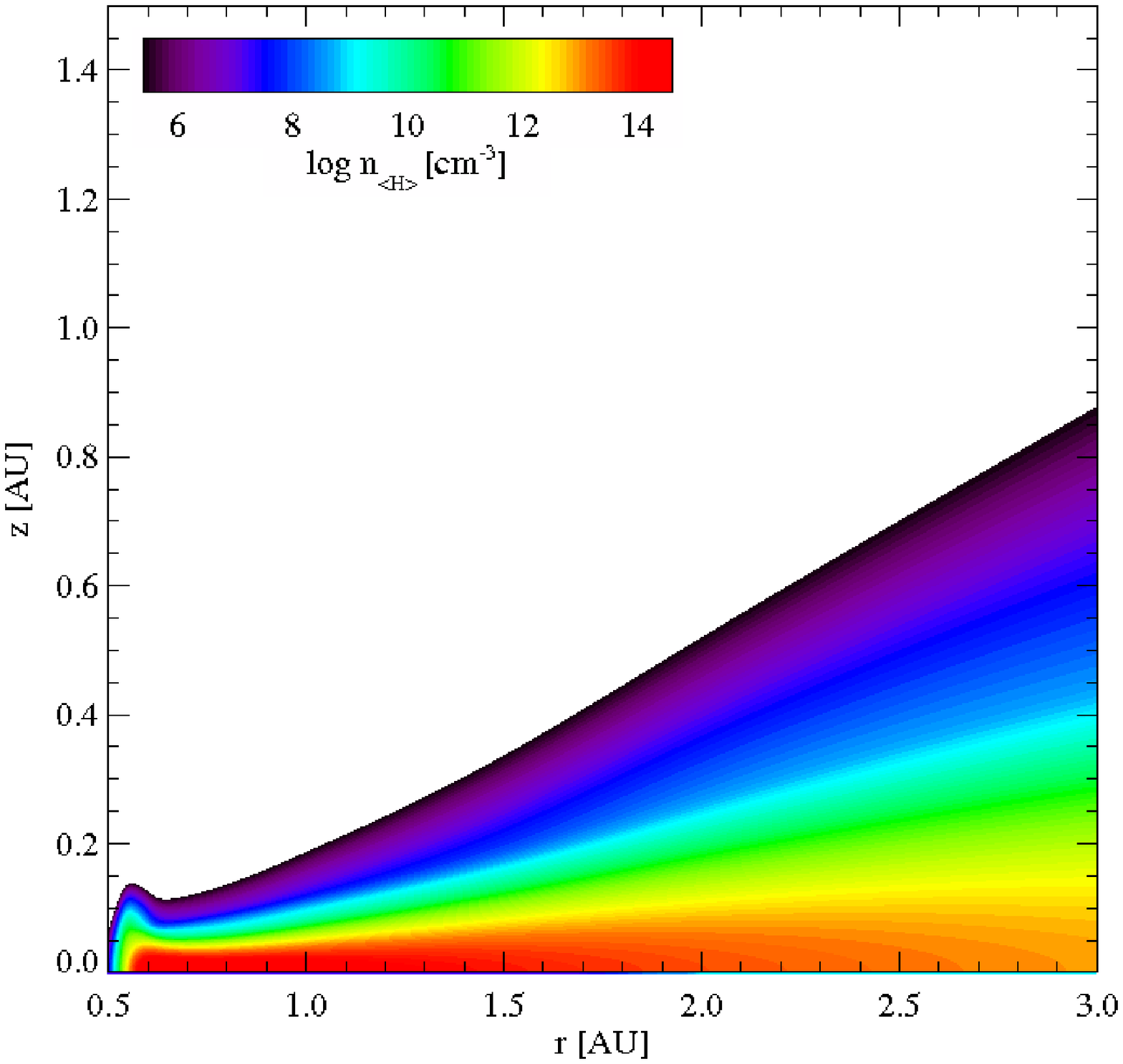} 
  \end{tabular}\\*[-3mm]  
  \caption{Density structure of the puffed-up inner rim. Regions with
           $\nH\!<\!10^{\,5.5}$ are suppressed for output (shown in
           white color). The red dashed contour line on the
           l.h.s. encircles hot regions $\Tg\!>\!1000\,$K.}
  \label{fig:InnerRim}
\end{figure*}

\subsection{Disk structure}

The physical structure of the disk is a consistent result of all model
components: dust radiative transfer, chemistry, and heating and cooling
balance. In order to explore how important the inclusion of the gas
heating and cooling balance is for the resulting disk structure, we
compare the full model (depicted on the l.h.s. of the following
figures) to a comparison model (r.h.s.) where we have assumed
$\Tg\!=\!\Td$ throughout the disk.

\subsubsection{Thermal structure}
\label{sec:Tstruc}

Figures~\ref{fig:Tgas} and \ref{fig:Tdust} show the resulting gas and
dust temperature structures of the models, respectively.
The most obvious feature in Fig.~\ref{fig:Tgas} is a hot surface layer
($\Tg\!\approx\!4000$\,--\,$7000\,$K) which bends around the inner rim
and continues radially to about 10\,AU.  This hot surface layer is
situated above $z/r\!\ga\!0.13$ in this model. Its lower edge is not
related to the vertical $A_V$ but rather to the position of the shadow
casted by the puffed-up inner rim. It coincides with the first
occurrence of CO and other molecules like OH (see
Fig.~\ref{fig:Chemistry}). The hot surface layer is optically thin,
predominantly atomic (molecule-free) and directly heated by the stellar
radiation in various ways (see Sect.~\ref{sec:HeatCool}).

The shielded and cold regions in the midplane ($z/r\!\la\!0.06$) are
characterized by small deviations between $\Tg$ and $\Td$, due to
effective thermal accommodation between gas and dust. However, beyond some
critical radius, here $\approx\!100\,$AU, even the midplane regions
become optically thin, and the interstellar UV irradiation causes an
increase of $\Tg$. We find midplane temperatures up to
$\Tg\!\approx\!2\Td$ around 400\,AU in this model. The critical radius
is related to $A_V\!=\!1$ and increases with disk mass.

The upper layers $z/r\!\ga\!0.1$ at $r\!\ga\!20\,$AU show no clear
trend, both $\Tg\!<\!\Td$ and $\Tg\!>\!\Td$ is possible, due to a
complicated superposition of various heating and cooling processes.

Apart from the thermally decoupled layers at the inner rim, the
surface and the very extended layers, the disk temperature is mainly
controlled by the dust continuum radiative transfer (see
Fig.~\ref{fig:Tdust}). $\Td(r,z)$ shows all the features typical for
protoplanetary disks \citep[see e.g.][]{Pascucci2004,Pinte2009}. The
midplane dust optical depth at $1\,\mu$m is about $1.8\cdot 10^5$ in this
model. The slightly different $\Td$-results for the two models are
caused by the different density structures (see Fig.~\ref{fig:dens})
which depend on $\Tg$. In case of the full model, the vertically
extended inner regions scatter the star light and thereby heat the
disk from above.

\subsubsection{To flare or not to flare}

Figure~\ref{fig:dens} shows the resulting density structures of both
models. The full model (l.h.s.) exhibits a remarkable vertical
extension (up to $z/r\!\approx\!1$) of both the inner rim and the
surface layers inward of $r\!\la\!10\,$AU. According to
Eq.\,(\ref{eq:hydrostat}), the vertical scale height $H$ is approximately
(assuming $z\!\ll\!r$, $c_T\!=\,$const) given by
\begin{equation}
  \left(\frac{H}{r}\right)^2 = \frac{2r\,\cT2}{GM_\star} \ ,
\end{equation}
where $H$ is defined as $\rho(z)\!\approx\!\rho(0)\exp(-z^2/H^2)$. The
temperature ratio $\Tg/\Td$ reaches values of about 10\,--\,30 in the
hot inner rim and the surface regions, and since $c_T^2\!\propto\!\Tg$,
the disk is vertically more extended by about the same factor in
comparison to the $\Tg\!=\!\Td$-model. This applies to the
inner rim in particular, because it is hot even at $z\!=\!0$, whereas
the enhancing effect only starts at $z/r\!\ga\!0.1$ in
general. However, in the regions $r\!\approx\!1\,$--\,7\,AU, $\Tg$
is almost constant in the hot surface layer 
($\approx\!4000\,$--\,7000\,K) and so $H/r$ increases further with
increasing radius, and the disk reaches its maximum vertical extension
here.

It is noteworthy that the vertical density structure $\rho(z)$ may be
locally inverted. Since Eq.\,(\ref{eq:hydrostat}) is a pressure
constraint, the density must locally re-increase if $\Tg$ drops quickly
with increasing height. This happens in the uppermost layers, in
particular around 10\,AU at $z/r\!\approx\!0.8$, a region which causes
the most numerical problems during the course of the global
iterations.

At larger radii $\ga\!30\,$AU, both models show a comparable vertical
extension, characterized by a generally flaring structure. The ``flaring''
(increase of $H/r$ with increasing $r$) is a natural consequence of
the radial dust temperature profile varying roughly like
$\Td(r)\!\propto\!r^{-p}$ with $p\!\approx\!0.25$ in the midplane and
$p\!\approx\!0.35\,$--\,0.45 in the optically thin parts, so
$H/r\propto r^{\,p}$.

\begin{figure*}
  \centering
  \begin{tabular}{cc}
    heating & cooling\\[-1mm]
    \hspace*{-8mm} \includegraphics[width=10.4cm,height=11cm]{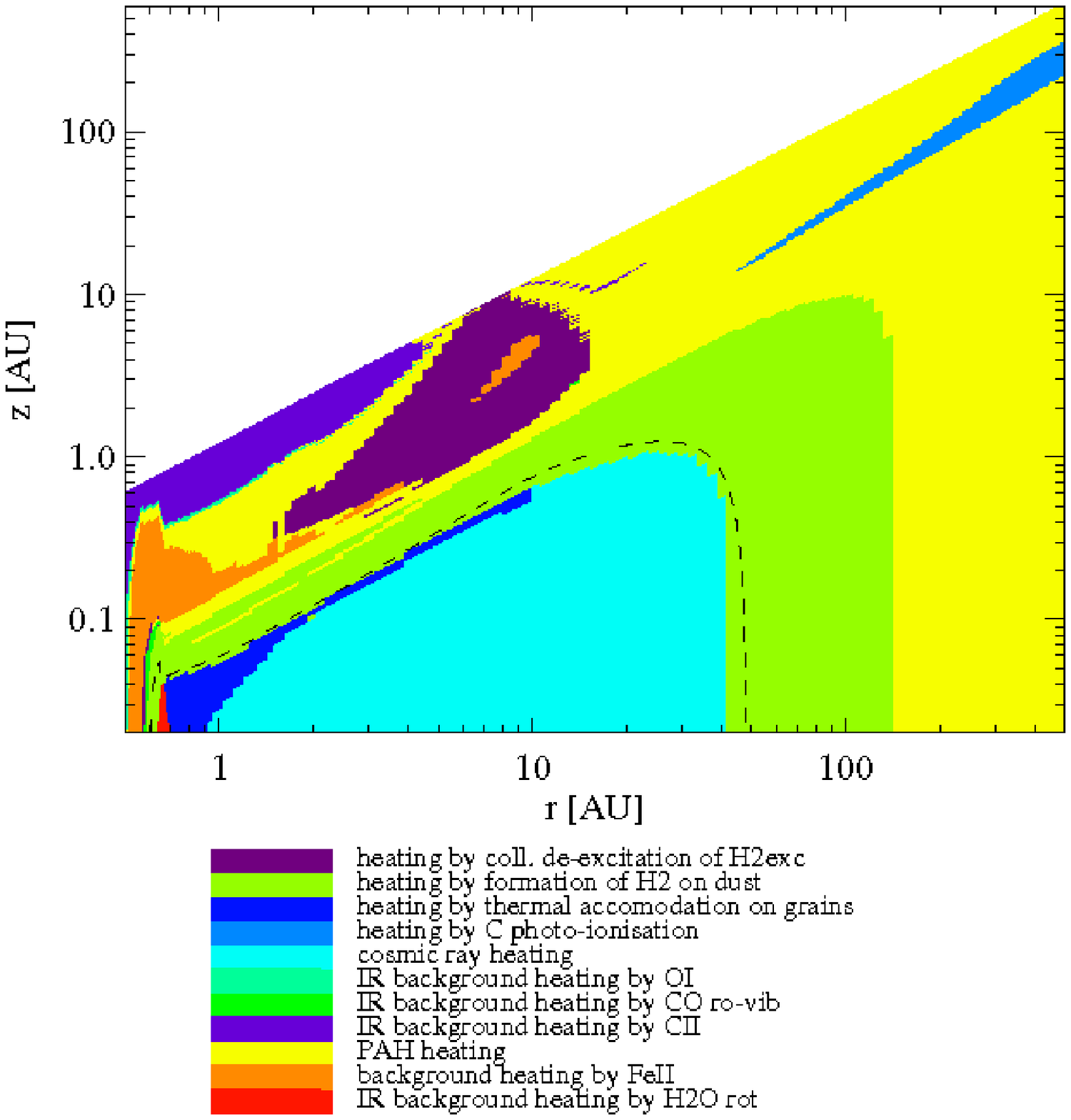} &
    \hspace*{-15mm}\includegraphics[width=10.4cm,height=11cm]{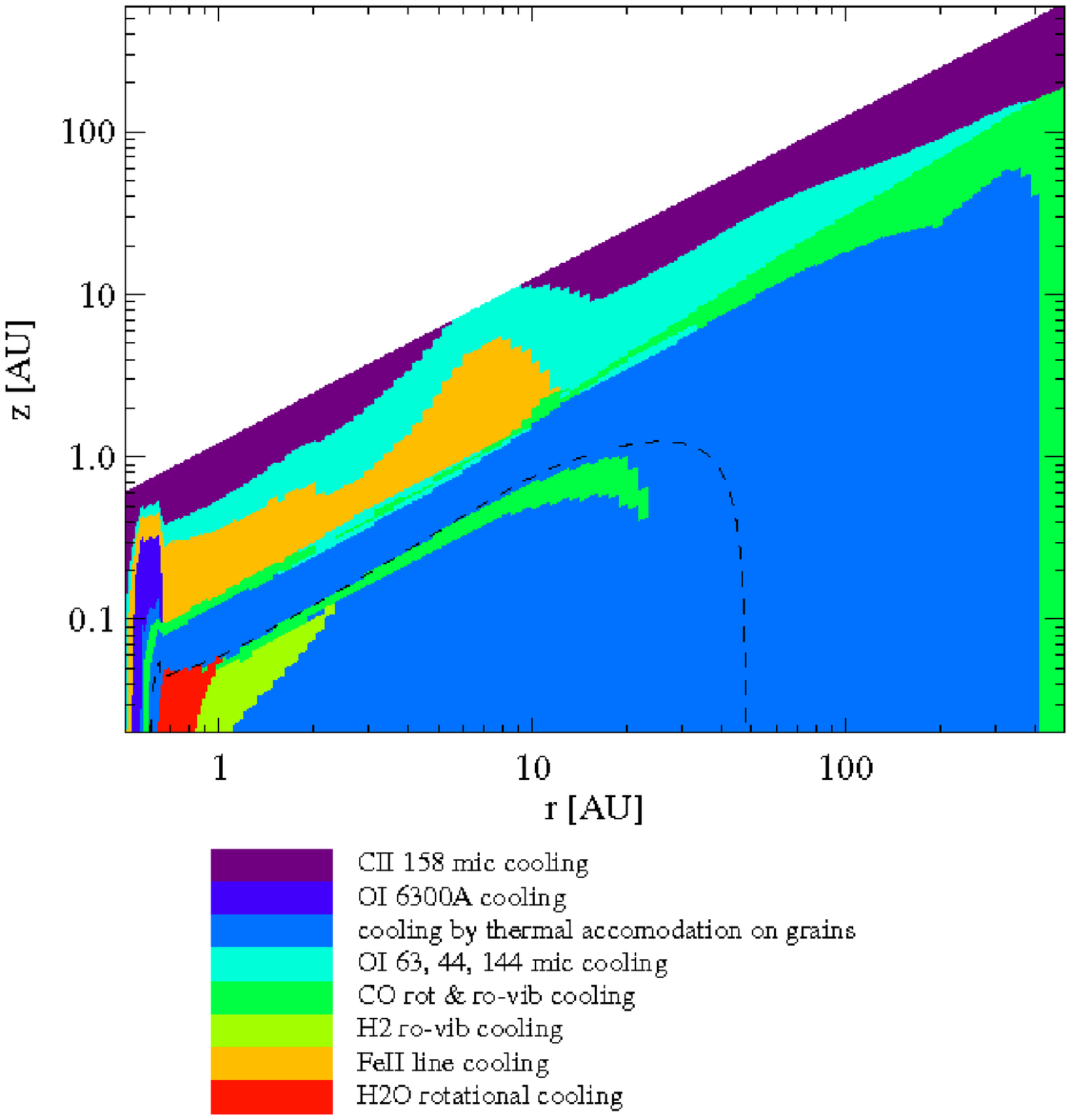} 
  \end{tabular}\\*[-3mm]  
  \caption{Leading heating process (l.h.s.) and leading cooling
  process (r.h.s.) of the model in gas thermal balance. The black
  dashed contour line indicates an optical extinction of $A_V\!=\!10$.}
  \label{fig:HeatCool}
\end{figure*}

\subsubsection{The puffed-up inner rim}

Figure \ref{fig:InnerRim} shows a magnification of the density
structure in the innermost regions. The figure demonstrates the large
impact of the treatment of the gas temperature in the model on the
resulting disk structure. There is a rapid decline of the density
between $\nH\!=\!10^9\rm\,cm^{-3}$ and $10^8\rm\,cm^{-3}$, which is
caused by the steep $\Tg$-increase at given pressure at the top of the
shadow at $z/r\!\approx\!0.13$ casted by the inner rim
(l.h.s.). Therefore, such densities merely exist in the model close to
the star, but the cool and dense midplane regions
$(\!>\!10^9\rm\,cm^{-3})$ are surrounded by an extended ``halo''
composed of thin hot atomic gas of almost constant density
($\nH\!=\!10^8$ to $10^7\rm\,cm^{-3}$) which extends as high up as
$z/r\!\approx\!0.5$. These results are astonishingly robust against
variation of the disk mass $M_{\rm disk}$ between $10^{-4}$ and
$10^{-1}\,M_\odot$ --- we always find the same kind of halo composed
of the same kind of gas with the same densities. Only the midplane
regions contain more or less cold matter, according to $M_{\rm disk}$.

The assumed position of the inner rim at 0.5\,AU in our model implies
maximum dust temperatures of about 500\,K, which is well below the
dust sublimation temperature, and the shape of the inner rim is
controlled by the radial force equilibrium at the inner edge which
implies a smooth density gradient, see Sect.~\ref{sec:SoftEdge}. In
contrast, \citet{Isella2005} investigated the effect of
pressure-dependent sublimation of refractory grains on the shape of
the inner rim.  In reality, different kinds of refractory grains will
be present which have not only different and pressure-dependent
sublimation temperatures, but the dust temperatures are strongly
dependent on dust kind due to dust opacity effects \citep[see][]{Woitke2006},
which can be expected to result in a highly complex chemical
structure of the inner rim.

In comparison, the $\Tg\!=\!\Td$-model does not possess the hot surface
layers and, consequently, shows a much flatter structure.
The inner rim is much less puffed-up causing the shadow
borderline to be situated deeper. The inner ``soft edge'' is likewise
less extended, only from 0.5\,--\,0.61\,AU in the
$\Tg\!=\!\Td$-model, whereas is extends from 0.5\,--\,0.8\,AU in the
full model, or about 40\% of the inner radius.

\subsubsection{Thermal balance}
\label{sec:HeatCool}

Figure~\ref{fig:HeatCool} shows the most important heating processes
(l.h.s.) and the most important cooling process (r.h.s.) in the full
model of the disk with the gas being in thermal balance. Again, there
is a clear dividing line at $z/r\!=\!0.13$ coinciding with the shadow
of the inner rim, which separates the directly illuminated hot surface
layers from the shielded and cold midplane regions.

The central midplane of the disk below $A_V\!\approx\!10$ is
dominated by thermal accommodation which assures $\Tg\!\approx\!\Td$
\citep[see also][]{Kamp2004,Nomura2005,Gorti2008}.
Since UV photons cannot penetrate into these layers, cosmic-ray ionization
is the only remaining heating process, mostly compensated for by thermal
accommodation cooling. In the central midplane $r\!\la\!1\,$AU, before
H$_2$O freezes out (see Fig.~\ref{fig:Chemistry}), there is
additionally H$_2$O rotational cooling, as well as some H$_2$
quadrupole and CO rotational cooling just below $A_V\!\approx\!10$.


Between $A_V\!\approx\!10$ and $z/r\!\approx\!0.13$, the UV radiation
can partly penetrate into the disk via scattering from above (see
Fig.~\ref{fig:chis}). This creates an active photon-dominated region
with a rich molecular chemistry, where most of the abundant molecules
like H$_2$, CO, HCN, OH and H$_2$O form, usually referred to as the
``intermediate warm molecular layer'' \citep{Bergin2007}. The layer is
predominantly heated by H$_2$ formation on grain surfaces and, with
increasing height, by photo-effect on PAH molecules. The gas
temperature increases upward in this layer, \eg from $\sim200\,$K to
$\sim700\,$K at 1\,AU, but the additional heating can still be
balanced by thermal accommodation in our model.

The upper edge of the warm molecular layer is characterized by a thin
zone of intensive CO ro-vibrational cooling. Above this zone, CO is
photo-dissociated -- below this zone, the CO lines become optically
thick. It is this CO ro-vibrational cooling that can counterbalance
the upwards increasing UV heating for a while, until the heating
becomes too strong even for CO. This happens just at the upper end of
the disk shadow $z/r\!\approx\!0.13$ where the direct stellar
irradiation becomes dominant.

Above the CO layer, the temperature suddenly jumps to about 5000\,K,
all molecules are destroyed (thermally and radiatively), and we enter
the hot surface layer described in the previous sections. This layer
is predominantly heated by collisional de-excitation of vibrationally
excited H$_2^\star$ (inner regions) and by PAH heating (outer
regions). Although H$_2$ is barely existent at these heights above the
disk (concentration is $10^{-4}$ to $10^{-7}$, see
Fig.~\ref{fig:Chemistry}), the few H$_2$ molecules formed on grain
surfaces can easily be excited by UV fluorescence, and these
H$_2^\star$ particles undergo de-exciting collisions. This heating is
balanced by various line cooling mechanisms. Since molecules are not
available, atoms and ions like O\,{\sc i} and Fe\,{\sc ii} are most
effective. The non-LTE cooling by the wealth of fine-structure,
semi-forbidden and permitted Fe\,I and Fe\,{\sc ii} lines has been
investigated in detail by \citep{Woitke1999}, who found that in
particular the semi-forbidden iron lines provide one of the most
efficient cooling mechanisms for warm, predominantly atomic gases at
densities $\nH\!=\!10^6$ to $10^{14}\rm\,cm^{-3}$.

Since the stellar optical to IR radiation can excite most of the
Fe\,{\sc ii} levels directly, radiative heating occurs. This
``background heating by Fe\,{\sc ii}'' as referred to in
Fig.~\ref{fig:HeatCool} (l.h.s.) turns out to contribute significantly
to the heating of the hot atomic layer close to the star
($r\!\la\!10\,$AU). In fact, further analysis shows that the gas
temperature in a large fraction of the hot atomic layer is regulated
by $\Gamma_{\rm FeII}\!\approx\,\Lambda_{\rm FeII}$, \ie by radiative
equilibrium of the gas with respect to the Fe\,{\sc ii} line
opacity. Similarly, we find a small zone in the midplane just behind
the inner rim where radiative equilibrium with respect to the water
line opacity is established. The regulation of the gas temperature via
radiative equilibrium is a typical feature for dense gases in strong
radiation fields, \eg in stellar atmospheres. This behavior is rather
unusual in PDR and interstellar cloud research from where most of the
other heating and cooling processes have been adopted.


The more distant regions $\ga\!50\,$AU are characterized by an
equilibrium between interstellar UV heating (photo-effect on PAHs)
and [C{\sc ii}] $158\,\mu$m, [O{\sc i}] 63 and $144\,\mu$m, CO rotational
line cooling, and thermal accommodation \citep[\eg][]{Kamp2001}.

\subsection{Chemical structure}

\begin{figure*}
  \centering
  \begin{tabular}{ccc}
    \hspace*{-5mm}\includegraphics[width=6.4cm]{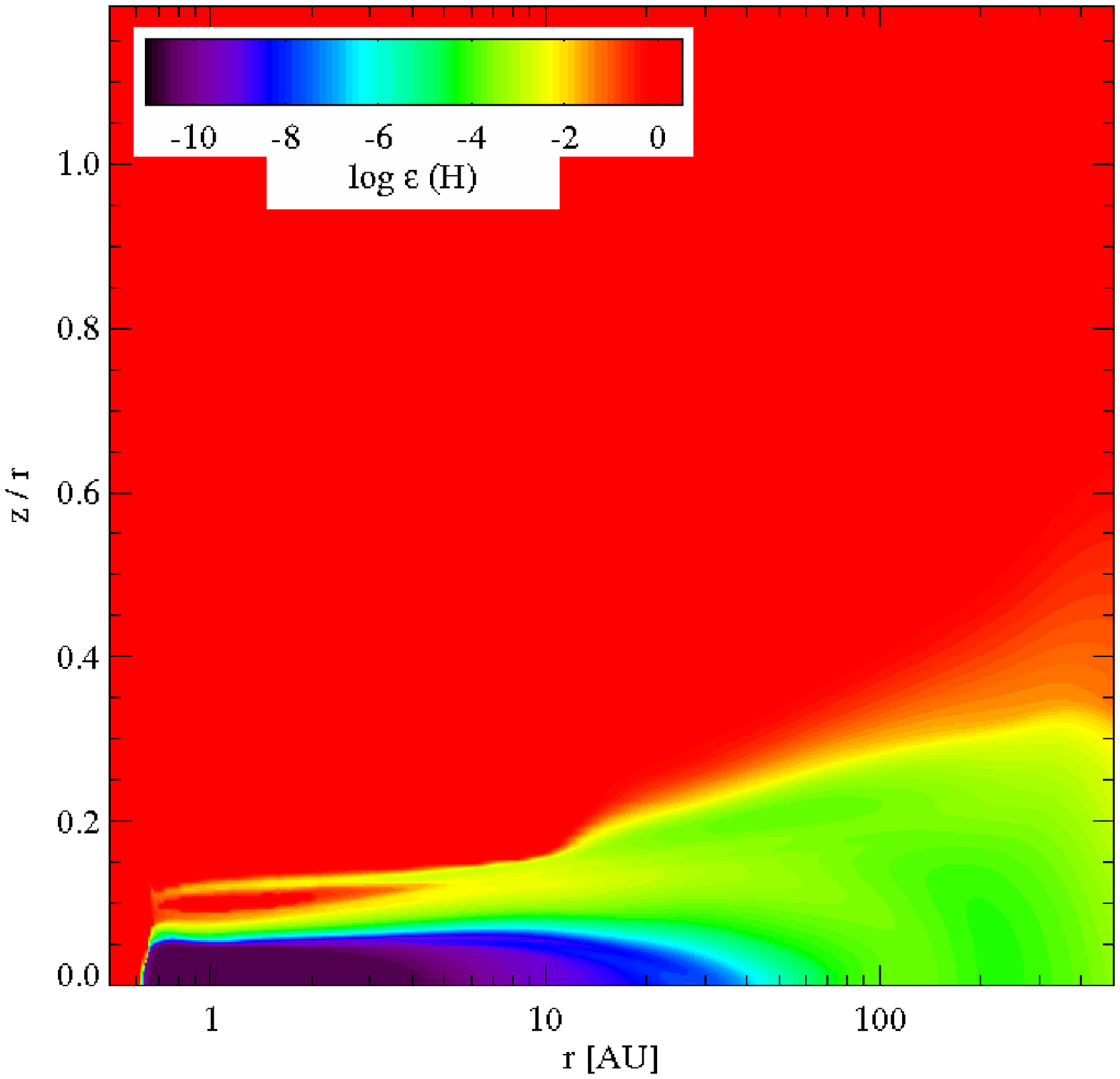} &
    \hspace*{-6mm}\includegraphics[width=6.4cm]{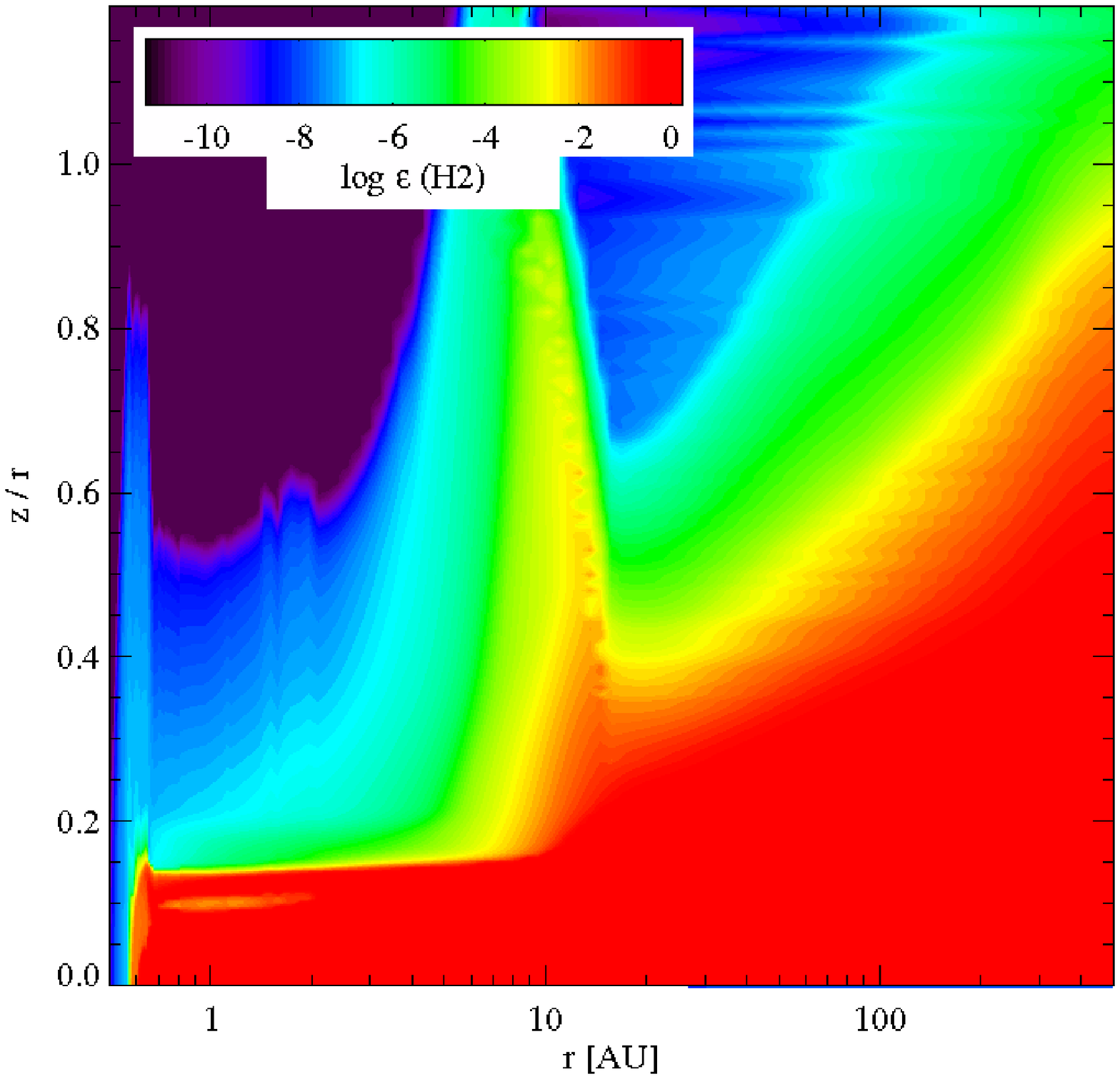} &
    \hspace*{-6mm}\includegraphics[width=6.4cm]{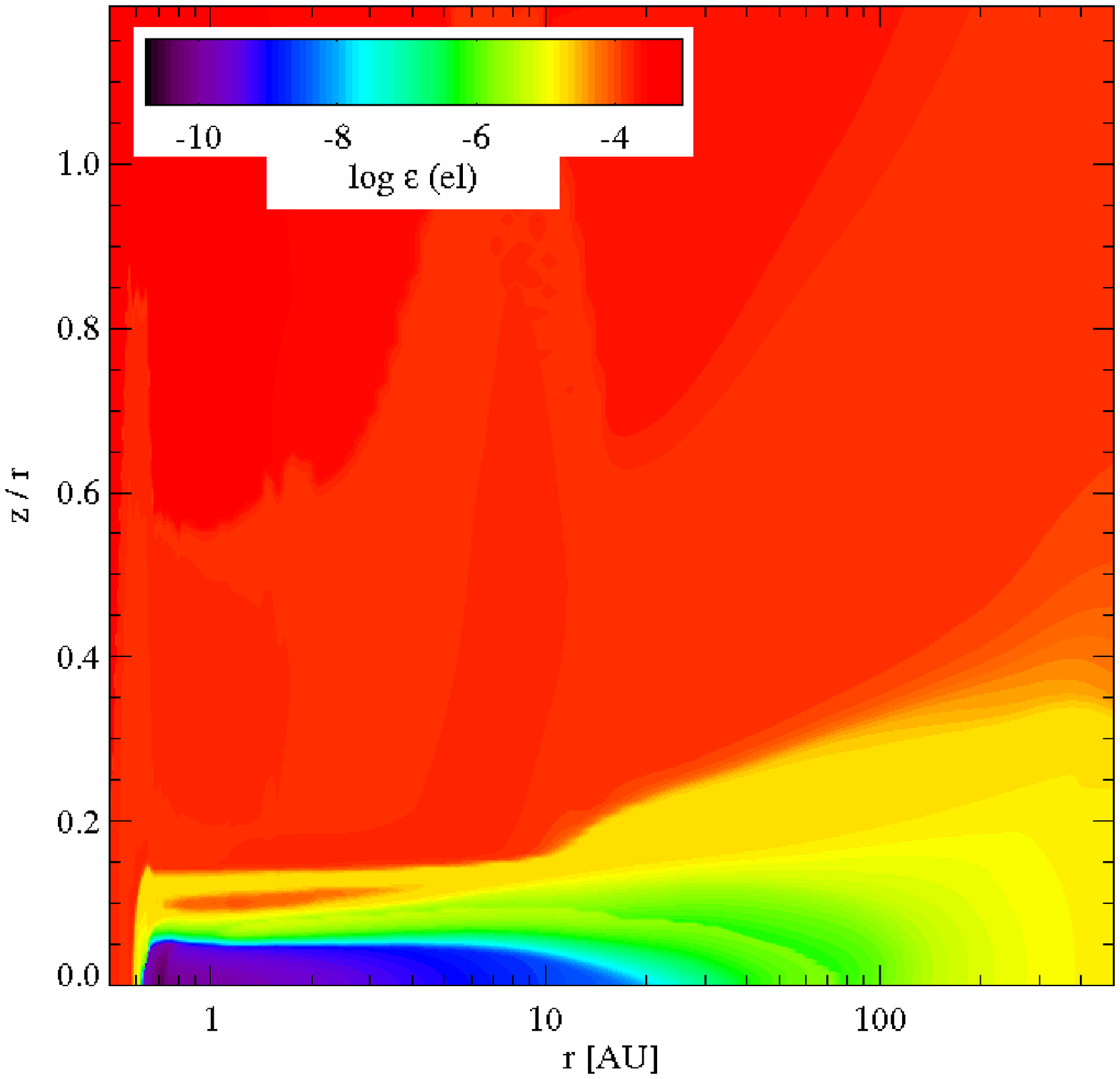} \\
    \hspace*{-5mm}\includegraphics[width=6.4cm]{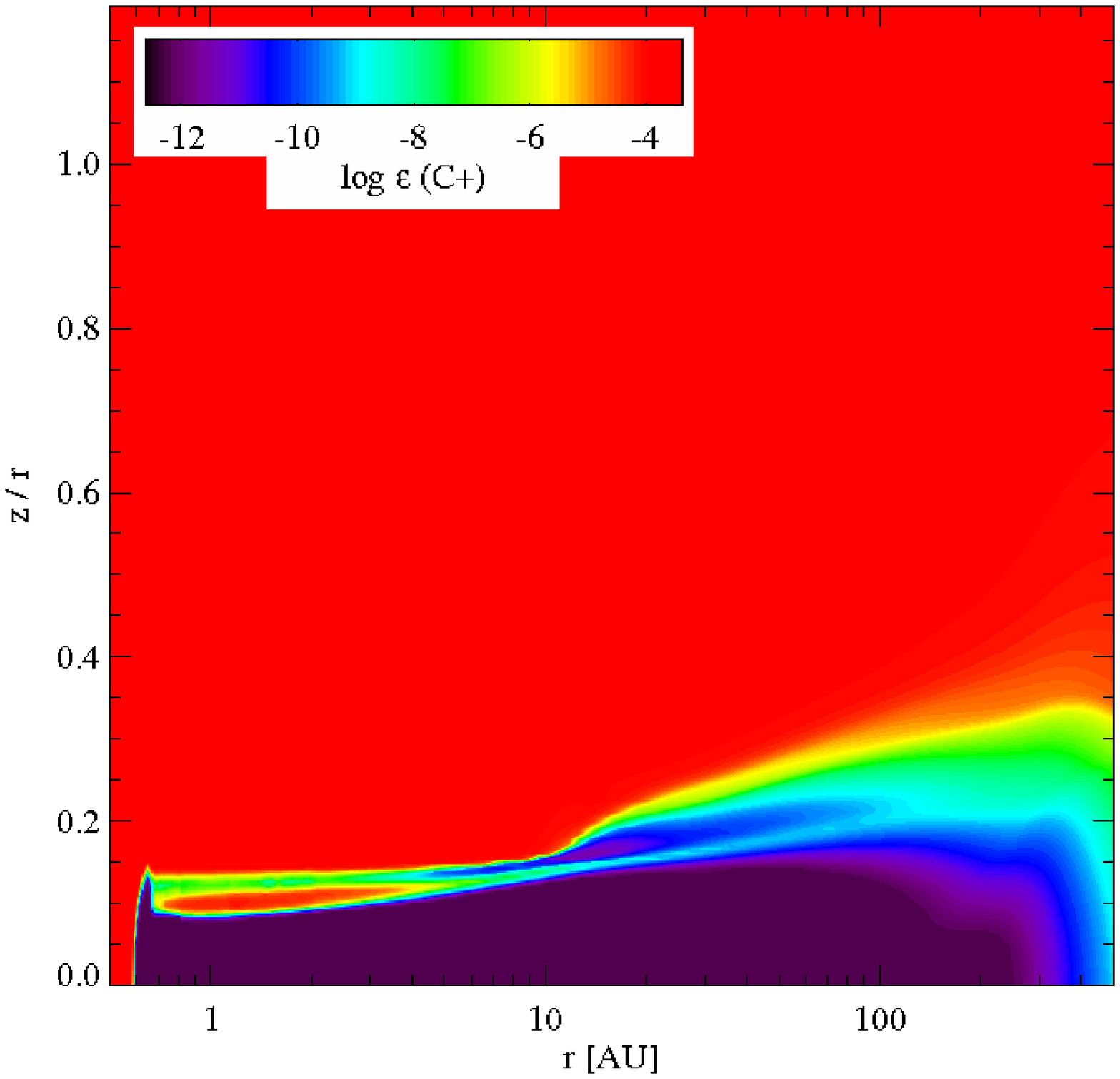} &
    \hspace*{-6mm}\includegraphics[width=6.4cm]{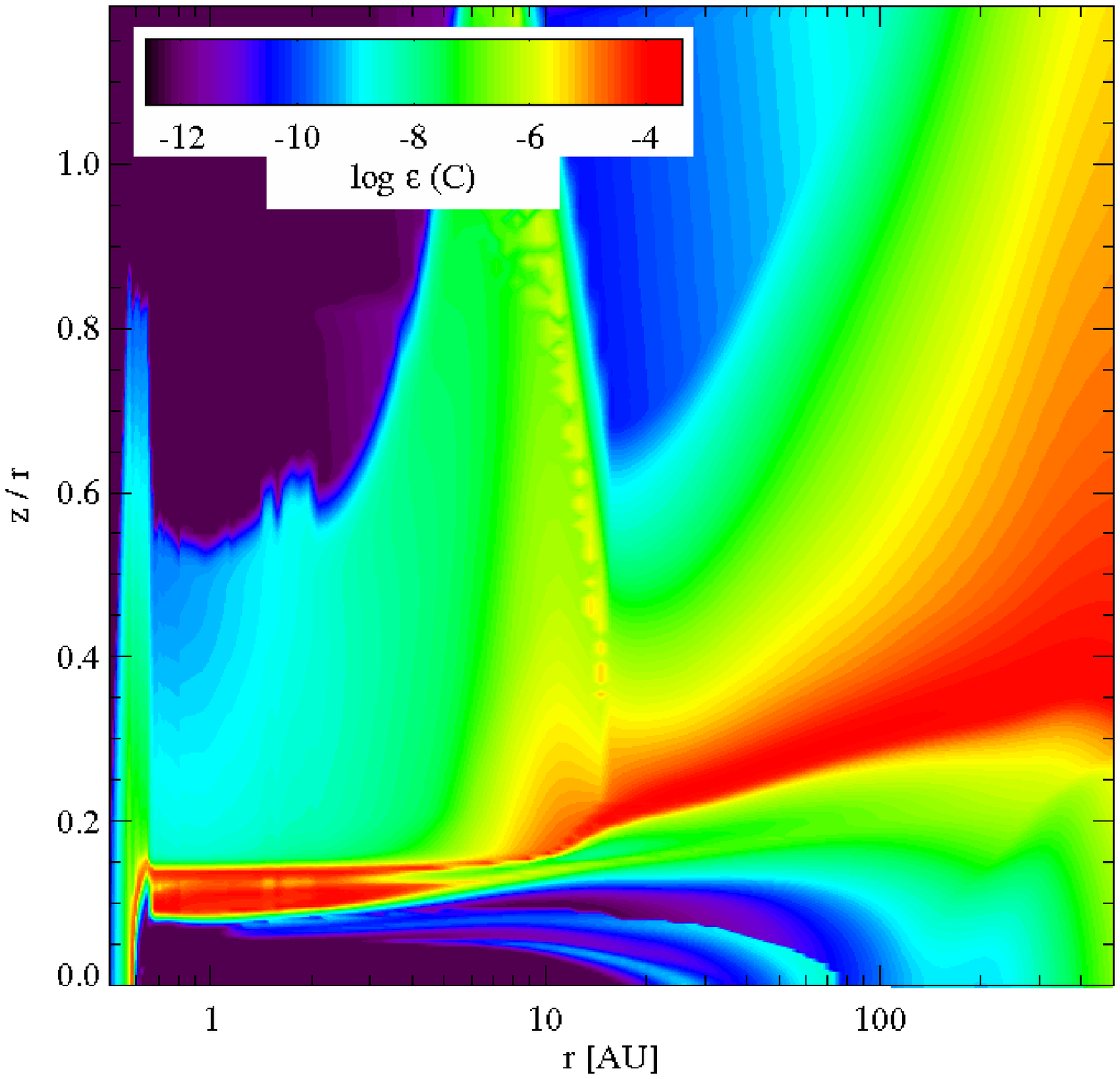} &
    \hspace*{-6mm}\includegraphics[width=6.4cm]{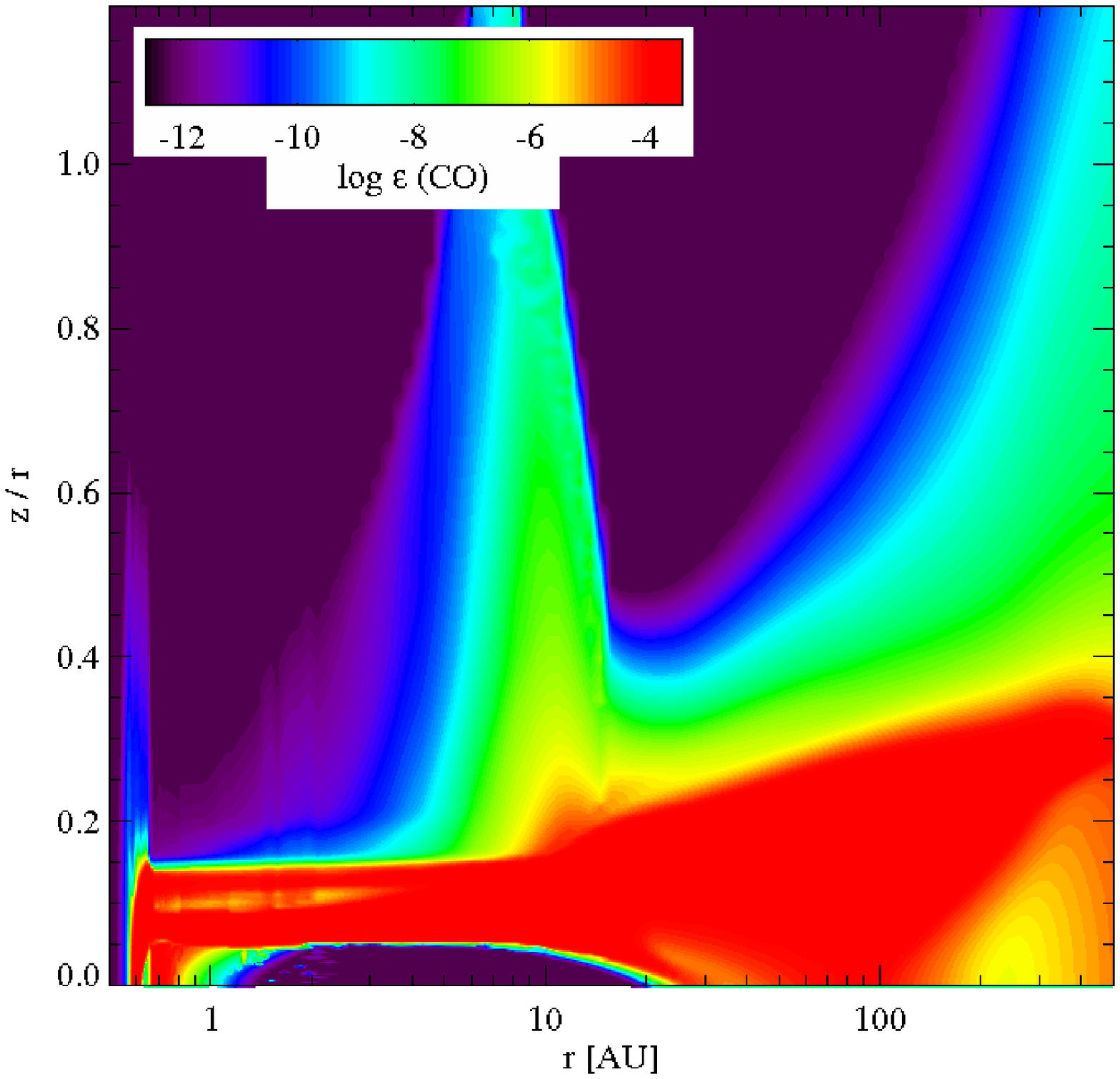} \\
    \hspace*{-5mm}\includegraphics[width=6.4cm]{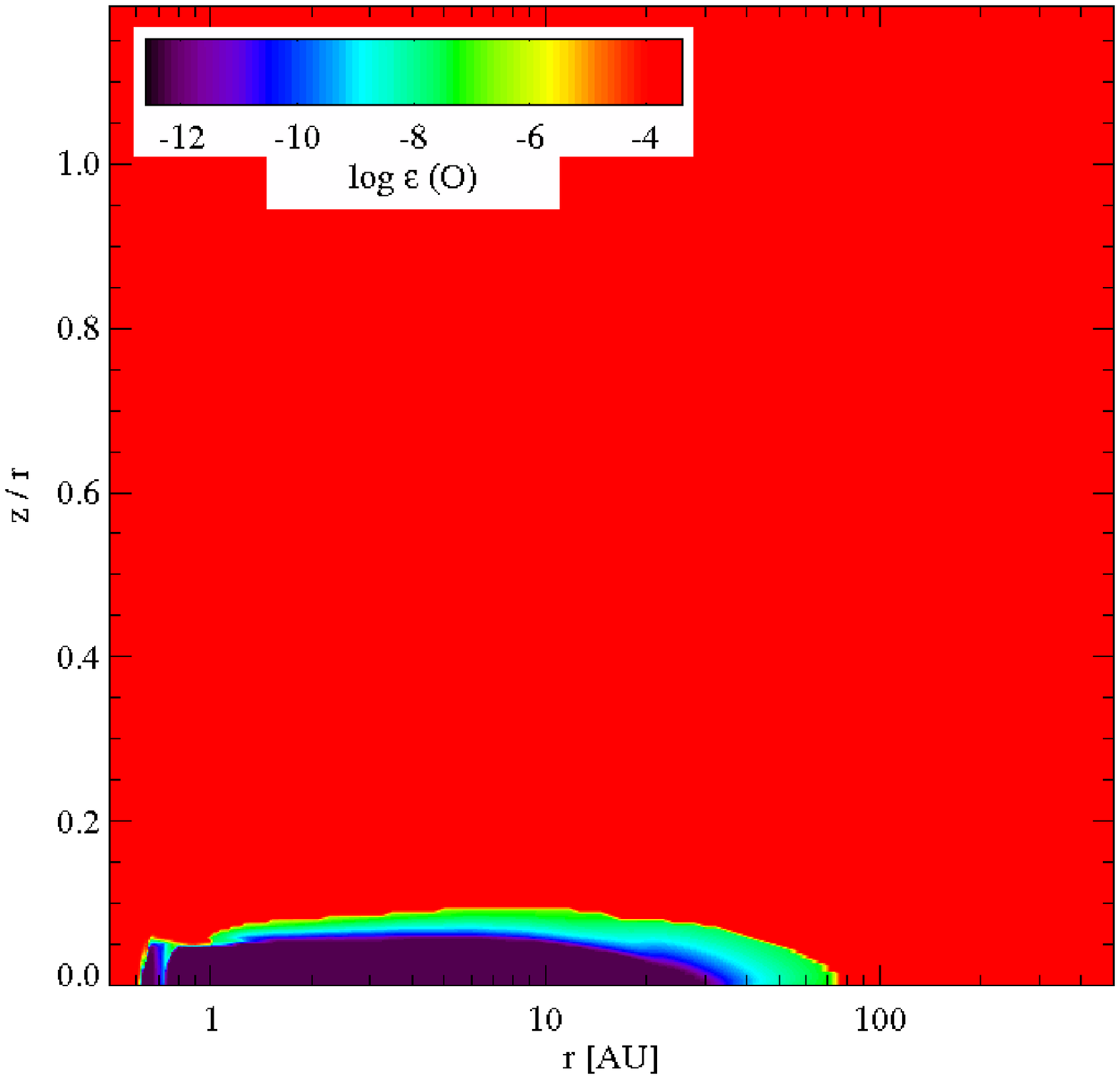} &
    \hspace*{-6mm}\includegraphics[width=6.4cm]{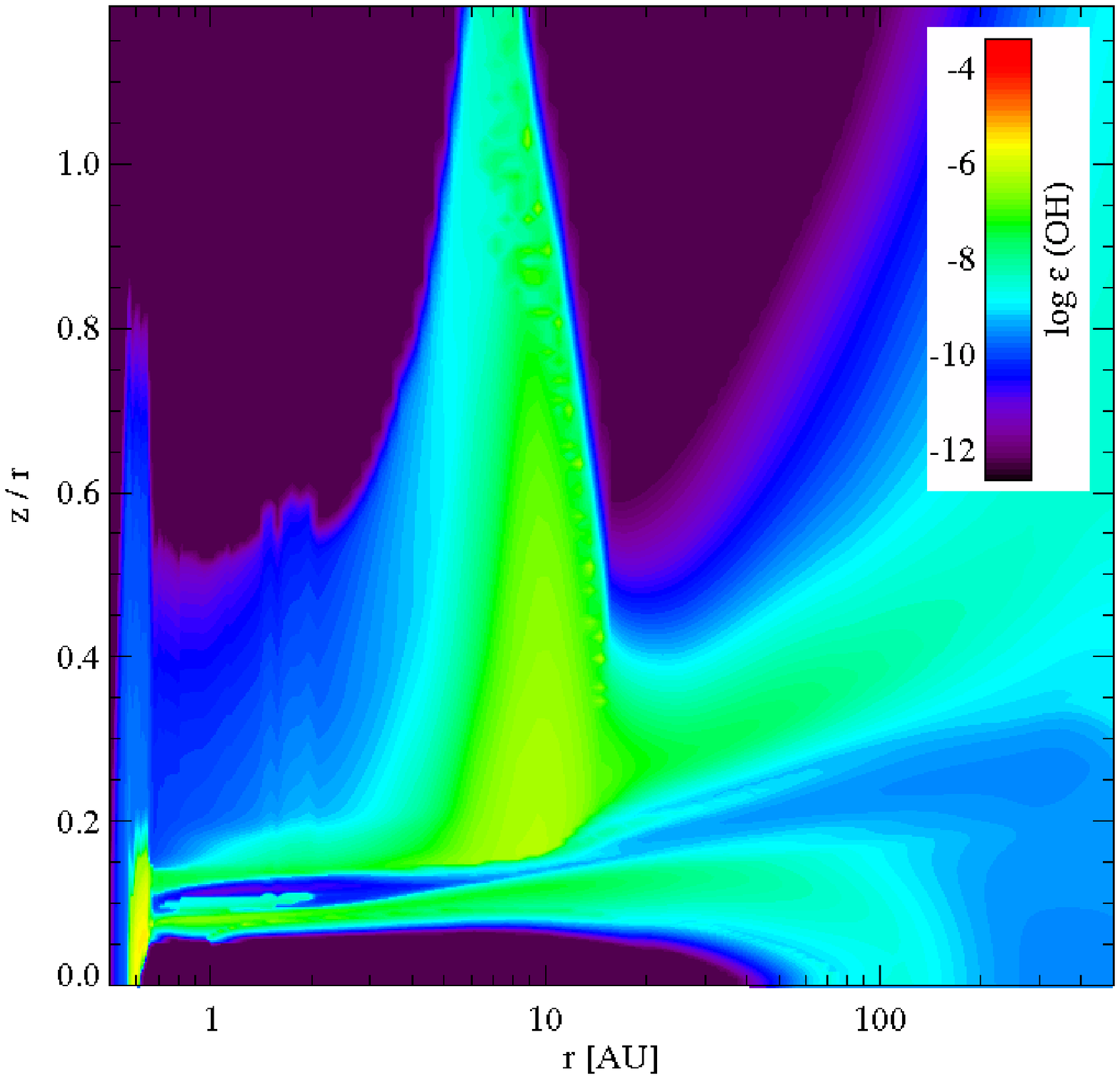} &
    \hspace*{-6mm}\includegraphics[width=6.4cm]{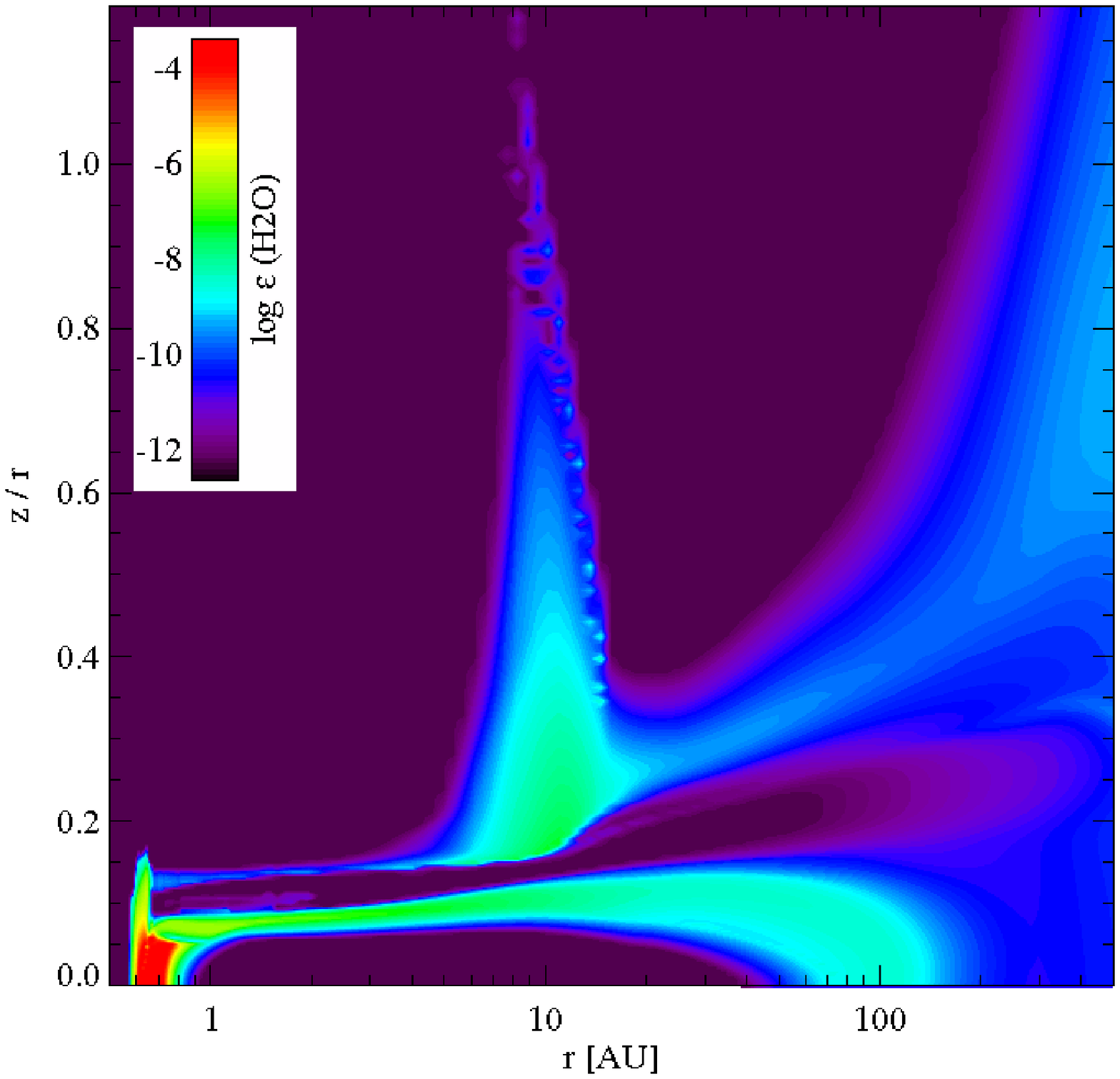} \\ 
    \hspace*{-5mm}\includegraphics[width=6.4cm]{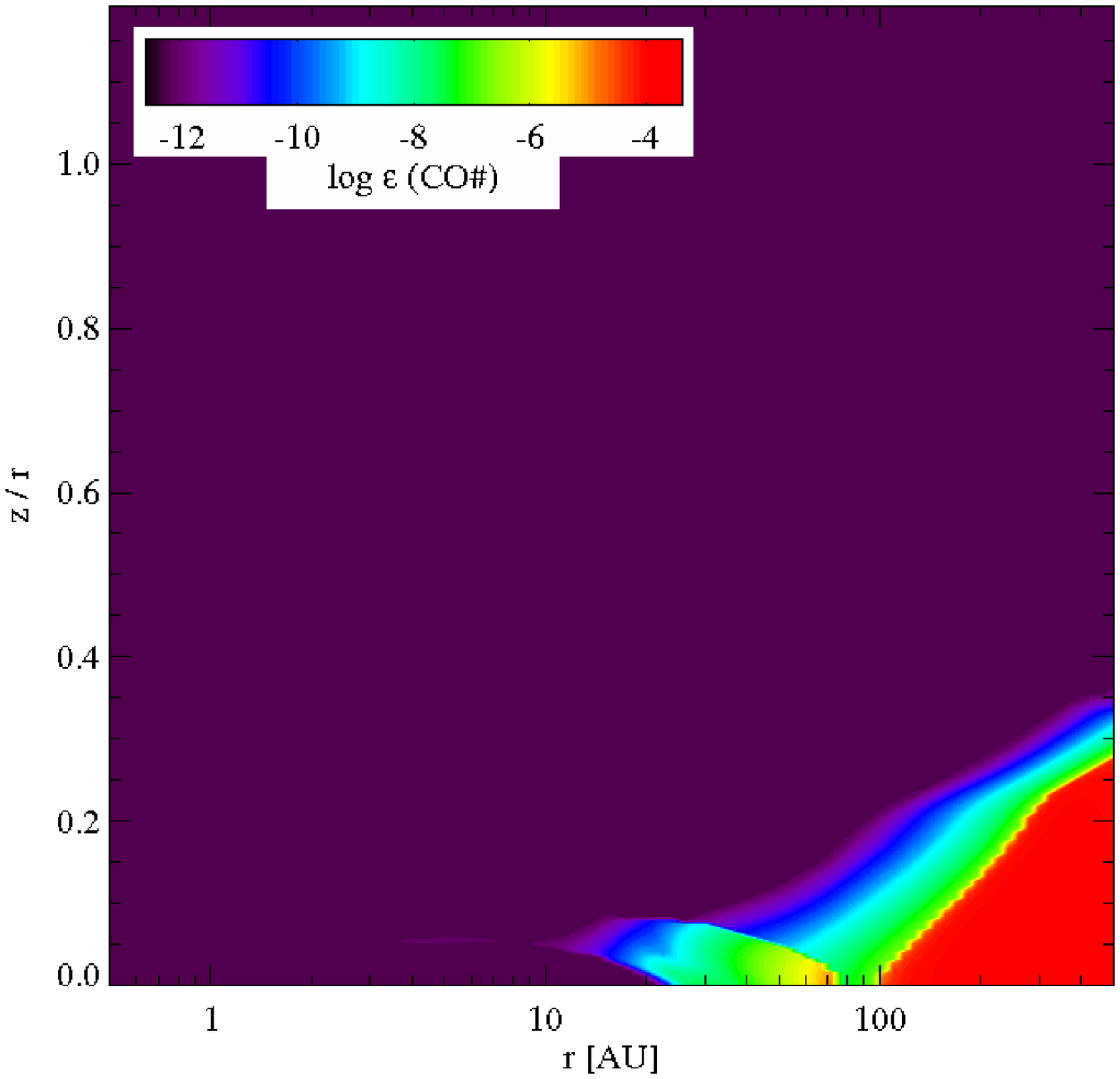} &
    \hspace*{-6mm}\includegraphics[width=6.4cm]{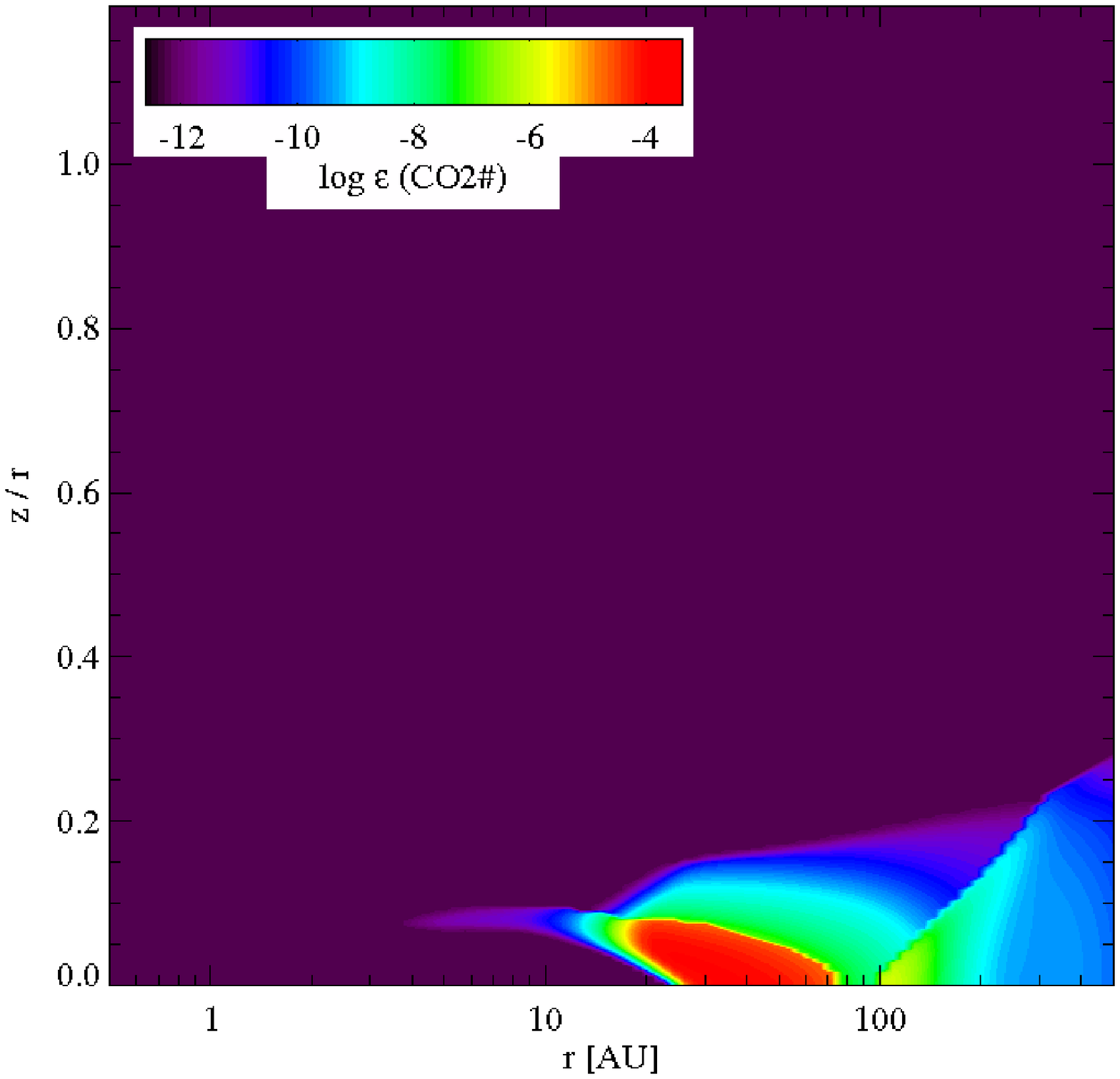} &
    \hspace*{-6mm}\includegraphics[width=6.4cm]{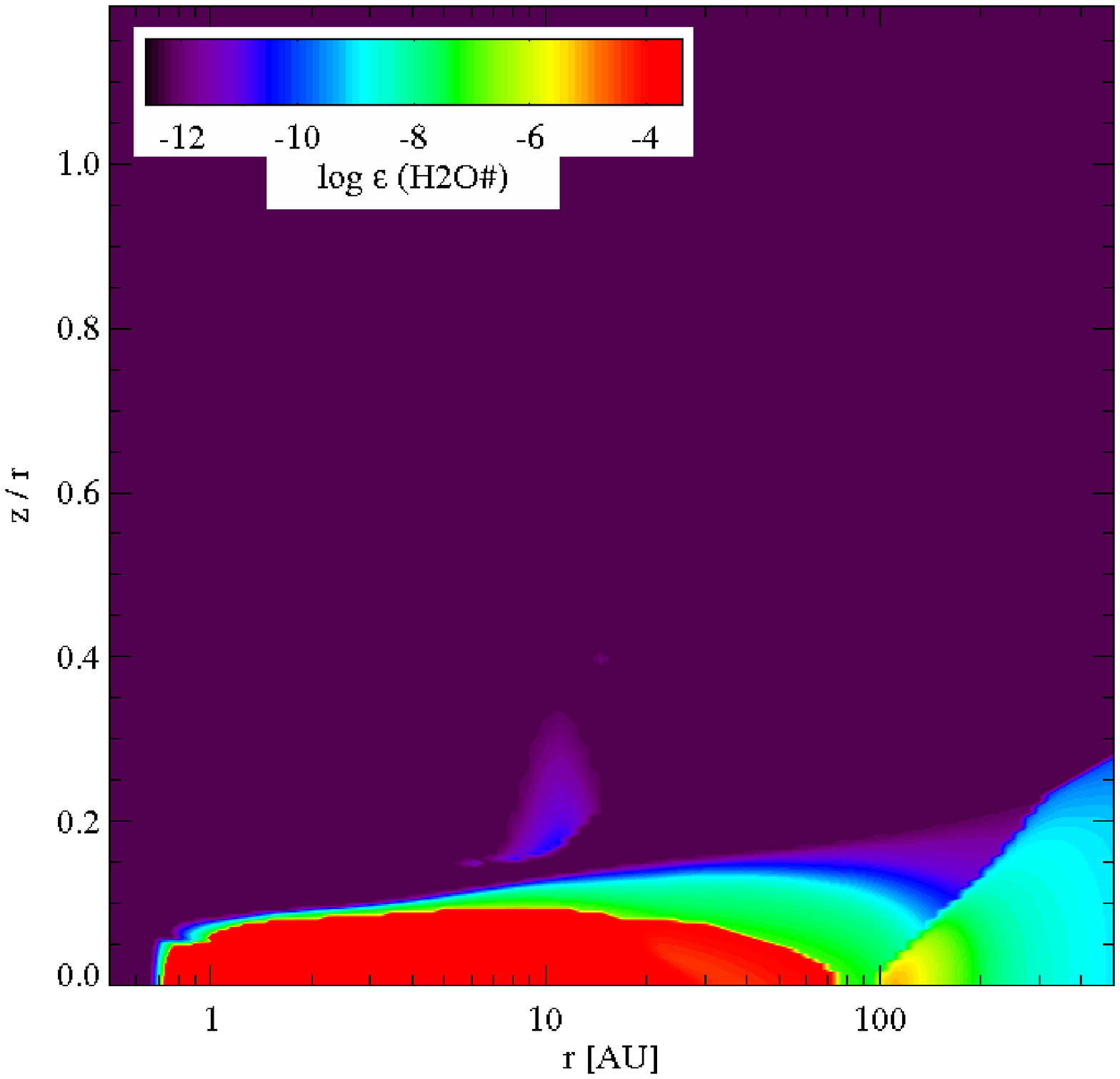} 
  \end{tabular}\\*[-3mm]  
  \caption{Chemical composition of the gas in a T\,Tauri type
  protoplanetary disk with $M_{\rm disk}\!=\!0.01\,M_\odot$, showing
  the concentrations $\epsilon_i\!=\!n_i/n_\HH$, {where $n_i$ is
  the particle density of kind $i$ and $n_\HH$ the total hydrogen
  nuclei density}. Upper row: H, H$_2$ and free electrons, second row:
  C$^+$, C and CO, third row: O, OH and H$_2$O and lower row: H$_2$O
  ice, CO$_2$ ice and CO ice. {Note the different scaling for H, H$_2$
  and e$^-$ in the first row.}}
  \label{fig:Chemistry}
\end{figure*}


The following discussion of the chemical results focuses on
aspects that are relevant for an understanding of the two-dimensional
disk structure. We restrict it to the most important atomic and
molecular cooling species and the species that trace
the dominant carriers of the abundant elements hydrogen, carbon and
oxygen throughout the disk (Fig.~\ref{fig:Chemistry}). A more
detailed discussion of particular chemical aspects and their relevance
to observations will be the topic of future work.

\subsubsection{Atomic and molecular hydrogen}

Inward of about 10\,AU, the H/H$_2$ transition occurs at the lower
boundary of the hot surface layer. There is a very sharp gradient of
UV field and gas temperature explained by the shadow casted by the dust in
the inner rim.  Above the shadow, the gas temperature is high enough
to efficiently destroy molecular hydrogen via $\rm H_2 + H \to 3H$,
and also by collisions with atomic oxygen.

At larger distances, H$_2$ can form on grain surfaces as soon as the
dust temperature drops to about 100\,K, where the formation efficiency 
$\epsilon(\Td)$ increases sharply. This happens primarily in
the secondary puffed-up regions around 10\,AU. The formation of
molecular hydrogen beyond this distance is mainly controlled by H$_2$
self-shielding, which is an intrinsically self-amplifying (\ie
unstable) process.  In addition, the gas density increases by a factor
of $\sim\!2$ when H$_2$ forms at given pressure, which causes
increased collisional H$_2$ formation rates in comparison to the
photo-dissociation rates. This H$_2$ formation instability leads to
local overdense H$_2$-rich regions in an otherwise atomic gas at high
altitudes at about 10\,AU in our model.  Other molecules like OH and
H$_2$O are also affected and these molecules can show even larger
concentration contrasts as compared to H$_2$ which causes the
instability.

\subsubsection{Electron concentration and dead zone}

The electron density in the upper part of the disk is set by the
balance between UV ionizations and electron recombinations of atoms
and molecules. In the UV obscured, cold and icy midplane below
$z/r\!\approx\!0.05$, extending radially from just behind the inner
rim to a distance of about 30\,AU, the electron concentration drops to
values below $10^{-8}$, but cosmic ray ionizations maintain a minimum
electron concentration of $\sim\!10^{-10}$ throughout the disk, because
the vertical hydrogen column densities in this model are insufficient
to absorb the cosmic rays ($N_{\rm H_2}\!\approx\!10^{25}\rm cm^{-2}$
at $r\!=\!3\,$AU). An electron concentration of
$\sim\!10^{-10}$ is two orders of magnitude larger than the minimum
value of $\sim\!10^{-12}$ required to sustain turbulence generation by
magneto-rotational instability (MRI), see \citep{Sano2002}.  Thus, our
model does not possess a ``dead zone'' in the planet forming region,
which is different from studies about massive and compact, actively 
accreting disks \citep[e.g.][]{Ilgner2006c}.

\begin{figure*}
  \centering
  \begin{tabular}{cc}
    \hspace*{-5mm}\includegraphics[width=9cm]{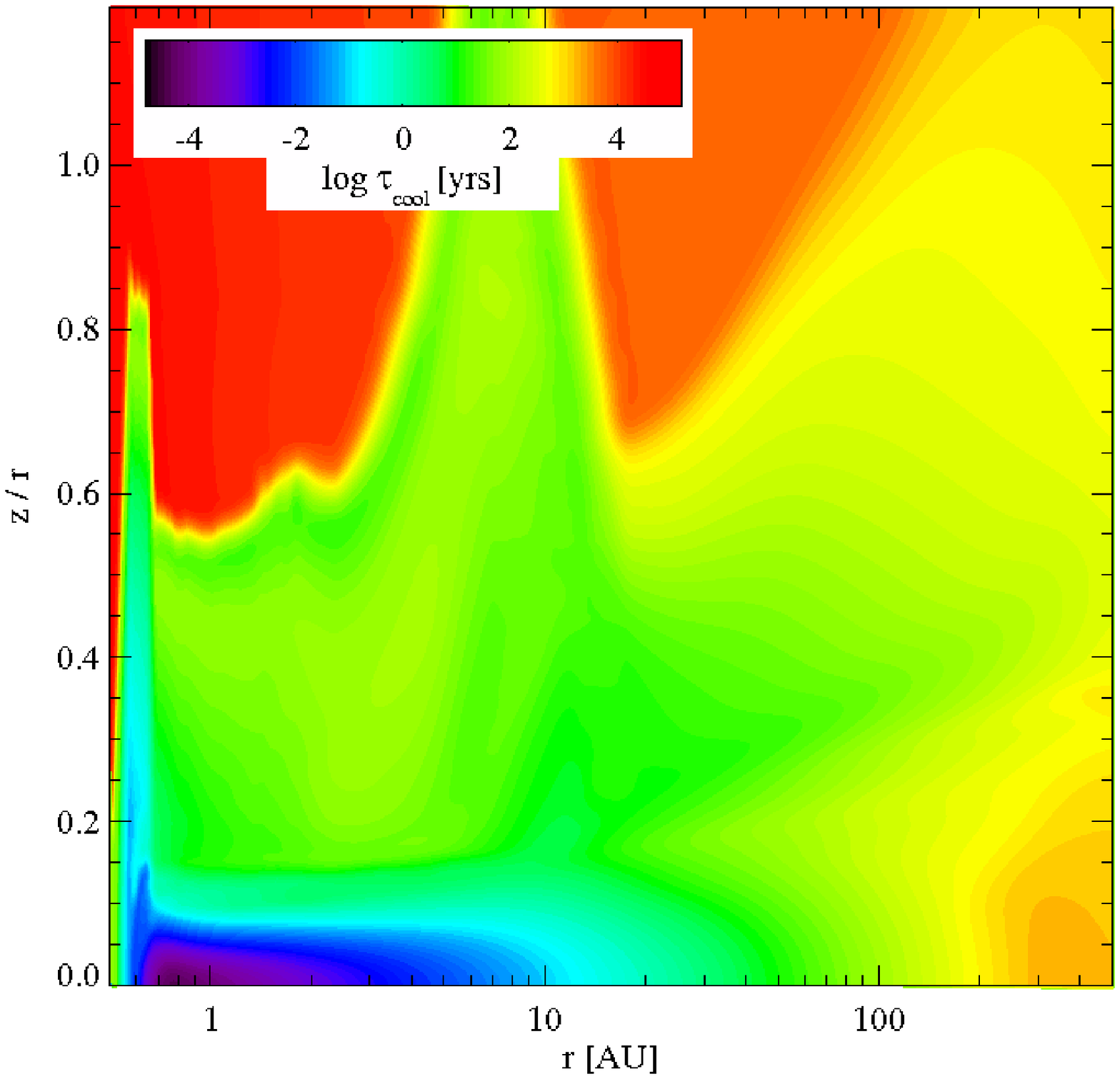} &
    \hspace*{-6mm}\includegraphics[width=9cm]{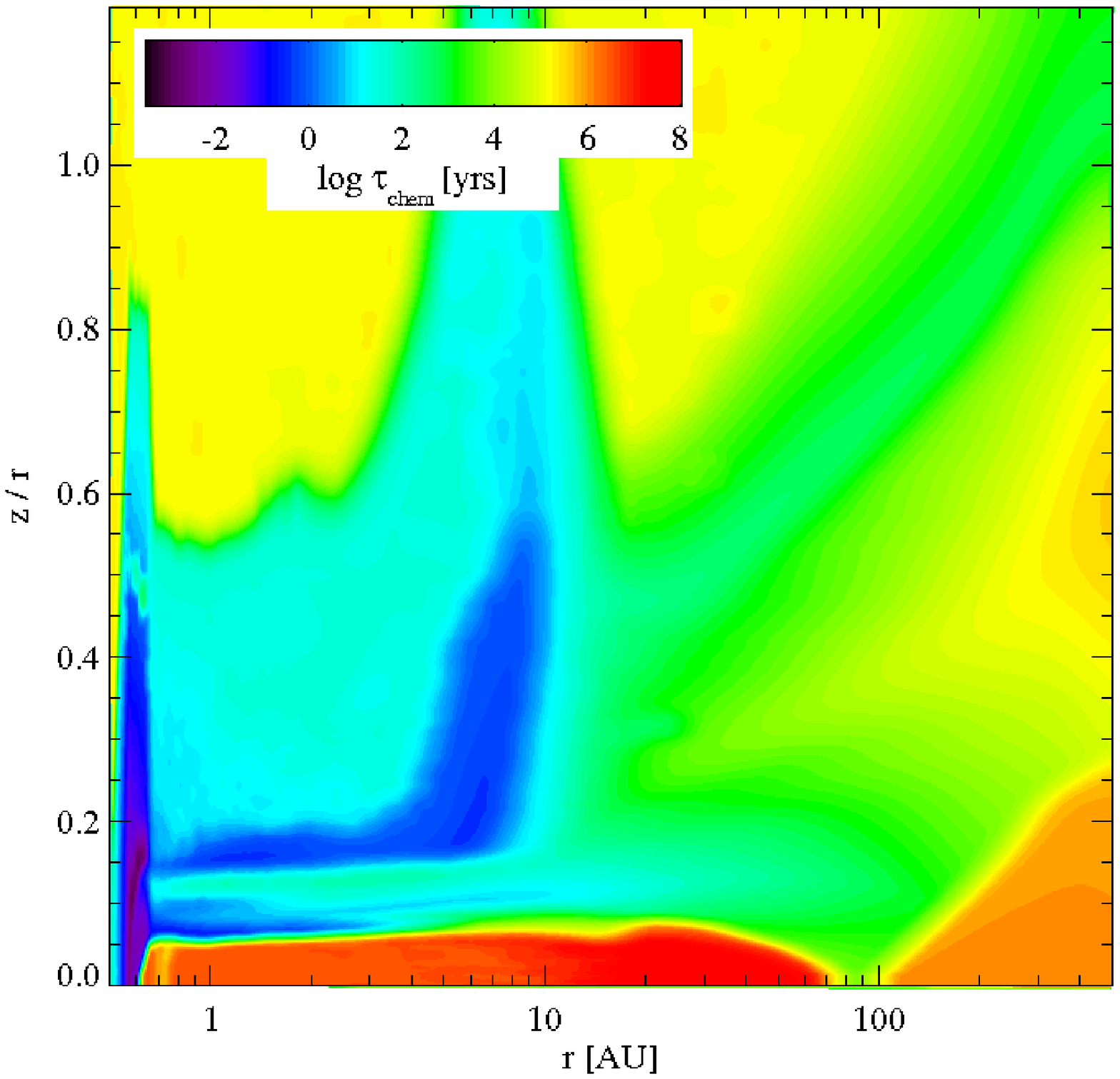} 
  \end{tabular}\\*[-3mm]  
  \caption{Cooling relaxation timescale $\tau_{\rm cool}$ (l.h.s.) and 
          chemical relaxation timescale $\tau_{\rm chem}$ (r.h.s.).}
  \label{fig:timescales}
\end{figure*}

\subsubsection{C$^+\!$, C, O, CO, OH, and H$_2$O}

Outside the shadowed regions, the models clearly show the classical
C$^+$/\,C\,/\,CO\,/\,CO-ice transition as expected from PDR chemistry
\citep[e.g.][]{Kamp2004,Jonkheid2004,Gorti2008}.  However, there are
some important differences to note in the 1-10\,AU range.  The
dominant form of carbon in the midplane is CH$_4$. At those high
densities, oxygen is locked up into H$_2$O-ice, leaving carbon to form
methane instead of CO.  Above the icy regions, in a belt up to
$A_V\!\approx\!10$, water molecules evaporate from the ice and CO
becomes again the dominant carbon and oxygen carrier.

At radial distances between $\approx\!0.8\!-\!10$\,AU in the warm
intermediate layer, the model shows a double layer with high
concentrations of neutral C, OH, H$_2$O and other, partly organic
molecules like CO$_2$ HCN and H$_2$CO (not depicted). This double
layer is a result of the full 2D radiative transfer modelling in
\ProDiMo. The radial UV intensities drop quickly by orders of magnitude
at the position of the inner rim shadow ($z/r\!\approx\!0.13$). The UV
radiation field then stays about constant, until $A_V\!\approx\!1$ is
reached, and also the vertical (+\,scattered) UV intensities
decrease. In combination with the downward decreasing gas temperatures
and increasing gas densities, this produces two layers of hot and cold
OH and H$_2$O molecules with a maximum of C$^+$ in between.
\citet{vanZadelhoff2003} have undertaken similar investigations
showing that dust scattering leads to a a deeper penetration and
redirection of the stellar UV into the vertical direction, with strong
impact on the photo-chemistry.

\subsubsection{Ice formation}

The ice formation is mainly a function of kind, gas density and dust
temperature. Hence, the location of the individual ``ice lines''
strongly depend on the disk dust properties assumed, such as total
grain surface area, disk shape and dust opacity.  Water and CO$_2$ ice
formation is mostly restricted to the midplane, where the densities
are in excess of $10^{10}$\,cm$^{-3}$, the reason being mainly the
reaction pathways leading to the formation of the gaseous molecules
that form these ice species.
In addition, UV desorption counteracts the freeze-out of molecules in
the upper layers at large distances from the star.

Inside 100\,AU, densities are high enough to form water in the gas
phase which subsequently freezes out onto the cold grains
($\Td\!\la\!100$\,K). This is a consequence of our stationary
chemistry that does not care about the intrinsically long timescale
for ice formation (see Fig.~\ref{fig:timescales}). As densities drop
and conditions for water formation in the gas phase become less
favorable, oxygen predominantly forms CO, which freezes out at dust
temperatures below $\sim 25$\,K at large distances. There is an
intermediate density and temperature regime ($20\!-\!100\,$AU), where
significant amounts of CO$_2$-ice are formed.

\subsection{Timescales}

An important question is whether our assumptions of gas energy balance
and kinetic chemical equilibrium are valid in protoplanetary
disks. The cooling relaxation timescale is calculated as
\begin{equation}
  \tau_{\rm cool} = \frac{3p}{2\Tg}\,\bigg|
                    \frac{\partial Q}{\partial \Tg}\bigg|^{-1}
\end{equation}
where $Q\!=\!\Gamma\!-\!\Lambda$ is the net heating rate.  The
chemical relaxation timescale is more complicated. We calculate it as
\begin{equation}
  \tau_{\rm chem} = \max\limits_{{\rm valid}\ n}\,\big|{\rm Re}\{\lambda_n\}^{-1}\big|
\end{equation}
where $\lambda_n$ are the valid eigenvalues of the chemical Jacobian
$\partial F_i/\partial n_j$ (see Eq.\,\ref{eq:chemEq}, without element
conservation). Since $F$ obeys $N_{\rm el}$ auxiliary conditions
(element conservation), there are $N_{\rm el}$ redundant modes which
have (mathematically) zero eigenvalues. Numerics yields extremely
small $\lambda$ for these modes, which must be disregarded.

Figure \ref{fig:timescales} (l.h.s.) shows that the cooling timescale
in the disk is smaller than typical evolution timescales by orders of
magnitude, and also smaller than typical mixing timescales
\citep[e.g.][]{Ilgner2004}, justifying our assumption of thermal
balance. In particular, the gas is thermally tightly coupled to the
dust via thermal accommodation in the midplane regions where the
cooling timescale scales as $\tau_{\rm cool}\!\approx\!1/\rho$.

The chemical relaxation timescales (r.h.s. of
Fig.~\ref{fig:timescales}) show that apart from the icy midplane,
where $\tau_{\rm chem}$ can be as large as $10^8$\,yrs, the chemical
relaxation timescale is typically $10^4$\,yrs or shorter, and as short
as $\approx\!1\!-\!100\,$yrs in the photon-dominated warm intermediate
layer, where most spectral lines form.

\begin{figure*}
  \centering
  \begin{tabular}{cc}
    \hspace*{-5mm}\includegraphics[width=9cm]{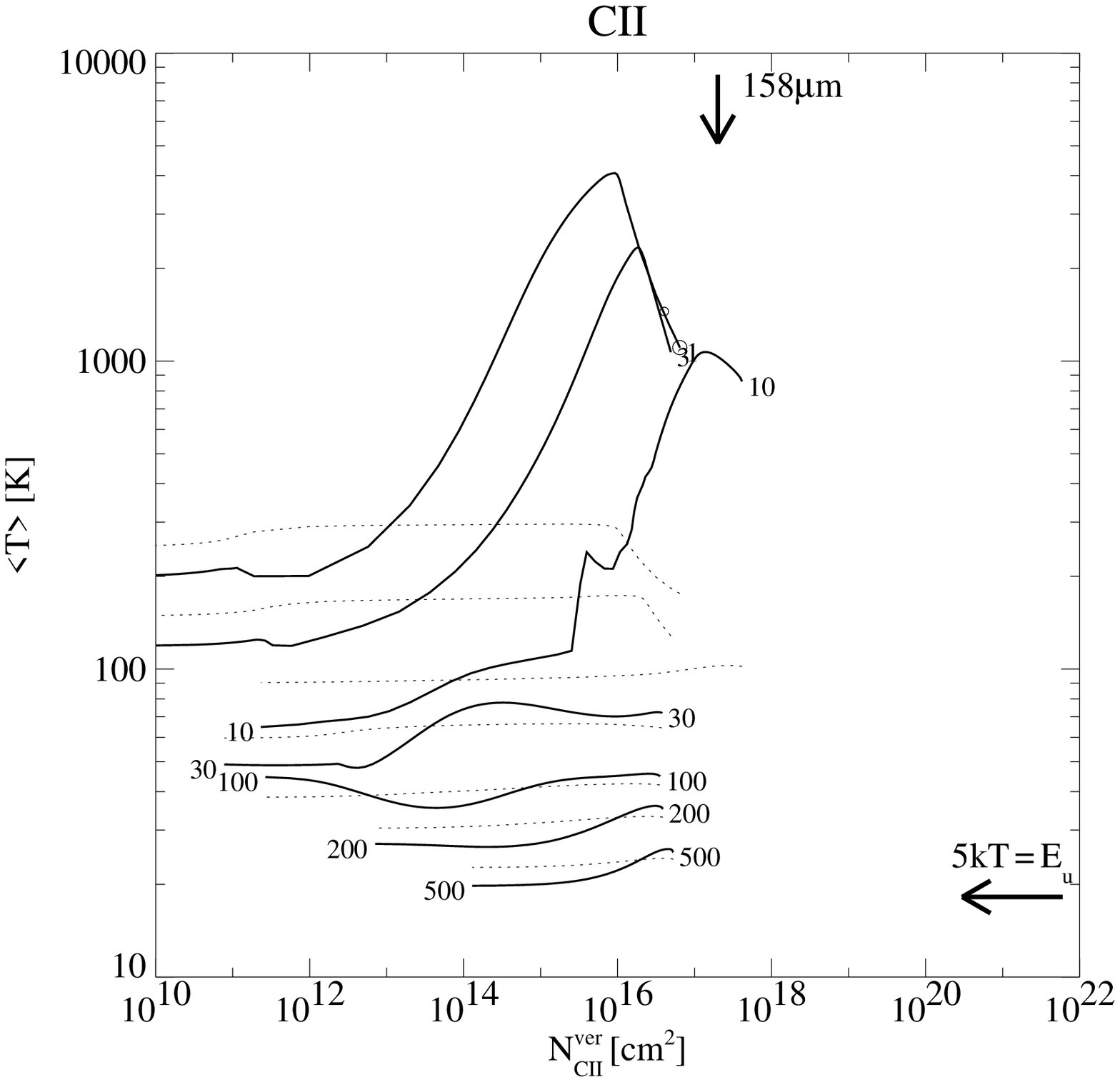} &
    \hspace*{-6mm}\includegraphics[width=9cm]{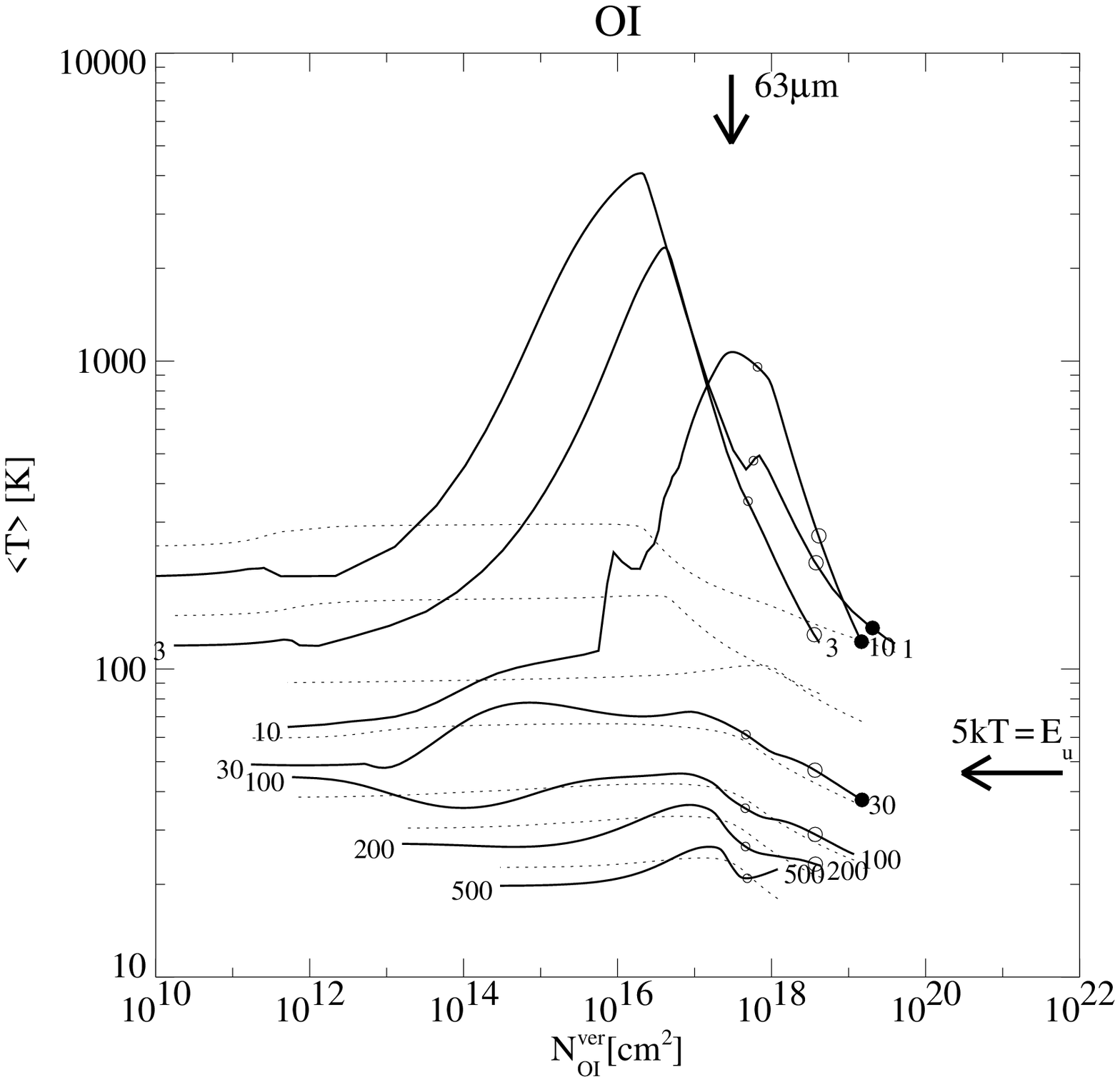} \\
    \hspace*{-5mm}\includegraphics[width=9cm]{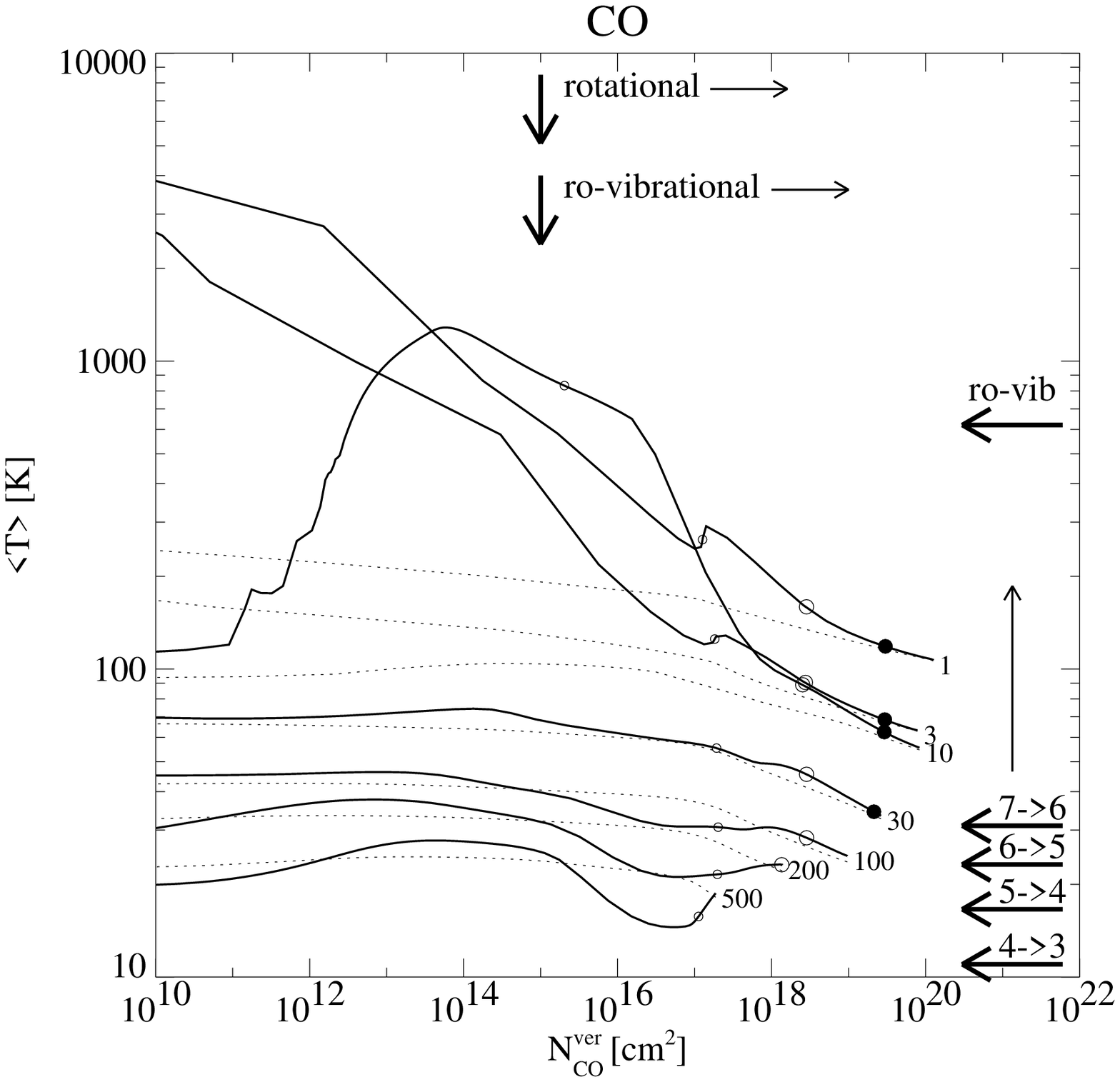} &
    \hspace*{-6mm}\includegraphics[width=9cm]{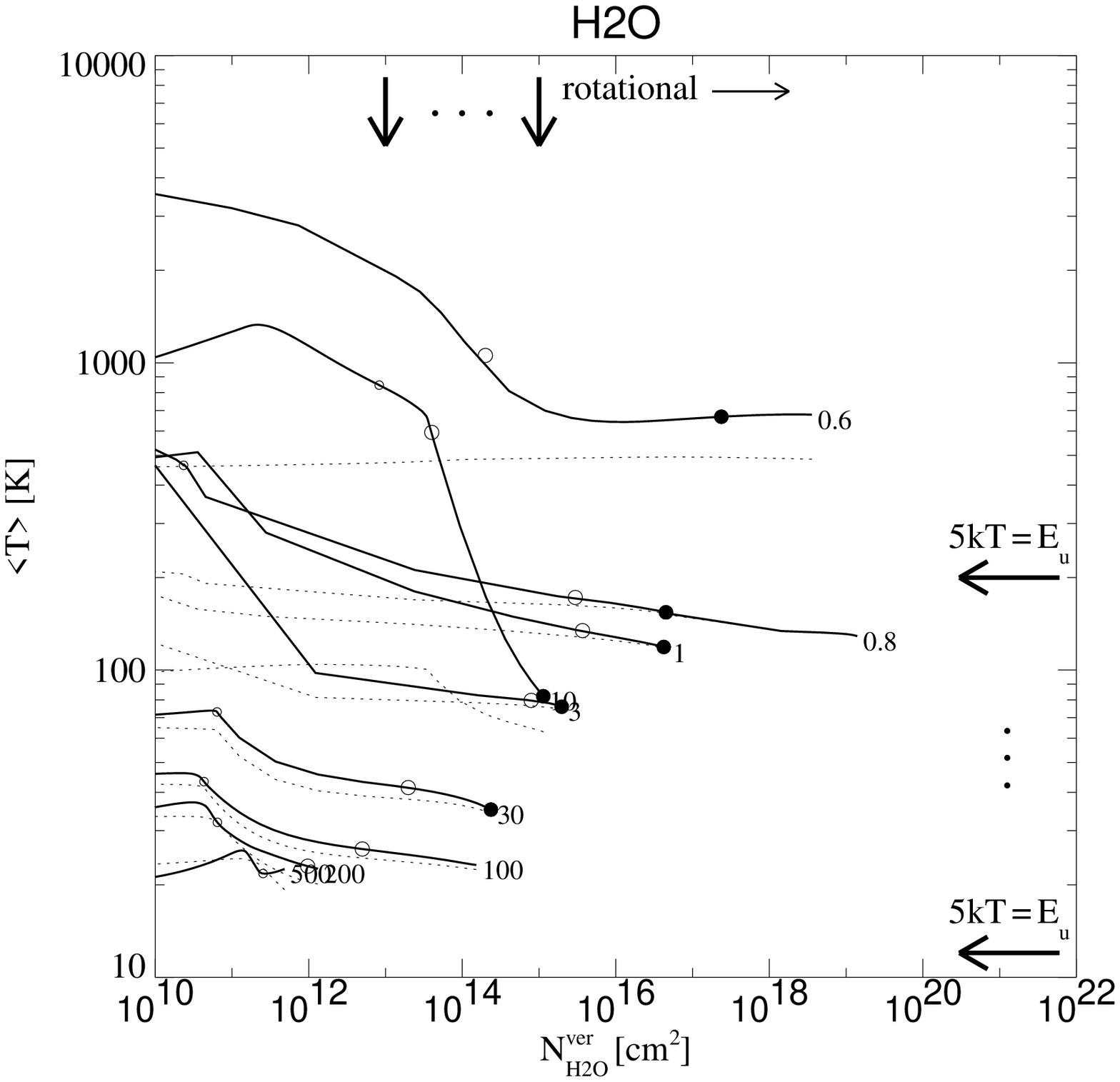} \\
  \end{tabular}\\*[-3mm]  
  \caption{Column density averaged gas temperatures $\langle\Tg\rangle$
    (Eq.\,\ref{eq:Tgcum}, full lines) over vertical species column
    density for different radii labelled by the numbers in AU. The
    thin dashed lines show the column density averaged dust temperatures
    $\langle\Td\rangle$ for comparison. The small open circles, large
    open circles and large full circles mark vertical $A_V\!=\!0.1$,
    $A_V\!=\!1$ and $A_V\!=\!10$, respectively. The vertical arrows at
    the top of the figures mark the species column densities $N_{\rm
    thick}$ where the indicated lines become optically thick
    (Eq.\,\ref{eq:Nthick}). We add some horizontal (rightward) arrows
    for rotational and ro-vibrational lines to demonstrate that
    $N_{\rm thick}$ is larger for high excitation lines. The leftward
    arrows on the r.h.s. of the plots indicate temperatures sufficient
    to thermally excite the upper level of the transitions
    $5\,kT\!\approx\!E_u$.}
  \label{fig:Tcum}
\end{figure*}

\subsection{Gas emission lines}

In order to discuss from which part of the disk the various gas
emission lines come from, we provide some simple estimates of
column densities and excitation temperatures in this section. Full 2D
non-LTE line transfer calculations will be covered in subsequent
papers.

The species column density $N_{\rm thick}$ required to achieve unit
line optical depth can be calculated from Eq.\,(\ref{eq:tauL}).  
Assuming maximum and vanishing population in the lower and upper
level, respectively ($n_l\!\approx\!n_{\rm sp}$, $n_u\!\approx\!0$), 
the result for $\tauL_{ul}\!=\!1$ is
\begin{equation}
  N_{\rm thick} = \frac{g_l}{g_u}\frac{8\pi\nu^3\Delta v_D}{A_{ul}\,c^3}
  \label{eq:Nthick}
\end{equation}
Table~\ref{tab:Nthick} shows some typical values of $N_{\rm thick}$
for permitted atomic resonance-lines, atomic fine-structure lines, and some
rotational and ro-vibrational molecular lines. The actual critical
column density is higher by a factor of $n_{\rm sp}/n_l$ if the
population of the lower level is less than maximum. 
 
Since the emission lines get saturated around $\tau\!\approx\!1$, the
majority of the observable line flux originates from a surface region
of thickness $N^{\rm ver}_{\rm sp}\!\approx\!N_{\rm thick}$. From the
full model, we calculate the vertical species column densities $N^{\rm
ver}_{\rm sp}$ and column density averaged gas temperatures $\langle
\Tg\rangle_{\rm sp}$ defined as
\begin{eqnarray}
  N^{\rm ver}_{\rm sp}(r,z) &=& 
        \int\limits_z^{z_{\rm max}(r)}\!\!n_{\rm sp}(r,z)\,dz  \\
  \langle \Tg\rangle_{\rm sp}(r,N^{\rm ver}_{\rm sp}) &=& 
        \frac{1}{N^{\rm ver}_{\rm sp}(r,z)}
        \int\limits_z^{z_{\rm max}(r)}\!\!\Tg(r,z)\,n_{\rm sp}(r,z)\,dz
  \label{eq:Tgcum}
\end{eqnarray}
Similar to Eq.\,(\ref{eq:Tgcum}), we define the column density averaged dust
temperature $\langle\Td\rangle_{\rm sp}$. The results are
shown in Fig.~\ref{fig:Tcum}. Treating the column above $N^{\rm
ver}_{\rm sp}\!=\!N_{\rm thick}$ as optically thin, ignoring deeper
layers for the line formation, and assuming LTE, the column density averaged
gas temperatures $\langle \Tg\rangle_{\rm sp}$ at depth $N^{\rm
ver}_{\rm sp}\!=\!N_{\rm thick}$ provide an estimate of the expected
excitation temperature of the observable line flux.

\begin{table}
\centering
\caption{Species masses in the disk and emission line characteristics.
  A velocity width of $\Delta v_D\!=\!1$\,km/s is assumed.}
\label{tab:Nthick}
\hspace*{-1mm}\resizebox{\columnwidth}{!}{
\begin{tabular}{cccccc}
\\[-4ex]
\hline
 & $\rm\!\!\!\!\!mass\,[M_{\displaystyle\oplus}]\!\!\!\!\!\!$& type 
                        & $\lambda\rm\,[\mu m]$  
                        & $N_{\rm thick}\rm\,[cm^{-2}]$ 
                        & $E_u\rm\,[K]$\\
\hline
\hline
&&\\[-2.2ex]
C{\sc ii}          & 0.055   & fine-struc.  & $158$      
                        & $2\times 10^{17}$   & 91    \\ 
C{\sc i}           & 0.079   & fine-struc.  & $370$     
                        & $2\times 10^{17}$   & 62    \\
O{\sc i}           & 4.6     & fine-struc.  & $63$    
                        & $3\times 10^{17}$   & 230   \\
CO                 & 4.2     & rotational   & $\!\!\!400-2600\!\!\!$ 
                        & $\sim\!10^{15}$      & $\sim\!5-300$ \\
                   &         & $\!\!$ro-vib. fund.$\!\!$& $4.6$
                        & $\sim\!10^{15}$      & $\sim\!3100$ \\
$\!\!\!\!$H$_2$O$\!\!$ & 0.007  & rotational   & $50-300$
                 & $\!\!\!\sim\!10^{13}-10^{15}$ & $\sim\!60-1000$ \\ 
H$_2$              & 2540    & $\!\!$ro-vib.$\!\!$   & $2-28$     
      & $\!\!\!\sim\!10^{23}-10^{24}$ & $\!\!\!\sim\!500-8000\!\!\!\!\!$ \\
$\!\!\!\!$MgII$\!\!$ & 0.15  & resonance k  & $0.28$   
                        & $6\times 10^{11}$   & 51300 \\
\hline
\end{tabular}}
\end{table}

\smallskip\noindent {\sffamily\itshape CII:}\ \ The most simple case in
Fig.~\ref{fig:Tcum} is the 157.7$\,\mu$m fine-structure line of C{\sc
ii}. The C$^+$ column density reaches a maximum value of
$\sim\!10^{17}\rm\,cm^2$, almost independent of radius $r$. This value
falls just short of $N_{\rm thick}$, meaning that the line remains
mostly optically thin throughout the entire disk. The column of
emitting C$^+$ gas is situated well above the optically thick dust in
the midplane as indicated by the missing $A_V$ marks in
Fig.~\ref{fig:Tcum}. Since the outer regions have a much larger
surface area as compared to the close regions, and the [C{\sc ii}]
157.7$\,\mu$m fine-structure line can easily be excited even in a cold
low density gas out to 500\,AU ($E_u\,{\rm [K]}\!=\!91$, $n_{\rm
cr}\!\approx\!3\times 10^3\,$cm$^{-3}$), the line flux is expected to
be dominated by the outer regions.\ \ \ {\sl The [C{\sc ii}] 157.7$\,\mu$m
line probes the upper flared surface layers of the outer disk.}

\smallskip\noindent {\sffamily\itshape OI:}\ \ The column of atomic oxygen gas
responsible for the [O{\sc i}] 63.2$\,\mu$m fine-structure line extends
deeper into the midplane and reaches column densities of
$\sim\!10^{19}\rm\,cm^{-2}$, \ie well beyond $N_{\rm thick}$.  Therefore,
the line can be expected to be optically thick in most cases. The line
saturates before a dust visual extinction of $A_V\!=\!0.1$ is reached,
\ie also this line mainly probes the conditions above the optically
thick midplane. Due to the higher energy $E_u$ required to excite the
upper level, the line is expected to originate mainly from regions
inward of about 100\,AU. We estimate this radius roughly by the
requirement that columns must provide temperatures $5\,k\Tg\!\ga\!E_u$
(see Fig.~\ref{fig:Tcum}, arrows on r.h.s.). At a column density of
$N^{\rm ver}_{\rm sp}\!=\!N_{\rm thick}$, the line intensity adopts an
excitation temperature of $\approx\!40\!-\!70\,$K ($r\!\ga\!30\,$AU)
and $\approx\!500\!-\!1000\,$K ($r\!\la\!10\,$AU). In these inner
regions, the [O{\sc i}] 63.2$\,\mu$m line partly originates from the hot
atomic surface layer of the disk (see Sect.~\ref{sec:Tstruc}) where
$\langle\Tg\rangle_{\rm O}$ peaks to about $1000\!-\!4000$\,K at column
densities $\sim\!10^{16}...10^{18}\rm\,cm^{-2}$, although eventually the
excitation temperature lowers again as the column of oxygen gas
extends into deeper, cooler layers. The final value of
$\approx\!500\!-\!1000\,$K is $\approx\!5\!-\!10$ times higher than
expected from $\Tg\!=\!\Td$ models (see dashed lines in
Fig.~\ref{fig:Tcum}), stressing the necessity to include the gas
energy balance in models for spectral interpretation.\ \ \ {\sl The
[O{\sc i}] 63.2$\,\mu$m line originates from the thermally decoupled
surface layers inward of about 100\,AU, above $A_V\!\approx\!0.1$.}

\smallskip\noindent {\sffamily\itshape CO:}\ \ Both the lowest rotational line
$1\!\to\!0$ of CO as well as the CO fundamental ro-vibrational line
$(v,J)\!=\!(1,1)\!\to\!(0,0)$ need a column density of about
10$^{15}\rm\,cm^{-2}$ to saturate.  These CO column densities are
exceeded by about $2.5\!-\!4.5$ orders of magnitude at all radii, \ie
these lines are optically thick throughout the entire disk. However,
high-$J$ rotational or high-$J$ ro-vibrational transitions test deeper
layers because $n_l\!\ll\!n_{\rm CO}$ and hence $N_{\rm
thick}\!\gg\!10^{15}\rm\,cm^{-2}$. The observable line intensities are
triggered mainly by the excitation criterion. For ro-vibrational
transitions, for example, temperatures of the order of a 1000\,K are
only available inward of about 10\,AU, where $\langle\Tg\rangle$ has a
strong negative gradient, \ie the emitting CO is situated just below
the borderline between the hot atomic and the warm intermediate disk
layer, and we find values of about $\langle\Tg\rangle_{\rm CO}(N^{\rm
ver}_{\rm CO}\!=\!N_{\rm thick}) \approx 1000\,$K. In the more distant
columns, the ro-vibrational lines are also optically thick but the
line source function $\SL_{ul}$ is very small. The lower rotational
lines until $5\!\to\!4$ originate from the entire disk, but for higher
rotational transitions, the emitting region shrinks remarkably.\ \ \ 
{\sl The CO rotational and ro-vibrational lines are mostly optically
thick and probe very different regions in the disk, depending on the
excitation energy $E_u$.}

\smallskip\noindent {\sffamily\itshape H$_2$O:}\ \ The situation
for the rotational water lines is quite different. Large amounts of
H$_2$O only form in deep layers, typically below $A_V\!=\!1$. {However, between 1\,AU and 100\,AU, the total H$_2$O column densities
are limited by about $10^{14}\!-\!10^{16}\rm\,cm^{-2}$ because of ice
formation in even deeper layers. Since} the rotational water
lines have larger $A_{ul}$ in general and get optically thick sooner,
these column densities seem sufficient to saturate the low-lying
rotational lines out to about 30\,AU. Because gas and dust
temperatures are coupled in the deep layers, the excitation
temperatures of the rotational water lines in LTE with the local dust
temperatures. Therefore, the line-to-continuum ratio might be a
problem for observations.

Higher rotational lines are difficult to excite and the origin of such
lines is probably limited to regions $r\!\la\!1\,$AU. In particular,
the inside of the inner rim is full of water and the temperatures here
are high enough to excite a wealth of high-excitation rotational and
probably also ro-vibrational lines, $\langle\Tg\rangle_{\rm
H_2O}(N^{\rm ver}_{\rm H_2O}\!=\!N_{\rm thick}) \approx 200\,{\rm
K}\!-\!2000\,$K between $r\!=\!0.6\,$AU and $r\!=\!0.8\,$AU. Another
interesting region is the vertically extended zone around 10\,AU (see
Figs.~\ref{fig:dens} and \ref{fig:Chemistry}) which contains some
amounts of hot water. {We are currently investigating the impact
of this hot water layer on the rotational lines as obversable with
Herschel \citep{Woitke2009a}.}\ \ \ {\sl The rotational H$_2$O lines
probe the conditions in the midplane regions $A_V\!\approx\!1\!-\!30$,
the inside of the inner rim, and possibly a vertically extended region
with hot water around $10\,{\rm AU}$ according to this model.}


\section{Conclusions and outlook}

This paper introduces a new code, {\sc ProDiMo}, to model the
physical, chemical and thermal structure of protoplanetary disks.  The
strength of the new code lies in a fully coupled treatment of 2D dust
continuum radiative transfer, gas phase and photo-chemistry, ice
formation, heating\plus cooling balance, and the hydrostatic disk
structure. In particular, we use the calculated radiation field as
input for the photo-chemistry and as background continuum for the
non-LTE modelling of atoms, ions and molecules.  The resulting gas
temperatures determine the vertical disk extension, which in turn
serves as input for the radiative transfer.  Another advantage of the
code is the robustness of its kinetic chemistry module which is applicable to
densities between $10^2$ to $10^{16}\rm\,cm^{-3}$; this makes possible
to model complete disks ranging from about $\sim$\,0.5\,AU to 500\,AU.

\smallskip\noindent {\sf Heating and cooling:}\ \ The heating\plus
cooling balance of the gas is the key to understand the vertical disk
extension. The stellar UV irradiation, both direct and indirect via
scattering, produces hot surface layers, in particular inside of
$\approx\!10\,$AU, even without X-rays. We have included large non-LTE
systems for Fe\,{\sc ii} and CO ro-vibrational line transitions to
counterbalance this heating in the disk regions with high $\rho$ and
high $\Tg$, but find that these transitions also
open new channels of radiative heating by the stellar optical\,--\,IR
irradiation. The disk surface layers close to the star are in fact
often stellar-atmosphere-like and characterized by radiative
equilibrium. More work is required to identify the important
heating\plus cooling processes in these layers and the gas inside
of the inner dust rim.

\smallskip\noindent {\sf Puffed-up inner rim and atomic halo:}\ \
Applying the new concept of ``soft edges'' to the inner rim of a
T\,Tauri disk, the models show a highly puffed-up inner rim extending
up to $z/r\!\approx\!0.7$, and an extended layer of hot
($\Tg\!\approx\!5000\,$K) and thin
($\nH\!\approx\!10^{\,7}\!-\!10^{\,8}\rm\,cm^{-3}$) atomic gas
reaching up to $z/r\!\approx\!0.5$ in the innermost 10\,AU. This
``halo'' is located above the intermediate warm molecular layer
that surrounds the compact, dense, cold and icy midplane. It seems
questionable that the highly puffed-up structures are hydrodynamical
stable and further investigations must show how these features are
related to mixing, winds and gas removal \citep{Alexander2008, StAndrews2008}.

\smallskip\noindent {\sf Scattering and photo-chemistry:}\ \ The dust
grains in the puffed-up inner rim and the halo scatter the stellar UV light
back onto the disk surface, which enhances the photo-chemistry and
the photo-desorption of ice at larger distances.

\smallskip\noindent {\sf Chemistry:}\ \ The surface regions of the
model reveal the classical PDR structure for H$_2$, H, C+, C and
CO. However, due to the full 2D UV radiation transfer in
{\sc ProDiMo}, a complicated multi-layered structure results for
H$_2$O and other organic molecules like CO$_2$ and HCN, which depend
sensitively on the model parameters.  The UV radiation field in the
intermediate warm layer is reduced in two steps, first the puffed-up
inner rim blocks the direct path of the radial photons from the star,
and second the vertical and scattered photons are absorbed in the deeper
layers. We find in particular two layers of hot and cold water
molecules. Mixing by hydrodynamical motions is likely to smooth out
such structures \citep{Ilgner2004, Semenov2006, Tscharnuter2007}.
     
\smallskip\noindent {\sf Cooling and chemical timescales:}\ \ From the
calculated relaxation timescales we conclude that the assumption of
gas thermal balance and kinetic chemical equilibrium should be
sufficient for the interpretation of most gas emission lines, although
mixing may play a role. In the midplane, the chemical timescale 
is as long as $10^8\,$yrs due to ice formation, but the spectral
lines form predominantly in higher layers where $\tau_{\rm
cool}\!\approx\!1\,...\,100\,$yrs and $\tau_{\rm
chem}\!\approx\!1\,...\,10^4\,$yrs, depending on distance.


\smallskip\noindent {\sf Emission lines:}\ \ Disk emission lines
originate mostly from the thermally decoupled surface layers, where
$\Tg\!>\!\Td$. A simple analysis of the line characteristics and
column densities shows that the [C{\sc ii}] 157.7$\,\mu$m
fine-structure line probes the outer flared surface of the disk,
whereas the [O{\sc i}] 63.2$\,\mu$m line originates from slightly
lower layers, but also probes the hot atomic gas inside of 10\,AU. We
identify this line as the key to test our models against
observations.  CO and H$_2$O rotational lines probe the conditions in
the intermediate warm molecular layer and, for low excitation lines,
the outer regions. The origin of most H$_2$O lines is possibly
restricted to regions $\la\!30\,$AU, partly because the water
abundances are higher there, and partly because the lines are
thermally excited only in these layers.

\smallskip\noindent {\sf Observations:}\ \ We intend to apply \ProDiMo
to a large sample of observational disk emission line data that will
be collected by the {\sc Pacs} spectrometer on the {\sc Herschel}
satellite, open time Key Program {\sc Gasps}. Based on calculated
chemical and thermal disk structures, detailed non-LTE line radiative
transfer calculations for the O\,{\sc i} and C\,{\sc ii}
fine-structure lines as well as some CO and H$_2$O molecular lines
will be carried out for analysis. We expect to be able to determine
the gas mass in disks from the line data, and to find tracers for hot
inner layers. In the next decade, \ProDiMo can be used to interpret
{\sc Alma} data which probe the physical conditions and the chemical
composition of the gas in the planet forming regions of protoplanetary
disks.

%
%

%

\bibliography{reference}


\end{document}